\documentclass[12pt, draftclsnofoot, onecolumn]{IEEEtran}
\usepackage[font={small}]{caption}
\usepackage{color}
\usepackage{graphicx}
\usepackage{gensymb}
\usepackage{epstopdf}
\usepackage{cite}
\usepackage{subfig}
\usepackage{amsthm}
\usepackage{amssymb}

\usepackage{epsfig}
\usepackage{amsfonts}
\usepackage{epstopdf}
\usepackage[]{algorithmic}
\usepackage[]{algorithm2e}
\usepackage{listings}
\usepackage{fancybox}
\usepackage{epsfig}
\usepackage{mathtools}
\usepackage{balance}
\usepackage[]{algorithm2e}

\usepackage{amsmath}
\ifCLASSINFOpdf
\else
\fi
\newcommand{\comm}[1]{}
\newcommand{\imj}{\mathsf{j}}
\newcommand{\std}{\mathrm{std}}
\newcommand{\fc}{f_{\mathrm{c}}}
\newcommand{\dx}{\mathrm{d}x}
\newcommand{\dy}{\mathrm{d}y}
\newcommand{\dr}{\mathrm{d}r}
\newcommand{\ds}{\mathrm{d}s}
\newcommand{\dth}{\mathrm{d} \theta}
\newcommand{\dOmg}{\mathrm{d}\Omega}

\newcommand{\Ntx}{N_{\mathrm{tx}}}

\newcommand{\Nsub}{N_{\mathrm{sub}}}

\allowdisplaybreaks

\begin{document}
\title{InFocus: A spatial coding technique to mitigate misfocus in near field LoS beamforming}
\author{{\IEEEauthorblockN{Nitin Jonathan Myers, {\it Member, IEEE} and Robert W. Heath Jr., {\it Fellow, IEEE}. }}
\thanks{ Nitin Jonathan Myers is with Samsung Semiconductor Inc., 5465 Morehouse Dr, San Diego, CA 92121 USA, email:
nitinjmyers@utexas.edu. Robert W. Heath Jr. is with the Department of Electrical and Computer Engineering, North Carolina State University, 890 Oval Dr, Raleigh, NC 27606 USA, email:  rwheathjr@ncsu.edu. This research was supported by the National Science Foundation under grant numbers NSF-CNS-1702800 and NSF-CNS-1731658. }}
\maketitle
\begin{abstract}
Phased arrays, commonly used in IEEE 802.11ad and 5G radios, are capable of focusing radio frequency signals in a specific direction or a spatial region. Beamforming achieves such directional or spatial concentration of signals and enables phased array-based radios to achieve high data rates. Designing beams for millimeter wave and terahertz communication using massive phased arrays, however, is challenging due to hardware constraints and the wide bandwidth in these systems. For example, beams which are optimal at the center frequency may perform poor in wideband communication systems where the radio frequencies differ substantially from the center frequency. The poor performance in such systems is due to differences in the optimal beamformers corresponding to distinct radio frequencies within the wide bandwidth. Such a mismatch leads to a misfocus effect in near field systems and the beam squint effect in far field systems. In this paper, we investigate the misfocus effect and propose InFocus, a low complexity technique to construct beams that are well suited for massive wideband phased arrays. The beams are constructed using a carefully designed frequency modulated waveform in the spatial dimension. InFocus mitigates beam misfocus and beam squint when applied to near field and far field systems.
\end{abstract}
\IEEEpeerreviewmaketitle
\section{Introduction}
\par The progress in circuit technology has allowed wireless communication at higher carrier frequencies where large bandwidths are available. For example, IEEE 802.11ay compliant millimeter wave (mmWave) radios operate at a carrier frequency of $60\, \mathrm{GHz}$ and use a bandwidth of up to $8\, \mathrm{GHz}$ \cite{11ay}. The bandwidth of next generation terahertz (THz) radios is expected to be in the order of several hundreds of $\mathrm{GHz}$ to support emerging applications such as holographic projection, virtual reality, augmented reality, and chip-to-chip communication \cite{THzintro_1}. Due to the small wavelengths in high carrier frequency-based systems, it is possible to integrate large antenna arrays in compact form factors. Beamforming is an important technology in these systems that allows the use of large antenna arrays to focus radio frequency (RF) signals. A good beamforming technique results in a sufficient link margin at the receiver and high data rates.
\par Beamforming in near field systems where the transceiver distance is smaller than the Fraunhofer distance \cite{Fraunhof}, is different from the conventional far field counterpart. While far field beams focus RF signals along a direction, the near field beams focus them in a spatial region \cite{NFbeams}. The center of this spatial region is defined as the focal point. Near field beamforming is useful for short range communication which is typical in data centers, wearable networks, and kiosk downloading stations. The phased array, i.e., an array of phase shifters, is a hardware architecture that is commonly used for beamforming. Similar to beam steering in a far field system, the phased array can steer the focal point electronically in a near field scenario \cite{NF_focus}. In this paper, we consider a line-of-sight (LoS) scenario and study near field beamforming using such phased arrays.
\par Standard near field beamforming solutions that are based on the center frequency suffer from the misfocus effect in phased arrays. In phased arrays, the phase shifts may be realized by first scaling the in-phase and the quadrature-phase signals using digitally controlled variable gain amplifiers, and then adding the scaled signals \cite{RF_Phase_shift_VGA}. In this case, the phase shifts are constant across the entire frequency band and the resultant beamforming weights are frequency flat. In large antenna arrays that operate at high bandwidths, however, the array response varies substantially with the frequency \cite{beamsquint_intro_1}. The standard beamforming approach tunes the beamforming weights according to the array response at the center frequency. Although such an approach maximizes the signal power at the center frequency, it results in a reduced beamforming gain at frequencies that are different from this frequency \cite{beamsquint_intro_2}. The poor gain is because the focal point in the standard beamforming method shifts with the frequency of the RF signal. We call this the misfocus effect. Misfocus limits the effective operating bandwidth of the phased array and leads to poor performance in wideband systems.
\par A common approach to mitigate misfocus is to use true time delay (TTD)-based arrays where the delay of the RF signal can be electronically controlled at each antenna \cite{TTD_intro_1}. One approach to realize a TTD-based array is using the Rotman lens \cite{Lens_rotman_1,Lens_rotman_2}. The beamforming weight realized with such an array is frequency selective and helps achieve large gains over wide bandwidths. Unfortunately, TTD-based arrays result in a higher implementation cost, occupy a larger area, and require a higher power consumption than typical phase shifter-based arrays \cite{mingming_gc,STBC_letters}. A delay-phase precoding (DPP) architecture that combines both TTD elements and phase shifters was proposed in \cite{tan2019delay} to alleviate beam squint. Although the TDD- and the DPP-based designs are interesting solutions proposed by the circuits community, the question is if it is possible to use just phase shifter-based arrays over wide bandwidths. In this paper, we answer this question by proposing a signal processing solution called InFocus to mitigate the misfocus problem in phase shifter-based arrays. 
\par InFocus constructs beams that achieve robustness to misfocus when compared to standard beamforming. Constructing such beams, equivalently the phase shifts, is a hard problem due to hardware constraints. For example, the constant magnitude constraint on the beamforming weights is common in phase shifter-based implementations. In addition, only a discrete set of phase shifts can be applied at the antennas due to the finite resolution of phase shifters. The near field characteristic of the channel response in short range scenarios further complicates the problem. To the best of our knowledge, prior work has not studied beamformer design for misfocus mitigation in near field phased arrays. We would like to mention that InFocus can also be used in far field systems for robustness to the beam squint effect. This is because far field systems can be interpreted as a special instance of a near field system where the transceiver distance is sufficiently large. 
\par The beam squint effect in far field phased arrays is analogous to the misfocus effect. Beam squint occurs because the beamforming direction of standard beams changes with the operating frequency \cite{beamsquint_intro_1}. As a result, it limits the effective operating bandwidth. Prior work has addressed the beam squint problem in far field system by designing denser beamforming codebooks \cite{mingming_gc}. A semidefinite relaxation (SDR)-based beam optimization technique was proposed in \cite{STBC_letters} for far field systems. Although the SDR-based approach in \cite{STBC_letters} can be applied for misfocus mitigation in the near field setting, it has a high complexity when compared to our method. The technique in \cite{STBC_letters} involves convex optimization over an $N^2$ dimensional variable for an $N$ element antenna array. In this paper, we consider planar arrays with $N=0.1$ million antennas. The procedure in \cite{STBC_letters} requires an unreasonably high complexity for such dimensions. Therefore, it is important to develop robust beam design techniques that are scalable to large antenna arrays.
\par In this paper, we develop InFocus to design misfocus robust beamformers for a short range LoS setting with a phased array-based transmitter (TX) and a single antenna receiver (RX). We summarize the main contributions of our work as follows.
\begin{itemize}
\item We investigate the beamforming capability of a short range LoS system which uses a circular planar phased array. Then, we determine conditions on the array size, the transceiver distance, and the bandwidth for which standard beamforming results in a misfocus effect. 
\item We construct a spatial phase modulation function to mitigate the misfocus effect for receivers on the boresight of the transmit array. Our construction uses a spatial frequency modulated continuous waveform (FMCW) chirp along the radial dimension of the transmit array. We show how the parameters of this chirp can be derived from the array geometry and the operating bandwidth.
\item For receivers that are not on the boresight of the TX array, we design a new phase modulation function to mitigate misfocus. The designed function has a non-linear frequency modulation profile which is determined using the stationary phase method. 
\item We evaluate the performance with InFocus-based beams and compare it with the standard beamforming method. Our results indicate that the beams designed with our approach result in a large and approximately flat beamforming gain over a wide bandwidth, and enable massive phased arrays to achieve a higher data rate than comparable techniques. 
\end{itemize}
InFocus does not require any iterative optimization and has a low complexity than the approach in \cite{STBC_letters}. In this paper, we assume that the location of the RX is known at the TX and focus on the misfocus robust transmit beam alignment problem. This location information may be available through near field channel estimation or localization algorithms \cite{NF_chest, Henk_nf_localization}. We do not model any reflections in the propagation environment. Extending our method to richer propagation scenarios is left for future work. An implementation of our technique is available on our github page \cite{Infocus_git}.
\par The rest of the paper is organized as follows. In Section \ref{sec:syschanmodel}, we describe the system and channel model in a phased array-based system. We discuss near field beamforming in Section \ref{sec:nearfieldintro} and the misfocus effect in Section \ref{sec:misfocusintro}. Section \ref{sec:mainsecInfocus} is the main technical section of the paper, where we explain the InFocus technique. We describe our FMCW chirp-based design and the use of the stationary phase method in Section \ref{sec:mainsecInfocus}. Simulation results are presented in Section \ref{sec:simulations}, before the conclusions and future work in Section \ref{sec:concl_future}.
\section{System and channel model} \label{sec:syschanmodel}
\par
\begin{figure}[h!]
\vspace{-2mm}
\centering
\subfloat[A circular planar array.]{\includegraphics[trim=0cm 0cm 0cm 0cm,clip=true,width=4cm, height=4cm]{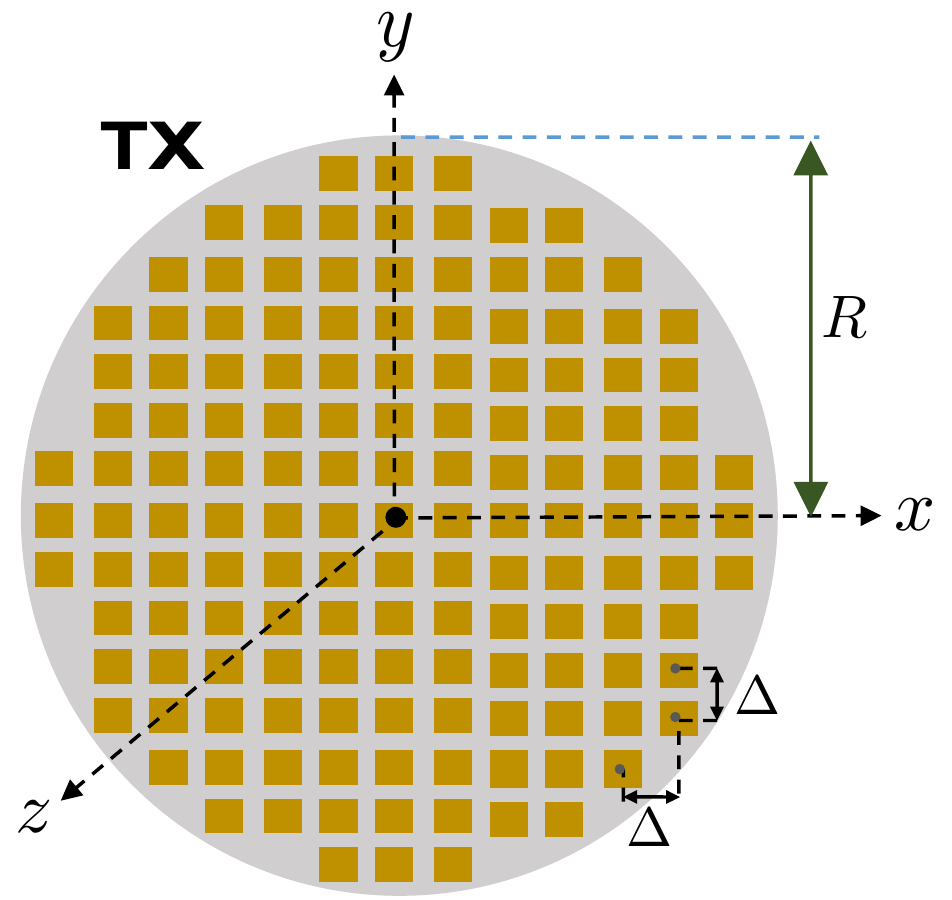}\label{fig:CUPA}}
\:\:\: \: \:\: \:
\subfloat[An LoS communication scenario]{\includegraphics[trim=0cm 0cm 0cm 0cm,clip=true,width=5.5cm, height=4cm]{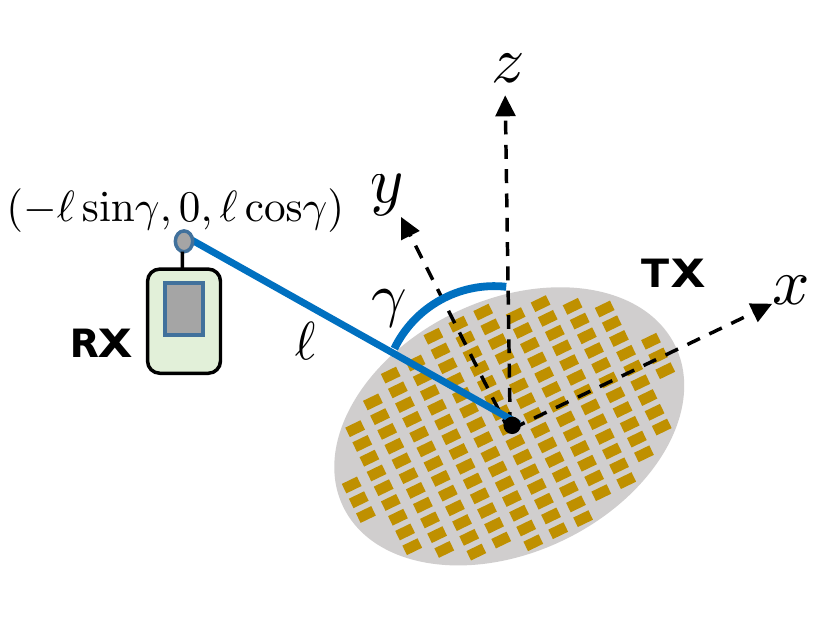}\label{fig:LOS}}
\caption{The figure shows an LoS communication system with a 2D-circular planar array of radius $R$ at the TX and a single antenna RX. We assume that the TX and the RX lie on the $xy$ and $xz$ planes. The line joining the center of the TX and the RX is of length $\ell$. This line makes an angle $\gamma$ with the boresight direction, i.e., the $z-$axis.
\normalsize}
\end{figure}

 We consider a wireless system operating over a bandwidth of $B$ around a carrier frequency of $f_{\mathrm{c}}$. The wavelength at this carrier frequency is $\lambda_{\mathrm{c}}=c/f_{\mathrm{c}}$ where $c$ is the speed of light. We consider a circular planar antenna array of radius $R$ at the transmitter (TX) as shown in Fig. \ref{fig:CUPA}. The TX array lies in the $xy$ plane and has $\Ntx$ isotropic antennas. We define $\Delta$ as the spacing between successive antenna elements along the $x$ and $y$ dimensions. The coordinate corresponding to the center of the TX array is defined as the origin $(0,0,0)$. The set of 2D coordinates associated with the antennas in the array is defined as $\mathcal{S}_{\mathrm{D}}$. The set $\mathcal{S}_{\mathrm{D}}$ has $\Ntx$ coordinates that satisfy $x^2+y^2 \leq R^2$ and $z=0$. The RF front end at the TX is comprised of a single RF chain and a network of $\Ntx$ phase shifters. By configuring the phase shifters in the phased array, the TX can direct its RF signals to maximize signal power at the RX.
\par We consider a single antenna receiver in an LoS scenario and focus on the transmit beamforming problem. Extending InFocus to mitigate beam misfocus in receivers with large arrays is an interesting research direction. We assume that the RX is at a distance of $\ell$ from the center of the TX array. The ray joining the origin and the RX makes an angle $\gamma$ with the normal to the transmit array, as shown in Fig. \ref{fig:LOS}. We also assume that the RX lies in the $xz$ plane without loss of generality. Such an assumption is reasonable due to the ``symmetric'' nature of the circular transmit array. The location of the RX is then $( -\ell \, \mathrm{sin} \,\gamma,\,0,\, \ell \, \mathrm{cos}\,\gamma)$. An LoS system is considered as near field if the transceiver distance is smaller than the Fraunhofer distance, i.e., $d \ll 8R^2/ \lambda_{\mathrm{c}}$ \cite{Fraunhof}. Communication in the near field regime is common in short range applications which use large antenna arrays and high carrier frequencies. One such application is kiosk downloading in the mmWave or the terahertz bands where the transceiver distance is about $50\, \mathrm{cm}$ \cite{kiosk_app}. This distance can be smaller than the far field limit of large phased arrays which operate at high carrier frequencies. For example, a $10 \, \mathrm{cm} \times 10 \, \mathrm{cm}$ antenna array in a $60\, \mathrm{GHz}$ kiosk results in a Fraunhofer distance of $400\, \mathrm{cm}$. Prior work on reflectarrays has experimentally demonstrated such an antenna array with $25600$ elements for near field beamforming \cite{NF_focus}. In this paper, we design beams to enable efficient wideband data transmission in such near field systems. 
\par Now, we explain the wideband multiple-input single-output (MISO) channel model for the near field LoS system shown in Fig. \ref{fig:LOS}. We use $h(x,y,f)$ to denote the channel between the TX antenna at $(x,y) \in \mathcal{S}_{\mathrm{D}}$ and the receive antenna for a frequency of $f$. The distance between these antennas is defined as 
\begin{equation}
\label{eq:ellxy_defn}
\ell(x,y)=\sqrt{(x+\ell\, \mathrm{sin} \,\gamma)^2+y^2+\ell^2\, \mathrm{cos}^2 \,\gamma}.
\end{equation}
For the LoS setting in Fig. \ref{fig:LOS}, $h(x,y,f)$ can be expressed as \cite{channel_model}
\begin{equation}
\label{eq:channel_freq}
h(x,y,f)=\frac{c}{2\pi f \ell(x,y)} e^{-\imj \, 2 \pi f\ell(x,y)/c},
\end{equation}
where $\imj =\sqrt{-1}$. The channel model in \eqref{eq:channel_freq} does not account for reflections or scattering in the propagation environment. Furthermore, the model ignores frequency dependent atmospheric absorption effects \cite{noise_psd}. These assumptions simplify the robust beamforming problem and lead to a closed form expression for the beamforming weights. 
\par We now discuss transmit beamforming in a short range LoS system. With the phased array architecture, the TX can control the RF signal transmitted from an antenna using a beamforming weight. We use $w(x,y)$ to denote the beamforming weight applied to the TX antenna at $(x,y) \in \mathcal{S}_{\mathrm{D}}$. The $\Ntx$ RF signals transmitted by the TX sum at the RX. In this case, we define $g(f)$ as the equivalent frequency domain single-input single-output (SISO) channel between the TX and the RX. The equivalent channel at frequency $f$ is given by
\begin{equation}
\label{eq:eqsiso}
g(f)=\sum_{(x,y) \in \mathcal{S}_{\mathrm{D}}} w(x,y) h(x,y,f).
\end{equation} 
We assume that the TX can only change the phase of the RF signals at the antenna. For a phase shift of $\phi(x,y)$, the beamforming weight with appropriate power normalization is 
\begin{equation}
\label{eq:w_phase_shift}
w(x,y)=\frac{e^{\imj \phi(x,y)}}{\sqrt{\Ntx}}.
\end{equation}
The phase profile used by the TX array is $\{\phi(x,y)\}_{(x,y) \in \mathcal{S}_{\mathrm{D}}}$ and the corresponding beamformer is  $\{w(x,y)\}_{(x,y) \in \mathcal{S}_{\mathrm{D}}}$. Practical phased array-based implementations only allow a coarse phase control of the RF signal due to the finite resolution of the phase shifters. In this paper, we first consider fine phase control to design phase profiles that result in misfocus robust beams. Then, the designed phase profiles are quantized according to the resolution of the phase shifters and are applied at the TX.
\par An ideal beamformer, equivalently the phase profile, is one that maximizes $|g(f)|$ at all frequencies within the desired bandwidth. Such a beamformer must be frequency selective as the channel in \eqref{eq:channel_freq} varies with the frequency. Unfortunately, the frequency independent constraint on $w(x,y)$ which is common in phase shifter-based arrays \cite{const_phase_shift,RF_Phase_shift_VGA}, does not allow the application of the ideal beamformer \cite{squint_60GHz_TMT}. Recent work has shown that frequency selective beamforming can be achieved with true time delay-based beamforming architectures \cite{TTD_paper_1,TTD_paper_2,TTD_paper_3}. Although TTD-based arrays are robust to misfocus and beam squint, they require a higher implementation complexity than traditional phase shifter-based arrays \cite{perera2018wideband}. Furthermore, TTD elements usually result in a higher insertion loss than phase shifters \cite{mingming_gc}. Recent work on mmWave beamforming at $28 \, \mathrm{GHz}$ has reported insertion losses of about $10\, \mathrm{dB}$ for a TTD element \cite{ttd_loss} versus $6\, \mathrm{dB}$ for a phase shifter \cite{phase_shift_loss}. In this paper, we focus on phase shifter-based arrays and construct new beams that achieve robustness to misfocus and beam squint. 
\par Our design ignores the orientation of the receive antenna and assumes that the transmit and receive antennas are co-polarized. Although such an assumption is made for simplicity of exposition, our solution can be applied in practical settings. In dual polarization beamforming systems, the misfocus robust beams derived in this paper can be used along both the horizontal and vertical polarization dimensions. Another approach to mitigate orientation mismatch is by using dynamic polarization control devices at the TX or the RX \cite{DPC}. These devices can electronically change the polarization angle of the transmitted or the received signal, and allow applying the proposed beamforming solutions even under an orientation mismatch. 
\section{Beamforming in near field systems demystified} \label{sec:nearfieldintro}
\par In this section, we explain the phase profiles associated with near field beamforming based on the center frequency. We also approximate the discrete sum associated with $g(f)$ in \eqref{eq:eqsiso} to an integral. In Section \ref{sec:mainsecInfocus}, we will use this approximation for a tractable design of the misfocus robust beamformers.  
\par
\begin{figure}[h!]
\vspace{-2mm}
\centering
\subfloat[Example of a boresight scenario]{\includegraphics[trim=0cm 0cm 0cm 0cm,clip=true,width=4.5cm, height=4.5cm]{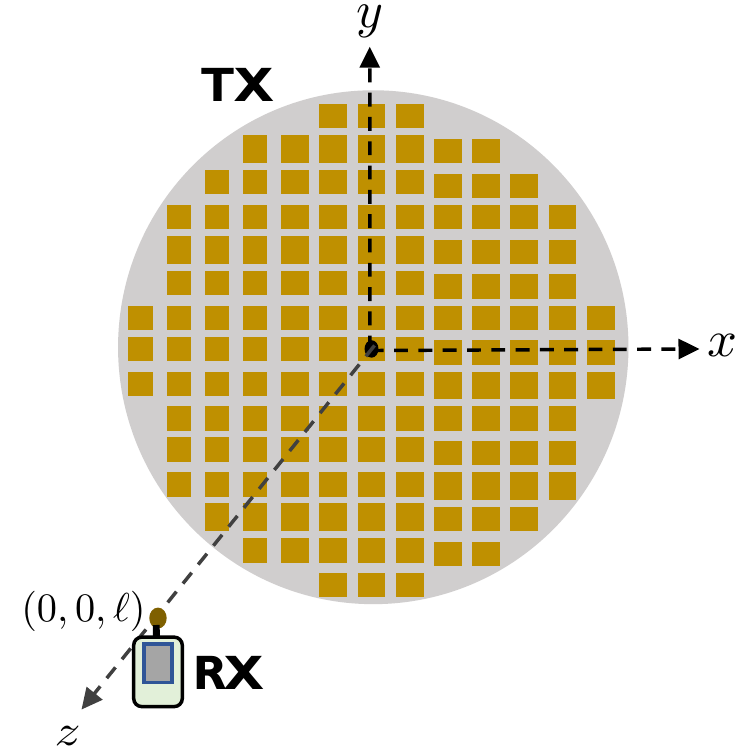}\label{fig:boresightpic}}
\:\:\:
\subfloat[Phase profile $\phi_{\std}(x,y)$]{\includegraphics[trim=1cm 6.25cm 2.5cm 7.5cm,clip=true,width=4.5cm, height=4.5cm]{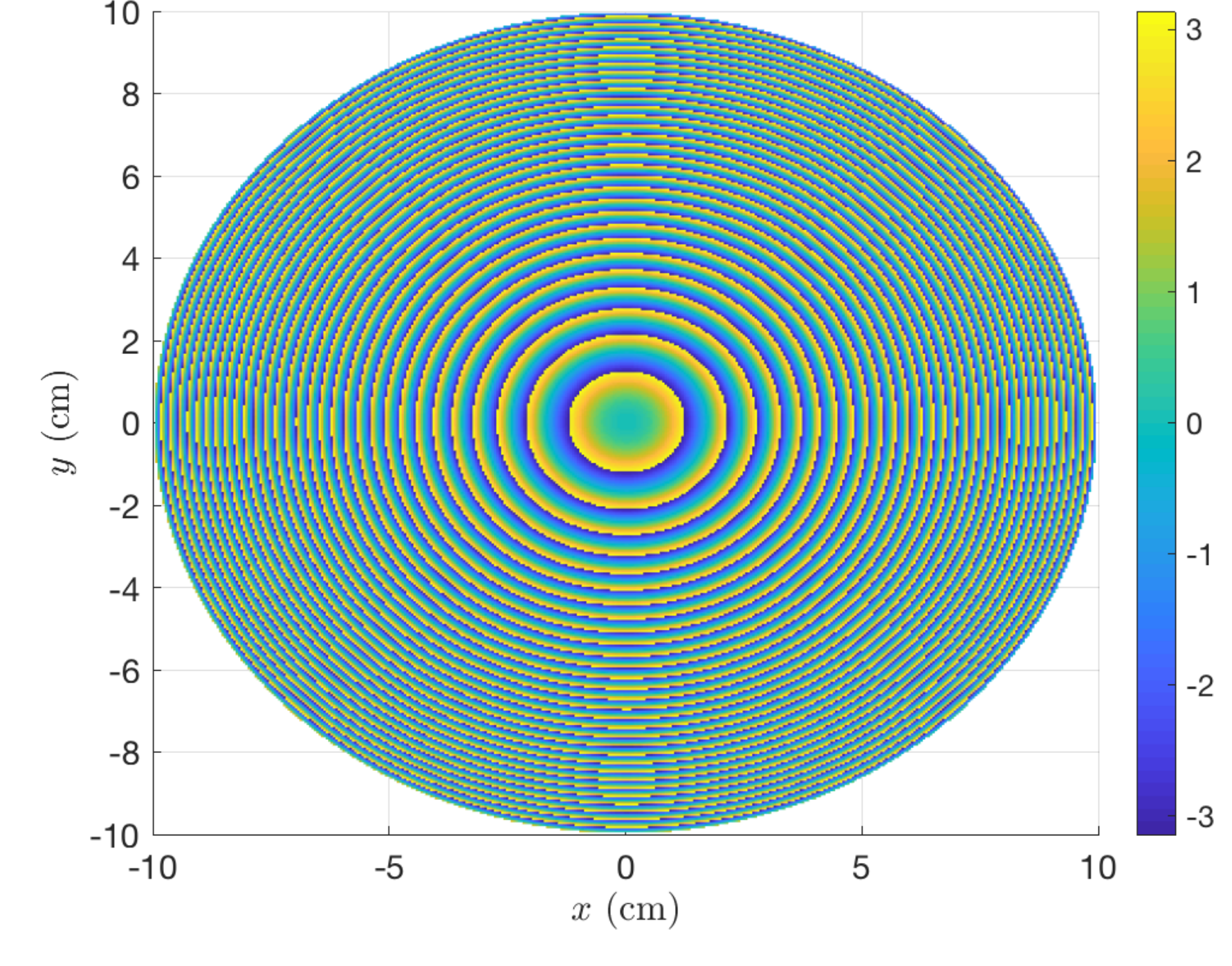}\label{fig:phi_std}}
\:\:\:
\subfloat[Received power with $\phi_{\std}(x,y)$]{\includegraphics[trim=1cm 6.25cm 2.5cm 7.5cm,clip=true,width=4.5cm, height=4.5cm]{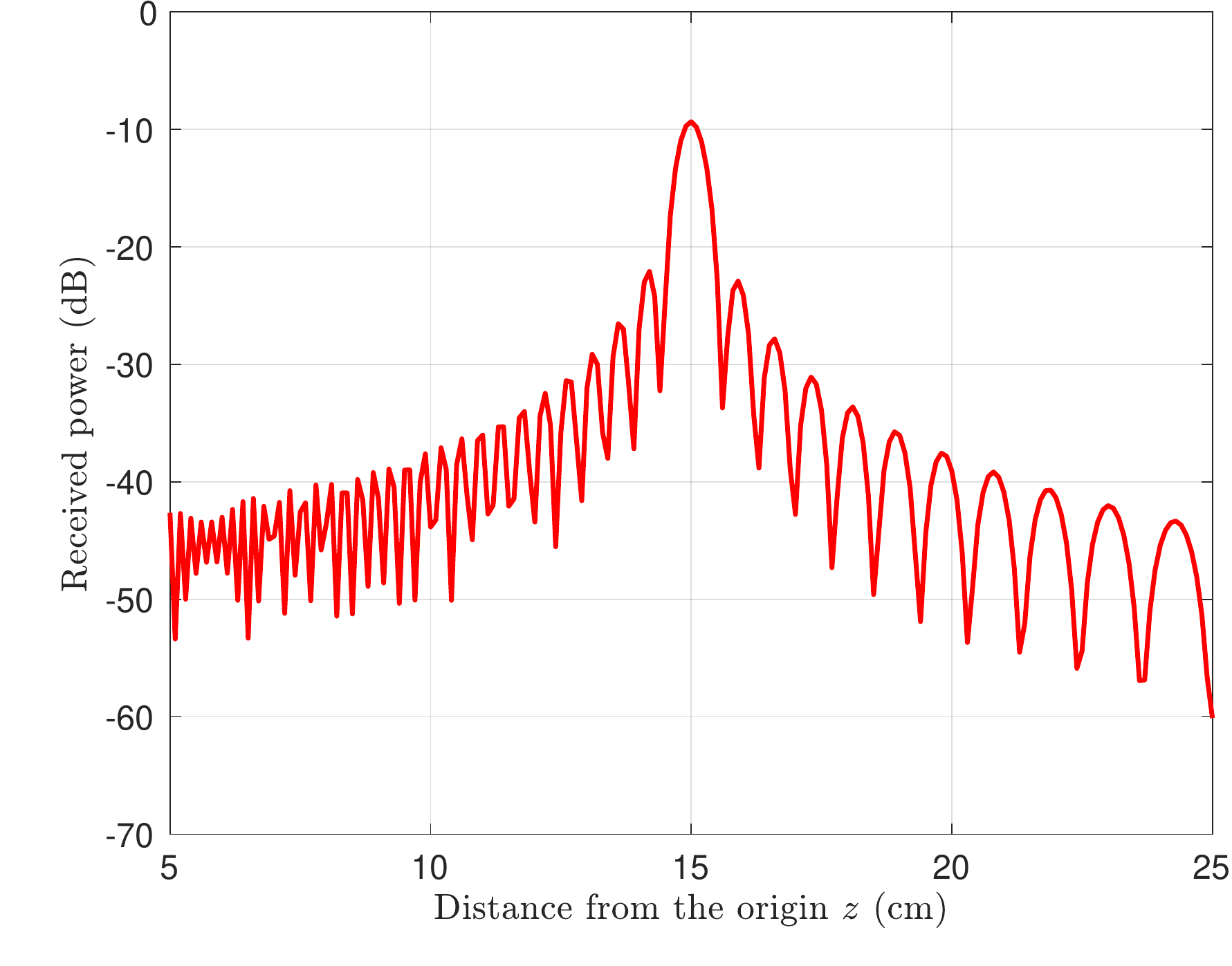}\label{fig:Power_z}}
\caption{We consider a boresight scenario with a half-wavelength spaced array at the TX. Here, $\fc= 300\, \mathrm{GHz}$, $R=10\, \mathrm{cm}$, $\Delta= \lambda_{\mathrm{c}}/2$ and $\ell=15 \, \mathrm{cm}$. The phase profile in standard beamforming is shown in Fig. \ref{fig:phi_std}. When this profile is applied at the TX, the RF signals are spatially concentrated around the RX at $15\, \mathrm{cm}$ as shown in Fig. \ref{fig:Power_z}. 
\normalsize}
\end{figure}

 The goal of beamforming is to adjust the phase shifts $\{\phi(x,y)\}_{(x,y) \in \mathcal{S}_{\mathrm{D}}}$ to maximize the received signal power. The standard beamforming method adjusts the phase shifts to maximize $|g(f_{\mathrm{c}})|^2$, i.e., the energy of the equivalent channel response at the center frequency $\fc$. We define these phase shifts as $\{\phi_{\std}(x,y)\}_{(x,y) \in \mathcal{S}_{\mathrm{D}}}$. The phase profile $\phi_{\std}(x,y)$ corresponding to the channel in \eqref{eq:channel_freq} is
\begin{equation}
\label{eq:phi_fc}
\phi_{\std}(x,y)=\frac{2\pi \fc \ell(x,y) }{c}.
\end{equation}
An example of this profile is shown in Fig. \ref{fig:phi_std} for $\fc= 300\, \mathrm{GHz}$, $\gamma=0  \degree$, $R=10\, \mathrm{cm}$ and $\ell=15 \, \mathrm{cm}$. The hyperbolic structure of such a phase profile allows near field systems to focus signals in a spatial region \cite{NFbeams}. To illustrate the spatial focusing effect, we evaluate the received power along the $z$ axis when the TX directs its signal to an RX at $(0, 0, 15 \, \mathrm{cm})$. It can be observed from Fig. \ref{fig:Power_z} that the received power is concentrated in a small region of $1\, \mathrm{cm}$ around the RX. 
\par Now, we approximate the discrete sum in $g(f)$ to an integral. The approximation considers an imaginary transmitter (ITX) with a continuous aperture of radius $R$. The set of coordinates within this aperture is defined as 
\begin{equation}
\label{eq:defns_set_S}
\mathcal{S}=\{ (x,y): x^2+y^2 \leq R^2\}.
\end{equation}
We observe from \eqref{eq:defns_set_S} that the ITX contains an uncountably infinite number of antennas. The concept of an ITX was discussed in \cite{holographic} under the label of hologram-based beamforming. For such an ITX, we assume that a continuous phase profile $\phi(x,y)$, a 2D-function supported on $\mathcal{S}$, can be applied to control the RF signals going out of the infinitesimally small antennas. In the discrete antenna setting with $\Ntx$ antennas, the energy of the beamforming weight profile in \eqref{eq:w_phase_shift}, i.e., $\sum_{(x,y) \in \mathcal{S}_{\mathrm{D}}} |w(x,y)|^2$, is $1$. Similarly, a unit energy beamforming weight profile at the ITX is defined as $w(x,y)=e^{\imj \phi(x,y)}/ \sqrt{\pi R^2}$ for $(x,y) \in \mathcal{S}$. Note that $\int_{\mathcal{S}} |w(x,y)|^2=1$. The channel between the ITX and the RX is modeled using \eqref{eq:channel_freq}. Analogous to $g(f)$ in \eqref{eq:eqsiso}, the equivalent SISO channel when the ITX applies a continuous phase profile $\phi(x,y)$ is defined as 
\begin{align}
\label{eq:ga_f_defn_1}
g_{\mathrm{a}}(f)&=\frac{1}{\sqrt{\pi R^2} \Delta }\int_{\mathcal{S}} h(x,y,f) e^{\imj \phi(x,y)}\dx \dy\\
\label{eq:ga_f_defn_2}
&=\frac{c}{2\pi^{3/2} Rf \Delta}\int_{\mathcal{S}}  \frac{1}{ \ell(x,y)} e^{\imj \left(\phi(x,y)-\frac{2 \pi f \ell(x,y)}{c} \right)} \dx \dy.
\end{align}
The function $g_{\mathrm{a}}(f)$ in \eqref{eq:ga_f_defn_2} is interpreted as a continuous approximation of $g(f)$ in \eqref{eq:eqsiso}. Equivalently, $g(f)$ is a Riemann sum approximation of $g_{\mathrm{a}}(f)$ \cite{graves1927riemann}. The error in the approximation, i.e., $|g(f)-g_{\mathrm{a}}(f)|$, is $\mathcal{O}(K/\Ntx)$ where $K$ is an upper bound on the second order partial derivatives of $h(x,y,f)w(x,y)$ along the $x$ and $y$ dimensions.
\par We would like to highlight that the approximate SISO channel $g_{\mathrm{a}}(f)$ is just a mathematical concept which aids robust beamformer design, and the model in \eqref{eq:ga_f_defn_1} may not be relevant in practical systems due to several constraints. First, it remains unclear how phase control at an infinitesimal level can be implemented using known RF components. Second, the design of a digital computer interface to control the aperture is challenging. Although holographic or metasurface beamforming mark a step towards such continuous aperture arrays, these technologies are still based on discrete components. It is important to note that the notion of ITX in our problem is adopted because integrals are easy to deal with than discrete summations. The idea underlying the beamformer design technique in this paper is to first design a continuous phase profile $\phi(x,y)$ that achieves robustness to beam misfocus. Then, the phase profile is sampled at the coordinates in $\mathcal{S}_{\mathrm{D}}$ for use in a practical system with $\Ntx$ antennas. The simulation results discussed in this paper are for the discrete antenna-based TX, while our analysis is for the ITX which has a continuum of antennas.
\section{Beam misfocus effect: How much bandwidth is too much?} \label{sec:misfocusintro}
\par In this section, we investigate the performance of the standard beamforming method when the receiver is along the boresight of the transmit array, i.e., $\gamma=0 \degree$. We explain how such a technique suffers from the beam misfocus effect in near field wideband systems. 
\par We now consider the system in Fig. \ref{fig:boresightpic} where the RX is along the $z-$axis and study beamforming with the phase profile in \eqref{eq:phi_fc}. For such a phase profile, we define $g_{\mathrm{a}, \std}(f)$ as the equivalent SISO channel between the ITX and the RX. We substitute $\phi(x,y)=\phi_{\std}(x,y)$ in \eqref{eq:ga_f_defn_2} to write
\begin{equation}
\label{eq:ga_f_std1}
g_{\mathrm{a}, \std}(f)=\frac{c}{2\pi^{3/2} R f  \Delta }\int_{\mathcal{S}}  \frac{1}{ \ell(x,y)}e^{-\imj \frac{2 \pi (f-\fc) \ell(x,y)}{c} } \dx\dy.
\end{equation}
Setting $\gamma=0 \degree$ in \eqref{eq:ellxy_defn}, we observe that $\ell(x,y)=\sqrt{x^2+y^2+\ell^2}$ for a boresight scenario. We define $r=\sqrt{x^2+y^2}$ and $\theta=\mathrm{tan}^{-1}(y/x)$, i.e., the polar coordinates associated with $(x,y)$, to rewrite the integral in \eqref{eq:ga_f_std1} as 
\begin{equation}
\label{eq:ga_f_std2}
g_{\mathrm{a}, \std}(f)=\frac{c}{2\pi^{3/2} R f  \Delta }\int_{r=0}^{R}\int_{\theta=0}^{2\pi}\frac{1}{ \sqrt{r^2+\ell^2}}e^{-\imj \frac{2 \pi (f-\fc)  \sqrt{r^2+\ell^2}}{c} } r \dr\dth.
\end{equation}
We define the spatial frequency, with units of $\mathrm{radians}/\mathrm{m}$, as 
\begin{equation}
\label{eq:defn_omega}
\omega= 2 \pi (f-\fc)/c,
\end{equation}
and the length
\begin{equation}
\label{eq:defn_davg}
d_{\mathrm{avg}}=(\ell+\sqrt{\ell^2+R^2})/2. 
\end{equation}
We observe that $\omega \in [- \pi B/c, \pi B/c]$ for $f \in [\fc-B/2, \fc+B/2]$. A closed form expression for the equivalent SISO channel is derived in Appendix-A. We define $\mathrm{sinc}(x)=\mathrm{sin} x/x$ and write $g_{\mathrm{a}, \std}(f)$ in compact form as
\begin{equation}
\label{eq:ga_f_std3}
g_{\mathrm{a}, \std}(f)=\frac{2c (d_{\mathrm{avg}}- \ell)e^{-\imj \omega d_{\mathrm{avg}}}}{\sqrt{\pi}Rf  \Delta }\times \mathrm{sinc}\left( \omega(d_{\mathrm{avg}}- \ell) \right).
\end{equation}
The equivalent SISO channel $g_{\mathrm{a}, \std}(f)$ in \eqref{eq:ga_f_std3} is a product of two terms which are frequency dependent.
\par
\begin{figure}[h!]
\vspace{-4mm}
\centering
\subfloat[Channel response at the RX for $\phi_{\std}(x,y)$.]{\includegraphics[trim=1cm 6.25cm 2.5cm 7.5cm,clip=true,width=6cm, height=5cm]{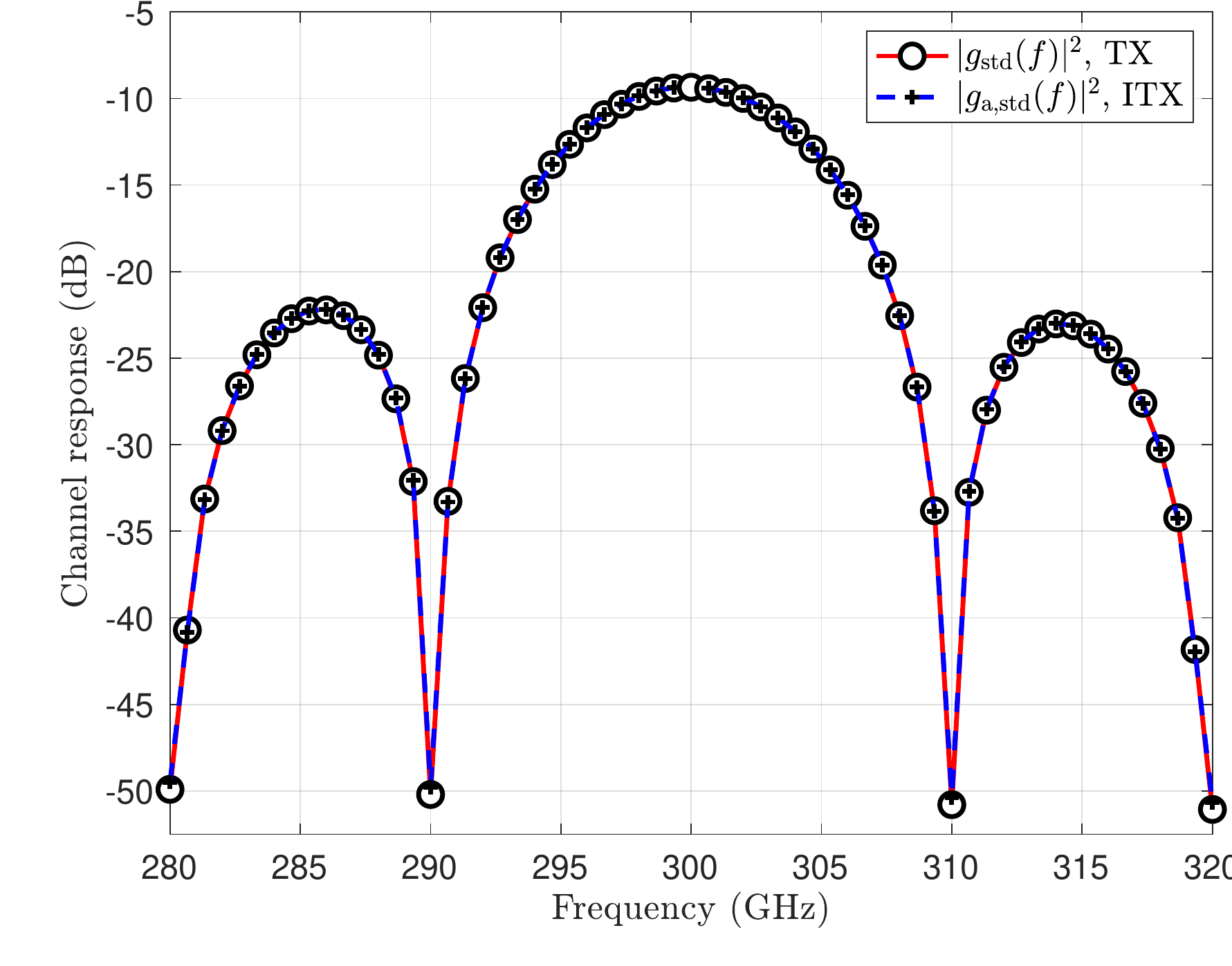}\label{fig:freq_resp_std}}
\:\:\:\:\:\:
\subfloat[Received power ($\mathrm{dB}$) with $\phi_{\std}(x,y)$.]{\includegraphics[trim=0cm 6.25cm 1cm 7.5cm,clip=true,width=6cm, height=5cm]{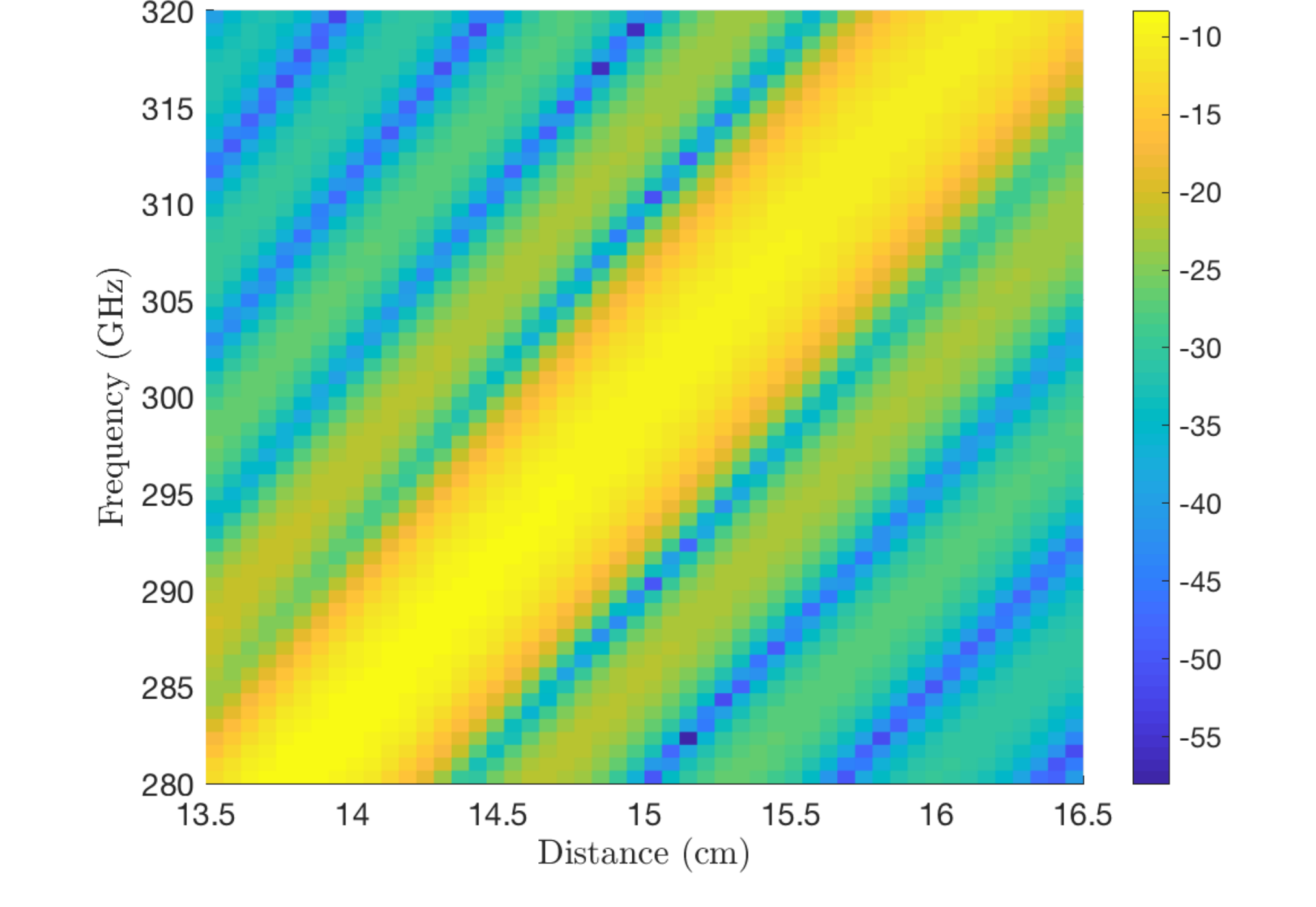}\label{fig:misfocus_surf}}
\caption{ For the system in Fig. \ref{fig:boresightpic}, we observe from Fig. \ref{fig:freq_resp_std} that the equivalent SISO channel with the standard beam has a low gain for $|f-\fc|>10\, \mathrm{GHz}$. Here, $\fc=300\, \mathrm{GHz}$, $\Delta=0.5\,\mathrm{mm}$, $R=10\,\mathrm{cm}$ and $\ell=15\, \mathrm{cm}$. Fig \ref{fig:misfocus_surf} shows that the focus point changes with the frequency of operation when the TX applies $\phi_{\std}(x,y)$ to its array.
\normalsize}
\end{figure}

 For wideband systems in the near field, the energy of the equivalent channel between the ITX and the RX, i.e., $|g_{\mathrm{a}, \std}(f)|^2$, varies substantially with the frequency. To illustrate this variation, we consider a boresight scenario with $\Delta=\lambda_{\mathrm{c}}/2$, $R=10\, \mathrm{cm}$, $\ell = 15 \, \mathrm{cm}$ and $\fc=300 \, \mathrm{GHz}$. In Fig. \ref{fig:freq_resp_std}, we show the energy in the channel response, i.e., $|g_{\mathrm{a},\mathrm{std}}(f)|^2$. We also show its discrete antenna counterpart defined as $|g_{\mathrm{std}}(f)|^2$. Here, $g_{\mathrm{std}}(f)$ is the equivalent SISO channel between the discrete antenna TX and the RX, and is obtained by setting $w(x,y)=e^{\imj \phi_{\std}(x,y)} / \sqrt{\Ntx}$ in \eqref{eq:eqsiso}. We notice from Fig. \ref{fig:freq_resp_std} that the normalized channel response with standard beamforming achieves it maximum at the center frequency. The gain at $310\, \mathrm{GHz}$, however, is about $41\,\mathrm{dB}$ lower than the maximum. The poor gain at frequencies that are far away from the center frequency is a disadvantage with standard beamforming in phase shifter-based arrays.
\par The poor gain with standard beamforming in near field wideband systems can be explained by the misfocus effect. An ideal beamformer is one that focuses RF signals with frequencies in $[\fc-B/2, \fc+B/2]$ at the desired receiver. The focus point of the standard beamformer, however, changes with the frequency leading to the beam misfocus effect. From Fig. \ref{fig:misfocus_surf}, the focus point is closer to the TX for $f<\fc$ and is away from the TX for $f>\fc$. Such an effect is analogous to chromatic abberation in optics where light rays of different wavelengths focus at different points \cite{chromatic}. Due to the misfocus effect, the receiver can only listen to signals within a small bandwidth around the center frequency. But, how small is this bandwidth? To answer the question, we consider the two terms in \eqref{eq:ga_f_std3} whose product is the equivalent channel $g_{\mathrm{a},\mathrm{std}}$. The first term in the product is inversely proportional to the frequency. The second term, which depends on the frequency through $\omega$, leads to a substantial decrease in the beamforming gain. For the system parameters in Fig. \ref{fig:freq_resp_std}, we compute the first and second terms in \eqref{eq:ga_f_std3}. These terms contribute to a beamforming loss of $0.3\, \mathrm{dB}$ and $41\, \mathrm{dB}$ at $310\, \mathrm{GHz}$, when compared to $|g(\fc)|^2$. The condition for which the loss due to the second term is less than $4\, \mathrm{dB}$ is expressed as
\begin{equation}
\label{eq:sinc_4db_case}
\left|   \frac{\omega(\sqrt{\ell^2+R^2}-\ell)}{2} \right| < \frac{\pi}{2}.
\end{equation}
We use $|\omega| \leq \pi B/c$ in \eqref{eq:sinc_4db_case} to conclude that the receiver observes ``all'' the RF signals when 
\begin{equation}
\label{eq:BoundB}
B < \frac{c}{\sqrt{\ell^2+R^2}-\ell}.
\end{equation}
It follows from \eqref{eq:BoundB} that the effective operating bandwidth with standard beamforming decreases as the TX antenna aperture radius $R$ increases.
\par The far field analogue of misfocus is the beam squint effect in which the direction of the beam changes with the frequency of the RF signal. Interestingly, far field beams that are directed along the boresight of an array do not suffer from the squint \cite{mingming_gc}. Near field beamforming with the standard design, however, suffers from misfocus even in a boresight scenario as seen in Fig. \ref{fig:misfocus_surf}. One approach to mitigate misfocus is to reduce the effective aperture of the TX array by turning off the antennas which are far from the TX center. The resultant array with fewer active antennas is called a thinned array. It can be observed from \eqref{eq:BoundB} that a smaller $R$ results in a larger effective operating bandwidth. Reducing the aperture, however, results in lower received power under the typical per-antenna power constraint. Is it possible to design new beams, equivalently phase profiles, that can mitigate the misfocus effect without turning off any antenna? In this paper, we propose InFocus to design such phase profiles. The proposed beams achieve close to uniform beamforming gain over wide bandwidths. This gain, however, is smaller than the maximum gain with the standard beamformer, i.e., $|g_{\mathrm{a},\std}(\fc)|^2$. In Section \ref{sec:simulations}, we show that the beams with InFocus lead to a higher rate than the standard beams in phased arrays.
\section{Misfocus robust beamforming with InFocus} \label{sec:mainsecInfocus}
\par The key idea underlying InFocus is to add a carefully designed phase profile to that of the standard beamformer for robustness to misfocus. We define $\psi_{\mathrm{des}}(x,y)$ as the phase profile added to $\phi_{\std}(x,y)$. The resultant phase applied at the TX is
\begin{equation}
\label{eq:phi_add1}
\phi(x,y)=\phi_{\mathrm{std}}(x,y)+\psi_{\mathrm{des}}(x,y),
\end{equation}
where $\phi_{\mathrm{std}}(x,y)=2 \pi \fc \ell(x,y)/c$. We consider the continuous aperture system with an ITX to design $\psi_{\mathrm{des}}(x,y)$. Substituting $\phi(x,y)=2 \pi \fc \ell(x,y)/c+ \psi_{\mathrm{des}}(x,y)$ in \eqref{eq:ga_f_defn_2}, the equivalent SISO channel for this system can be expressed as
\begin{align}
\label{eq:ga_basic_1}
g_{\mathrm{a}}(f)&=\frac{c}{2\pi^{3/2} Rf \Delta}\int_{\mathcal{S}}  \frac{1}{ \ell(x,y)} e^{\imj \psi_{\mathrm{des}}(x,y)} e^{-\imj \frac{2 \pi (f-\fc) \ell(x,y)}{c}} \dx \dy\\
\label{eq:ga_basic_2}
&=\frac{c}{2\pi^{3/2} Rf \Delta}\int_{\mathcal{S}}  \frac{1}{ \ell(x,y)} e^{\imj \psi_{\mathrm{des}}(x,y)} e^{-\imj \omega \ell(x,y)} \dx \dy.
\end{align} 
The question at this point is if it is possible to determine a 2D-phase function $\psi_{\mathrm{des}}(x,y)$ that results in a ``uniform'' beamforming gain for $f \in [\fc-B/2, \fc+B/2]$. 
\par For a tractable design of the phase profile, we ignore the $1/f$ scaling in \eqref{eq:ga_basic_2} and define
\begin{equation}
\label{eq:gtild_def_1}
\tilde{g}(f)=\int_{\mathcal{S}}  \frac{1}{ 2 \pi \ell(x,y)} e^{\imj \psi_{\mathrm{des}}(x,y)} e^{-\imj \omega \ell(x,y)} \dx \dy.
\end{equation}
We observe that $g_{\mathrm{a}}(f)=c\tilde{g}(f)/(\sqrt{\pi} Rf \Delta)$. InFocus constructs $\psi_{\mathrm{des}}(x,y)$ such that $|\tilde{g}(f)|^2$ is large and approximately flat over the desired bandwidth. Our design ignores the $1/f$ term in $g_{\mathrm{a}}(f)$ as it leads to a smaller variation in $|g_{\mathrm{a}}(f)|^2$ when compared to the integral in \eqref{eq:ga_basic_2}. For example, the variation in  $|g_{\mathrm{a}}(f)|^2$ due to the $1/f$ term is $20\, \mathrm{log}_{10}(320/280)\approx 1.2\, \mathrm{dB}$ for a $40\,\mathrm{GHz}$ system at $300\, \mathrm{GHz}$. The variation in $|g_{\mathrm{a}}(f)|^2$ due to $\tilde{g}(f)$, however, is about $40 \, \mathrm{dB}$ with the standard design where $\psi_{\mathrm{des}}(x,y)=0$. In Section \ref{sec:boresight_section}, we first explain the construction of $\psi_{\mathrm{des}}(x,y)$ for misfocus robust beamforming in a boresight scenario. Then, we extend our solution to a general setting in Section \ref{sec:bf_arb}.
\subsection{Beamforming along the boresight}\label{sec:boresight_section}
\par We simplify $\tilde{g}(f)$, the scaled version of the equivalent SISO channel $g_{\mathrm{a}}(f)$ when $\gamma=0 \degree$. In the polar coordinate representation, $\ell(x,y)=\sqrt{r^2+\ell^2}$ when $\gamma=0 \degree$. Furthermore, $\psi_{\mathrm{des}}(x,y)$ can be expressed as $\psi_{\mathrm{des}}(r \mathrm{cos}\, \theta, r \mathrm{sin}\, \theta)$. Then, $\tilde{g}(f)$ in \eqref{eq:gtild_def_1} is
\begin{equation}
\label{eq:gtild_def_2}
\tilde{g}(f)=\int_{r=0}^{R} \int_{\theta=0}^{2 \pi}  \frac{1}{ 2\pi \sqrt{r^2+\ell^2} } e^{\imj \psi_{\mathrm{des}}(r \mathrm{cos}\, \theta, r \mathrm{sin}\, \theta)} e^{-\imj \omega \sqrt{r^2+\ell^2}} r \dr \dth.
\end{equation}
We define $s=\sqrt{r^2+\ell^2}$. In this case, $\ds= r\dr / \sqrt{r^2+\ell^2}$ and \eqref{eq:gtild_def_2} can be simplified to 
\begin{equation}
\label{eq:gtild_def_3}
\tilde{g}(f)=\int_{s=\ell}^{\sqrt{\ell^2+R^2}} \frac{1}{2 \pi}\int_{\theta=0}^{2 \pi} e^{\imj \psi_{\mathrm{des}}(r \mathrm{cos}\, \theta, r \mathrm{sin}\, \theta)} e^{-\imj \omega s } \ds \dth.
\end{equation}
Due to the radial symmetry in the boresight scenario, it is reasonable to design a $\psi_{\mathrm{des}}(r \mathrm{cos}\, \theta, r \mathrm{sin}\, \theta)$ that varies only with $r$ and is independent of the angle $\theta$. As $s=\sqrt{r^2+\ell^2}$ is directly related to $r$, we model the variation in $\psi_{\mathrm{des}}(r \mathrm{cos}\, \theta, r \mathrm{sin}\, \theta)$ through a $1\mathrm{D}$-function $\psi(s)= \psi_{\mathrm{des}}(r \mathrm{cos}\, \theta, r \mathrm{sin}\, \theta)$ where $r=\sqrt{s^2-\ell^2}$. With this definition, we rewrite \eqref{eq:gtild_def_3} as
\begin{equation}
\label{eq:gtild_simp_1}
\tilde{g}(f)=\int_{s=\ell}^{\sqrt{\ell^2+R^2}}e^{\imj \psi(s)} e^{-\imj \omega s } \ds.
\end{equation}
The problem now is to design a 1D-function $\psi(s)$ for robustness to misfocus. 
\par We investigate the design of the phase function $\psi(s)$ to achieve a large and approximately flat $|\tilde{g}(f)|^2$ over the desired frequency range $[\fc-B/2, \fc+B/2]$. As $f=\fc+c \omega/(2\pi)$, this requirement on $|\tilde{g}(f)|^2$ is equivalent to a uniform gain over $|\tilde{g}(\fc + c \omega/(2\pi))|^2$ for $\omega \in [-\pi B/c, \pi B/c]$. For ease of notation, we define $\hat{g}(\omega)=\tilde{g}(\fc + c \omega/(2\pi))$, $d_{\mathrm{max}}=\sqrt{\ell^2+R^2}$, and $\mathbb{I}_{\ell,d_{\mathrm{max}}}(s)$ as an indicator function which is $1$ for $s\in [\ell, d_{\mathrm{max}}]$. Now, $\hat{g}(\omega)$ is
\begin{align}
\label{eq:FT_eqn_pre}
\hat{g}(\omega)&=\int_{s=\ell}^{\sqrt{\ell^2+R^2}}e^{\imj \psi(s)} e^{-\imj \omega s } \ds\\
\label{eq:FT_eqn}
&=\int_{s=- \infty}^{\infty}e^{\imj \psi(s)} \mathbb{I}_{\ell,d_{\mathrm{max}}}(s)e^{-\imj \omega s } \ds.
\end{align}
An interesting observation from \eqref{eq:FT_eqn} is that $\hat{g}(\omega)$ is the Fourier transform of $e^{\imj \psi(s)} \mathbb{I}_{\ell,d_{\mathrm{max}}}(s)$. Now, is it possible to design $\psi(s)$ such that $e^{\imj \psi(s)} \mathbb{I}_{\ell,d_{\mathrm{max}}}(s)$ has a constant magnitude Fourier transform over $\omega \in [-\pi B/c, \pi B/c]$? No, because signals that are localized in the $s-$domain cannot be localized in the Fourier representation \cite{uncert_prin}. For example, a rectangular function which is localized in the $s-$domain has a $\mathrm{sinc}$ representation in the Fourier domain which is spread over all frequencies. As a compromise, we seek to construct $\psi(s)$ such that $|\hat{g}(\omega)|^2$ has a high spectral concentration in $[-\pi B/c, \pi B/c]$ and is approximately uniform over this band. 
\par
\begin{figure}[h!]
\vspace{-2mm}
\centering
\subfloat[Instantaneous frequency of a chirp.]{\includegraphics[trim=1cm 6.25cm 2.5cm 7.5cm,clip=true,width=4.9cm, height=4.9cm]{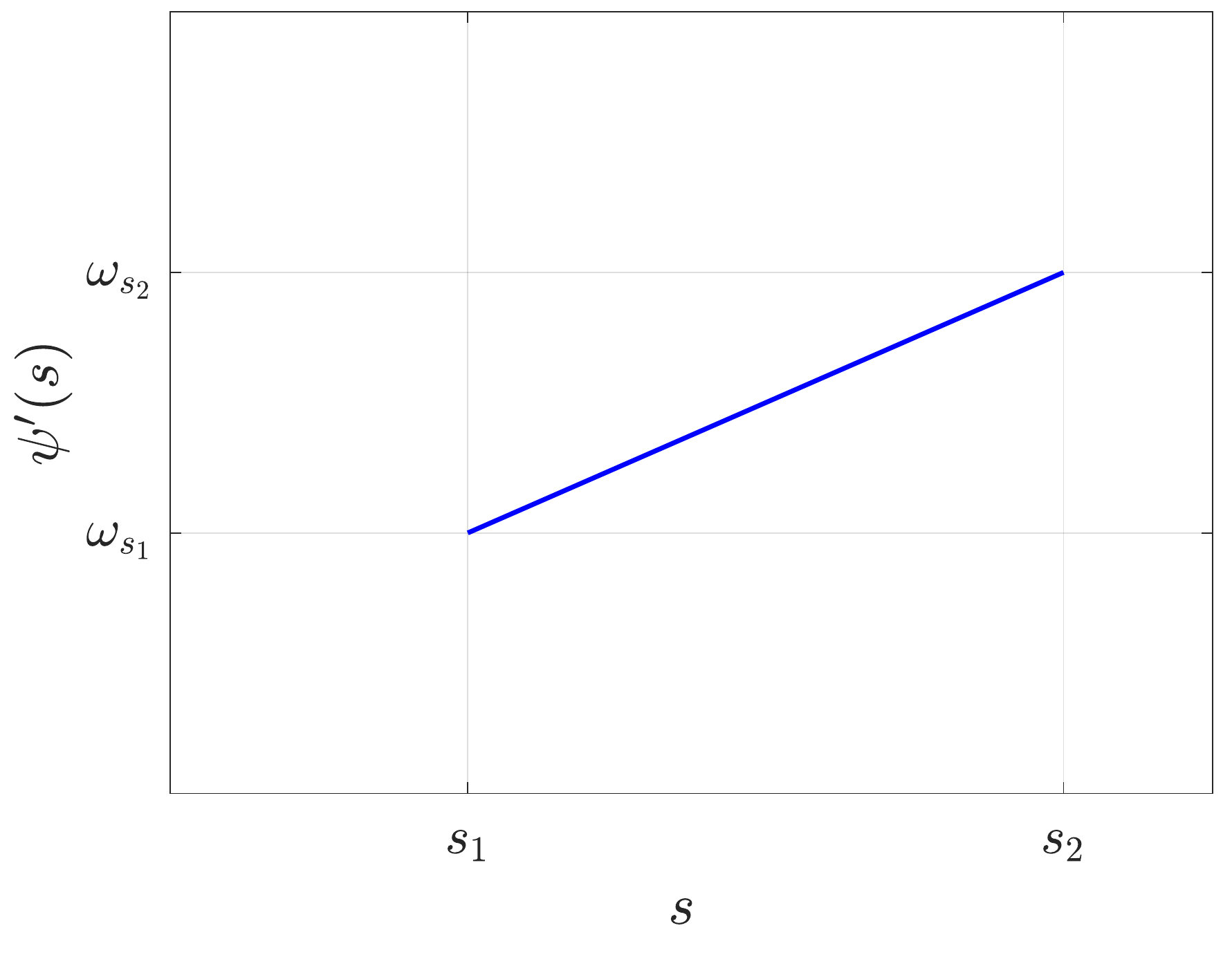}\label{fig:chirpfreq}}
\:\:\:
\subfloat[A chirp defined over ${[ s_1,s_2 ]}$.]{\includegraphics[trim=1cm 6.25cm 2.5cm 7.5cm,clip=true,width=4.9cm, height=4.9cm]{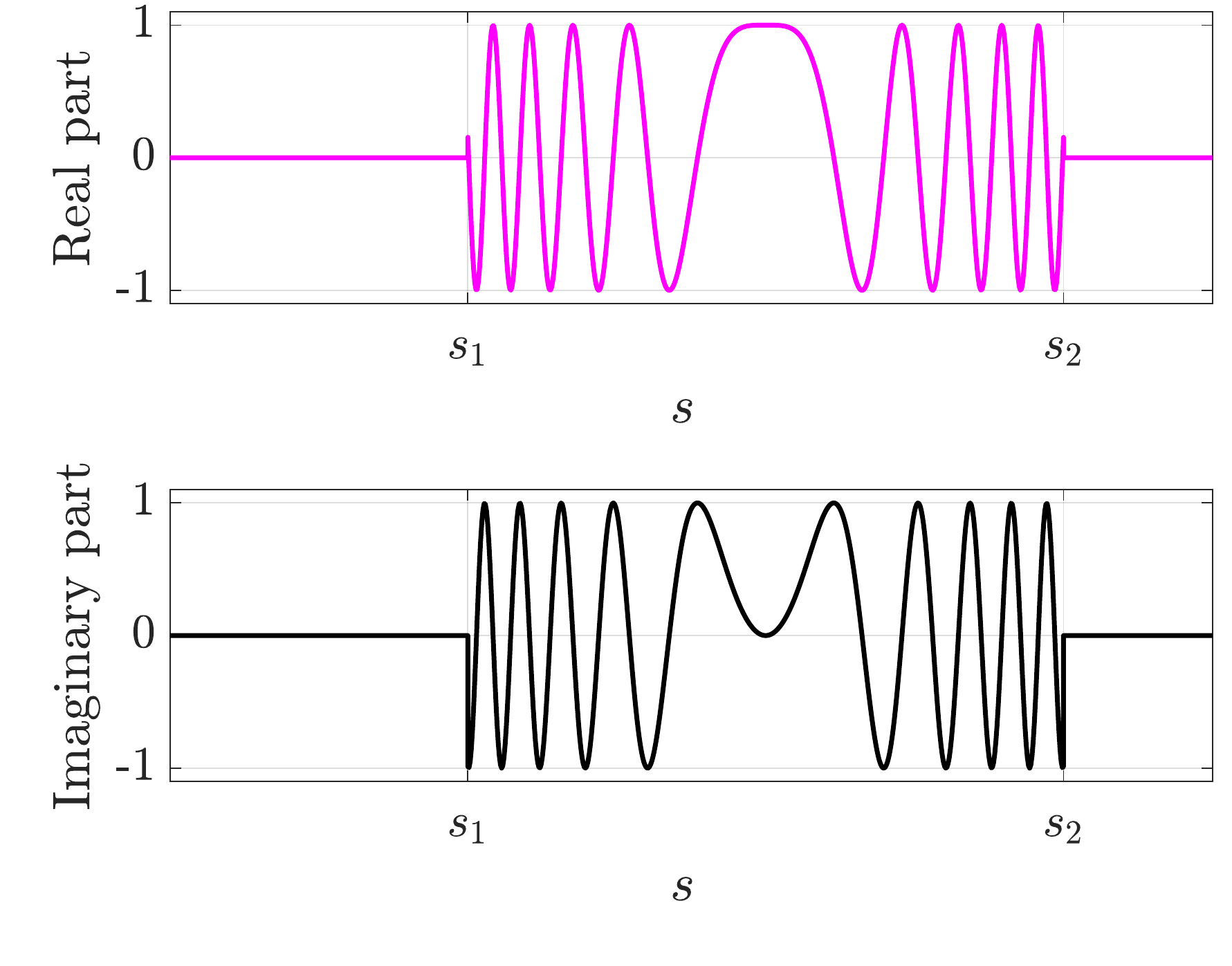}\label{fig:chirpsig}}
\:\:\:
\subfloat[Fourier transform of the chirp]{\includegraphics[trim=1cm 6.25cm 2.5cm 7.5cm,clip=true,width=4.9cm, height=4.9cm]{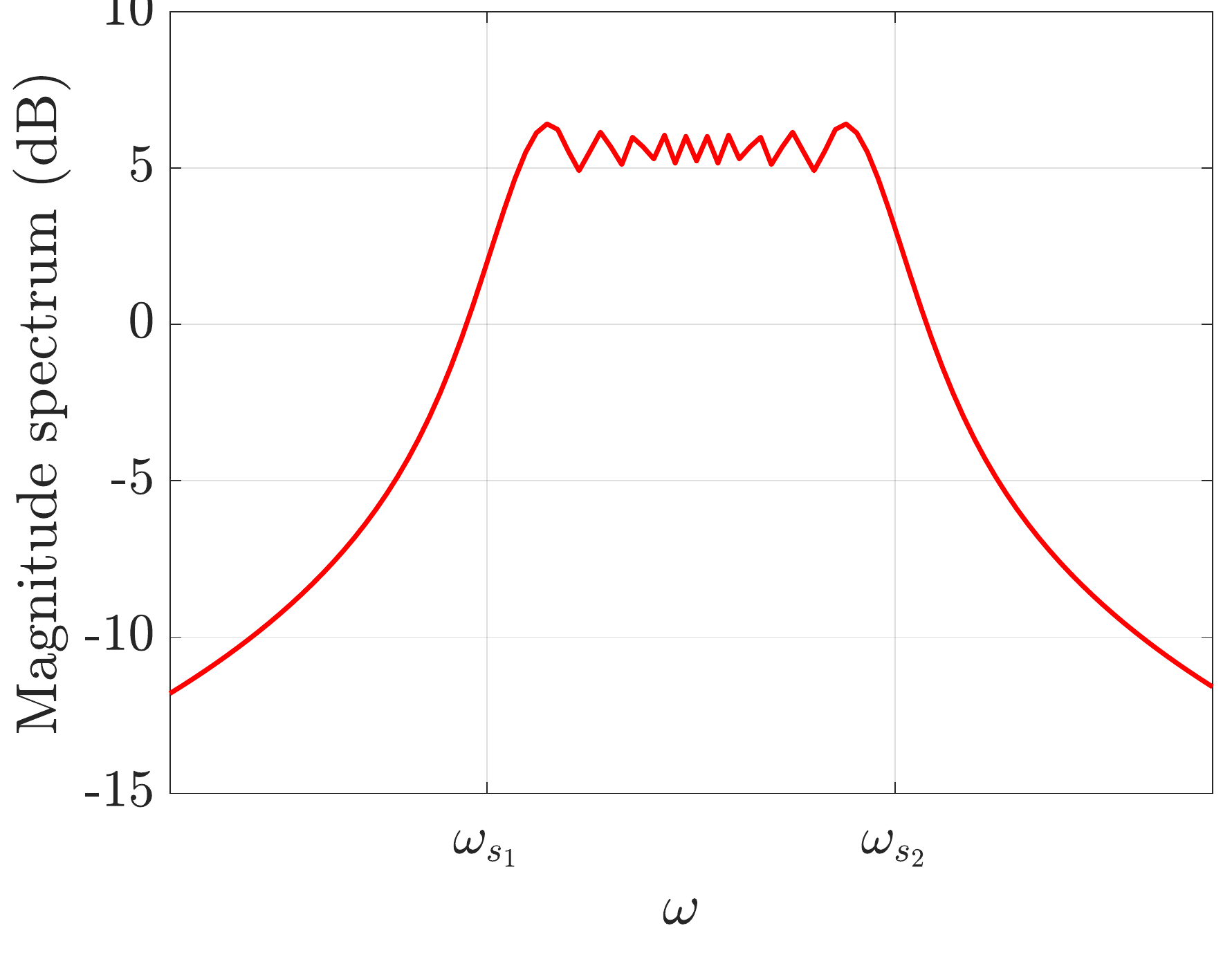}\label{fig:chirpspec}}
\caption{The instantaneous frequency of a chirp over $s\in [s_1,s_2]$ varies linearly with $s$ as shown in Fig. \ref{fig:chirpfreq}. Here, the start and end frequencies are $\omega_{s_1}$ and $\omega_{s_2}$. In Fig. \ref{fig:chirpsig}, we show the real and imaginary components of the chirp signal. The frequency spectrum of this chirp is concentrated within $[\omega_{s_1}, \omega_{s_2}]$ and is approximately flat in this band. 
\normalsize}
\end{figure}

 Now, we discuss a linear FMCW chirp in the $s-$domain and explain why it is a good candidate for $e^{\imj \psi(s)} \mathbb{I}_{\ell,d_{\mathrm{max}}}(s)$. A linear FMCW chirp which starts from $s_1$ and ends at $s_2>s_1$ is a complex exponential signal whose instantaneous frequency linearly increases with $s \in [s_1, s_2]$. In this paper, we use an FMCW chirp along the spatial dimension, different from the common application along the time dimension. The chirp signal has the form $e^{\imj \psi(s)}$ for $s \in [s_1, s_2]$ and is $0$ otherwise. The instantaneous frequency of the chirp signal is defined as $\psi'(s)$, the derivative of $\psi(s)$, where
\begin{equation}
\psi'(s)=\frac{\mathrm{d}\psi(s)}{\mathrm{d}s}.
\end{equation} 
We define $\omega_{s_1}=\psi'(s_1)$ and  $\omega_{s_2}=\psi'(s_2)$ as the start and end frequencies of the chirp. The linear variation in the frequency profile is shown in Fig. \ref{fig:chirpfreq}. Prior work in radar has shown that the Fourier transform magnitude of a chirp is approximately flat over $[\omega_{s_1}, \omega_{s_2}]$ and is nearly zero outside this band. A closed form expression for the spectral magnitude of a chirp can be found in \cite{bell_fmcw}. It was shown in \cite{bell_fmcw} that the spectral leakage outside $[\omega_{s_1}, \omega_{s_2}]$ is less than $5 \%$ of the energy of the chirp when $(\omega_{s_2}-\omega_{s_1})(s_2-s_1)>20 \pi$. The product $(\omega_{s_2}-\omega_{s_1})(s_2-s_1)$ is called the dispersion factor of the chirp. Examples illustrating a chirp and its spectrum are shown in Fig. \ref{fig:chirpsig} and Fig. \ref{fig:chirpspec}. The ``uniform'' spectral characteristic of a chirp within a band can be exploited to construct an appropriate $\psi(s)$ for misfocus robust beamforming.
\par We derive $\psi(s)$ using the properties of a linear FMCW chirp. First, we observe from \eqref{eq:FT_eqn_pre} that the chirp to be designed must start at $s_1=\ell$ and end at $s_2=\sqrt{\ell^2+R^2}$. Second, as the spectrum of this chirp must be concentrated in $[-\pi B/c, \pi B/c]$ for misfocus robust beamforming, we require $\omega_{s_1}=-\pi B/c$ and  $\omega_{s_2}=\pi B/c$. Third, the phase profile of the chirp must take the form $\psi(s)=\alpha s + \beta s^2$ for some constants $\alpha$ and $\beta$. Such a phase variation results in a linear instantaneous frequency profile with $s$ which leads to a ``uniform'' spectral characteristic over the desired band.  We observe that $\psi'(s_1)$, the instantaneous frequency of the chirp at $s_1$ is $\omega_{s_1}$. Similarly, $\psi'(s_2)=\omega_{s_2}$. Note that $\psi'(s)=\alpha + 2 \beta s$. We put together these observations to write 
\begin{align}
\label{eq:alph_bet_eq1}
\alpha + 2 \beta \ell& = \frac{-\pi B}{c}\,\,\, \mathrm{and}\\
\label{eq:alph_bet_eq2}
\alpha + 2 \beta \sqrt{\ell^2+R^2}&= \frac{\pi B}{c}.
\end{align}
Solving the linear equations \eqref{eq:alph_bet_eq1} and \eqref{eq:alph_bet_eq2}, we get $\beta= \pi B/(c\sqrt{\ell^2+R^2}-c\ell)$ and $\alpha=-\pi B (\sqrt{\ell^2+R^2}+\ell)/(c\sqrt{\ell^2+R^2}-c\ell)$. The proposed $1\mathrm{D}$ phase function is then 
\begin{equation}
\label{eq:psi_opt1D}
\psi(s)=\frac{-\pi B(\sqrt{\ell^2+R^2}+\ell) }{c(\sqrt{\ell^2+R^2}-\ell)}s + \frac{\pi B}{c(\sqrt{\ell^2+R^2}-\ell)} s^2.
\end{equation}
As $s=\sqrt{x^2+y^2+\ell^2}$, the $2\mathrm{D}$ phase profile $\psi_{\mathrm{des}}(x,y)$ can be derived from \eqref{eq:psi_opt1D} using $\psi_{\mathrm{des}}(x,y)=\psi(\sqrt{x^2+y^2+\ell^2})$. The phase profile applied at the TX is $\phi(x,y)=\psi_{\mathrm{des}}(x,y)+\phi_{\std}(x,y)$. 
\par
\begin{figure}[h!]
\vspace{-2mm}
\centering
\subfloat[Phase profile $\psi_{\mathrm{des}}(x,y)$.]{\includegraphics[trim=1cm 6.25cm 2.5cm 7.5cm,clip=true,width=4.5cm, height=4.5cm]{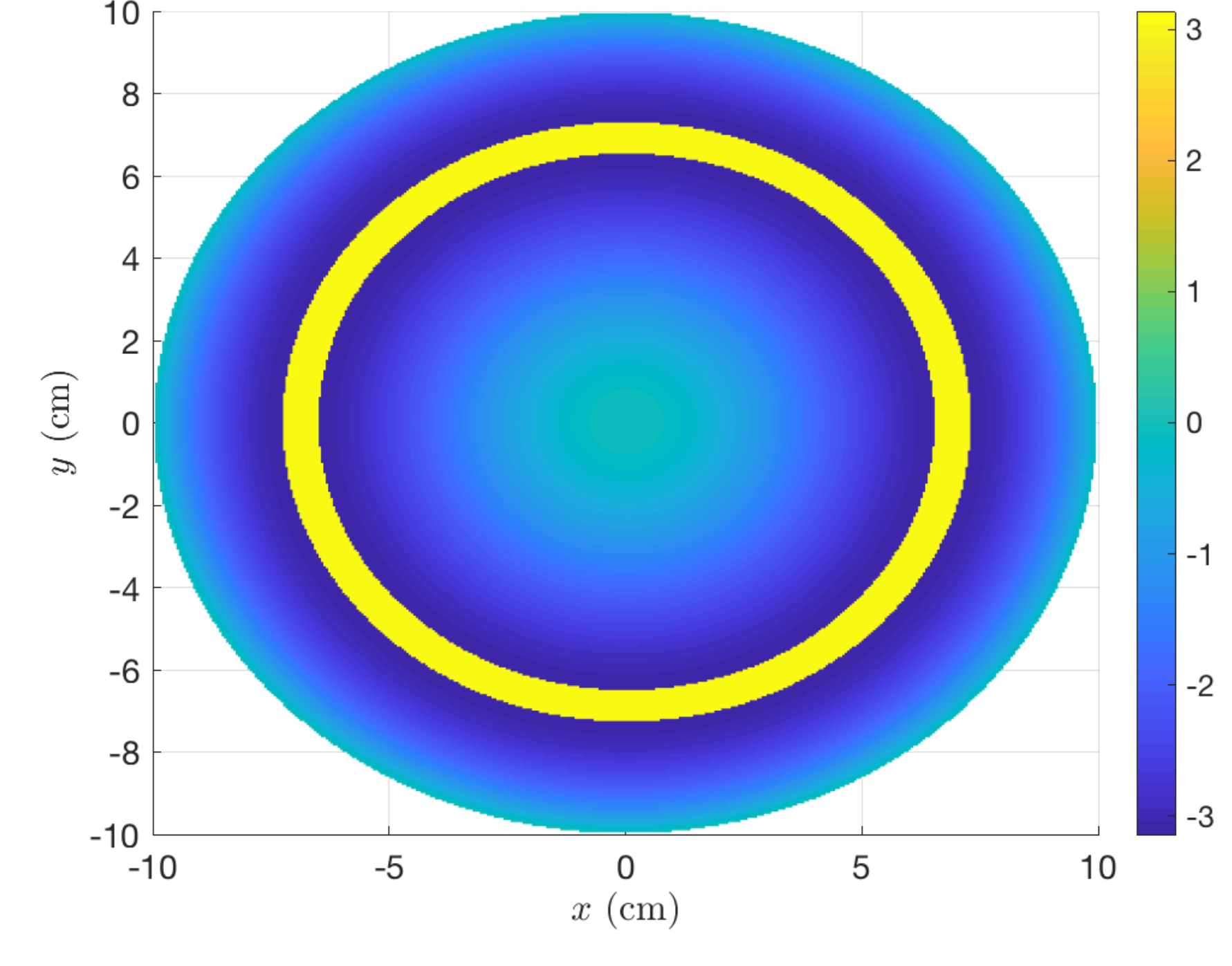}\label{fig:psi_des}}
\:\:\:
\subfloat[$\phi_{\std}(x,y)+\psi_{\mathrm{des}}(x,y)$.]{\includegraphics[trim=1cm 6.25cm 2.5cm 7.5cm,clip=true,width=4.5cm, height=4.5cm]{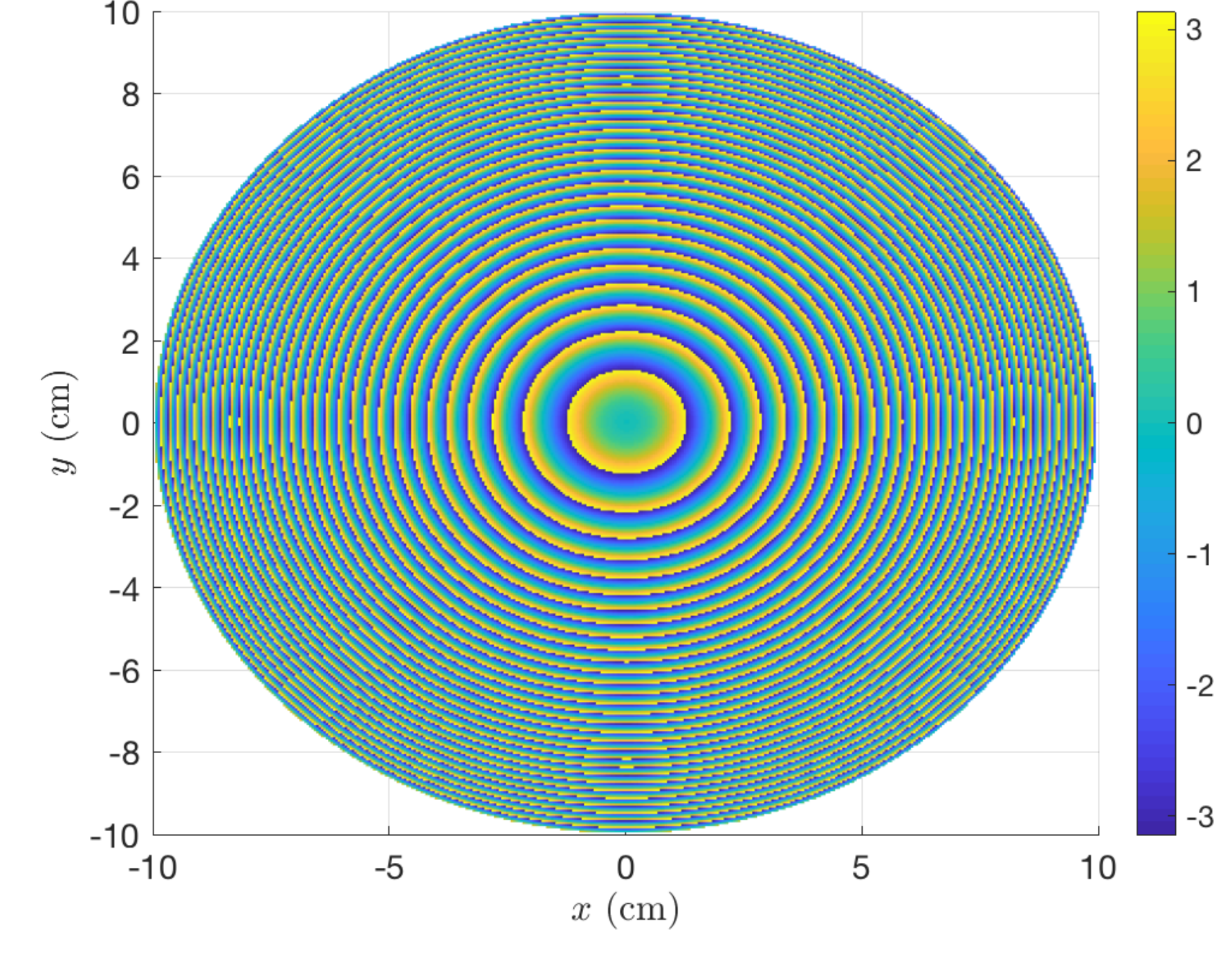}\label{fig:combined_phase_boresight}}
\:\:\:
\subfloat[$20\, \mathrm{log}_{10}|g(f)|$ with frequency.]{\includegraphics[trim=1cm 6.25cm 2.5cm 7.5cm,clip=true,width=4.5cm, height=4.5cm]{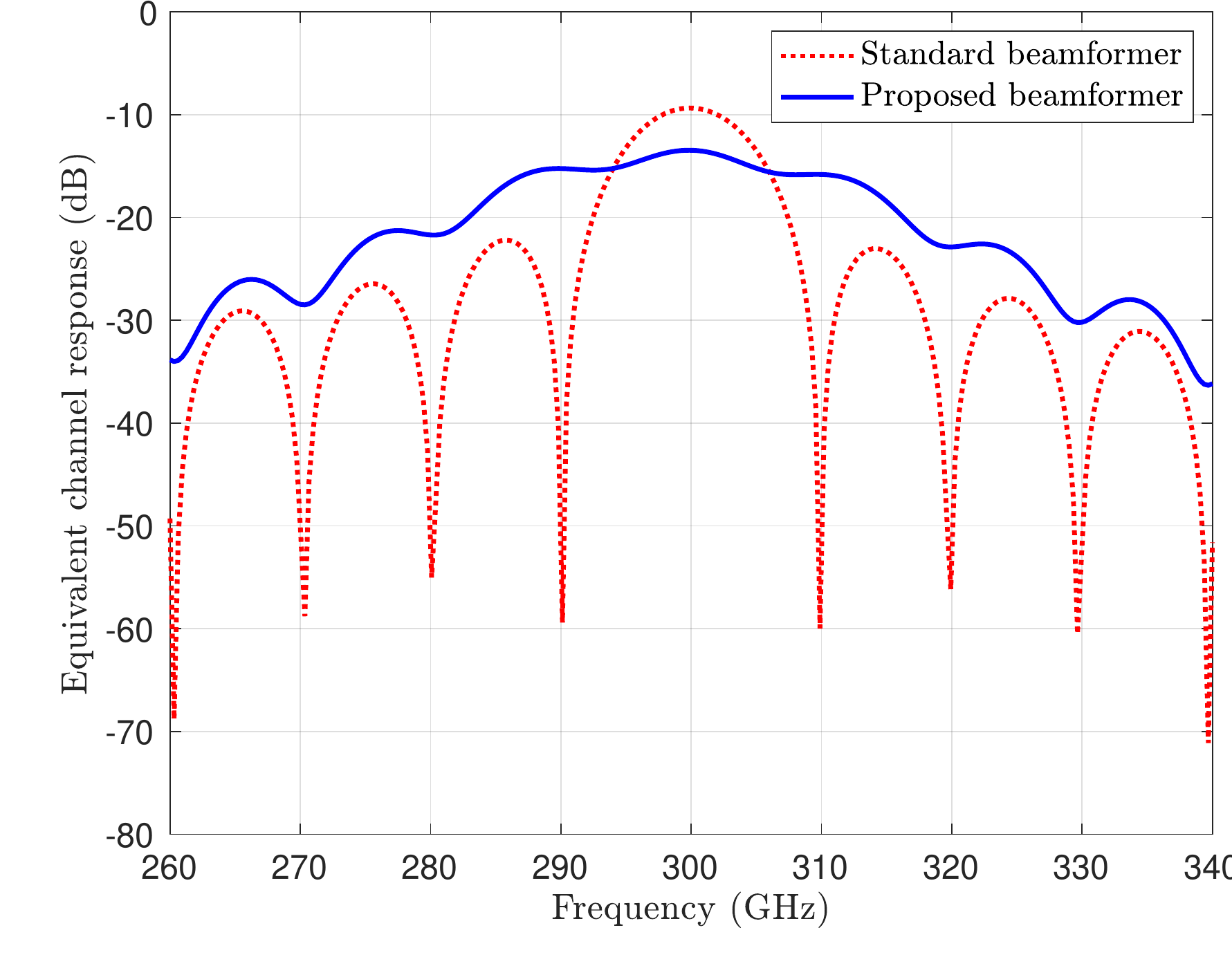}\label{fig:boresightInFocus}}
\caption{The chirp-based phase profile constructed with InFocus for $B=40\, \mathrm{GHz}$ is shown in Fig. \ref{fig:psi_des}. Here, we consider a boresight scenario with $\fc= 300\, \mathrm{GHz}$, $\ell=15\, \mathrm{cm}$ and $R= 10 \, \mathrm{cm}$. For $\Delta=0.5\, \mathrm{mm}$, the proposed phase profile applied at the TX is shown in Fig. \ref{fig:combined_phase_boresight}. The equivalent SISO channel response with the proposed method is large and approximately flat over the desired bandwidth, i.e., $[280\, \mathrm{GHz}, 320\, \mathrm{GHz}]$, when compared to standard beamforming.
\normalsize}
\end{figure}

  We now explain beamforming using a discrete version of the proposed phase profile and compare the performance of our design with standard beamforming which uses $\phi_{\std}(x,y)$. In Fig. \ref{fig:psi_des}, we plot $\psi_{\mathrm{des}}(x,y)$ for a near field system with $R=10\, \mathrm{cm}$, $\fc= 300\, \mathrm{GHz}$ and $B=40\, \mathrm{GHz}$. In this example, the RX is at a distance of $\ell=15 \, \mathrm{cm}$ along the boresight of the TX array. The phase profile used at the TX, i.e., the sum of $\phi_{\std}(x,y)+\psi_{\mathrm{des}}(x,y)$, is shown in Fig. \ref{fig:combined_phase_boresight}. We observe from Fig. \ref{fig:boresightInFocus} that such a phase profile achieves a ``uniform'' gain over the desired bandwidth. The equivalent SISO channel response shown in Fig. \ref{fig:boresightInFocus}, i.e., $20\, \mathrm{log}_{10}(|g(f)|)$, has components outside the desired frequency band. This is due to the fact that $\hat{g}(\omega)$, the Fourier transform of the chirp signal, has spectral components outside $[-\pi B/c, \pi B/c]$. The spectral leakage is determined by the dispersion factor $2 \pi B(\sqrt{\ell^2+R^2}-\ell)/c$. In this example, the dispersion factor is $25.3$, which is less than $20\, \pi$ for $95\%$ energy containment within the desired band. In practice, the effective channel is the product of $g(f)$ and the Fourier transform of the pulse shaping filter. Although an appropriate pulse shaping filter can mitigate the spectral leakage in $g(f)$, it is important to design phase profiles which lead to a large $|g(f)|^2$ within the desired band. Our chirp-based construction is one solution that achieves a large gain. 
\par The closed form nature of the proposed phase profile is favorable from an implementation perspective when compared to optimization techniques. Optimization techniques usually involve an iterative approach to solve for the $\Ntx$ variables $\{w(x,y)\}_{(x,y)\in \mathcal{S}_{\mathrm{D}}}$ under the constraint that $|w(x,y)|=1/\sqrt{\Ntx}$. Such techniques, however, may result in a high complexity when applied to large antenna array systems. For example, in a half-wavelength spaced array of $R=10\, \mathrm{cm}$ that operates at $300\, \mathrm{GHz}$, $\Ntx$ is about $125,000$. InFocus is a low complexity solution that is well suited to massive phased array-based systems.

\subsection{Beamforming at an arbitrary location in the near field} \label{sec:bf_arb}
\par In this section, we design misfocus robust beamformers when the RX is not along the boresight direction. We first derive the beamformer when the projection of the RX on the $xy$ plane, defined as $\mathbb{P}_{\mathrm{RX}}$, falls outside $\mathcal{S}$. Then, we extend our derivation to the other case. 
\subsubsection{Solution when $\mathbb{P}_{\mathrm{RX}} \notin \mathcal{S}$} \label{sec:projection_outside}
\par
\begin{figure}[h!]
\vspace{-2mm}
\centering
\subfloat[A near field scenario where $\mathbb{P}_{\mathrm{RX}} \notin \mathcal{S}$.]{\includegraphics[trim=0cm 0cm 0cm 0cm,clip=true,width=7.5cm, height=3.5cm]{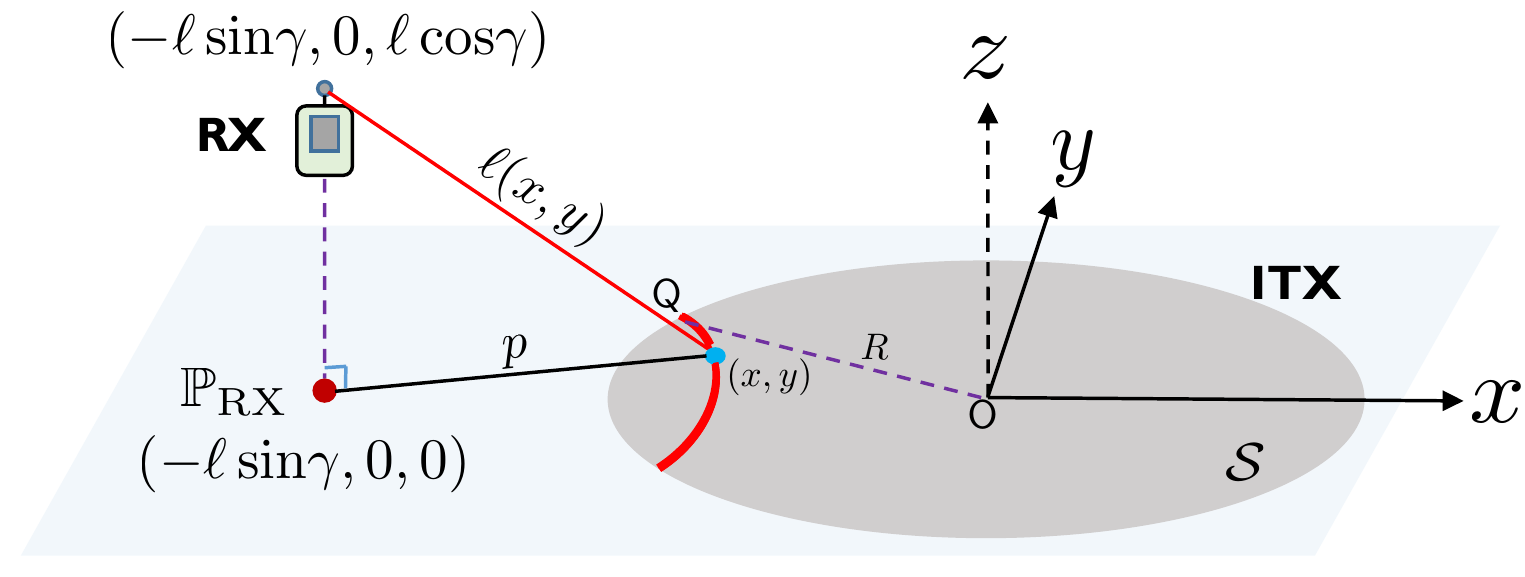}\label{fig:outside_sideview}}
\:\: \:\: \:\:
\subfloat[Top view of the system in Fig. \ref{fig:outside_sideview}.]{\includegraphics[trim=0cm 0cm 0cm 0cm,clip=true,width=5.5cm, height=3.5cm]{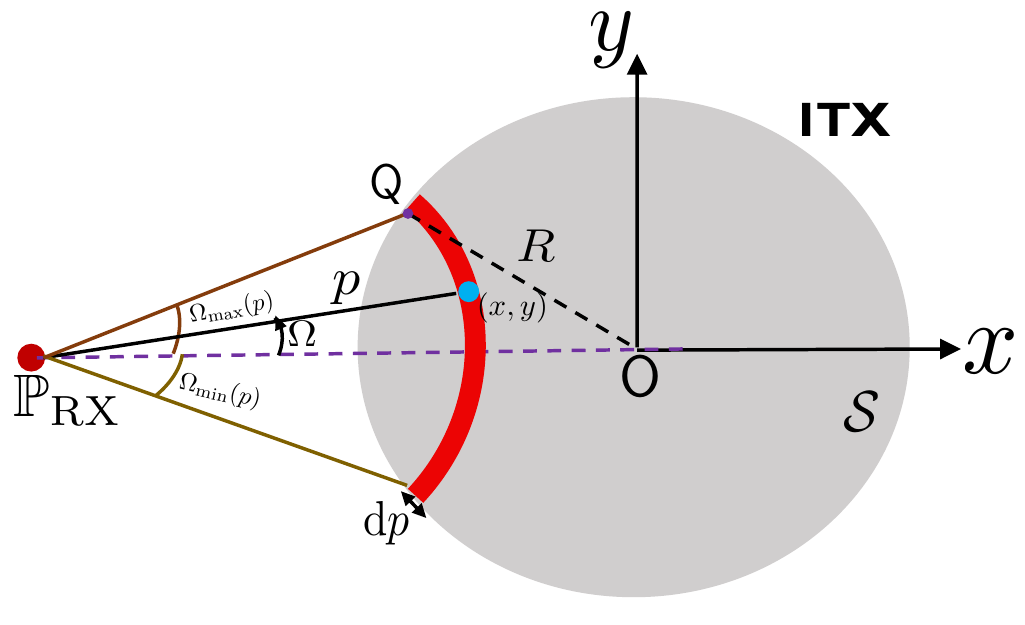}\label{fig:outside_topview}}
\caption{A scenario where $\mathbb{P}_{\mathrm{RX}}$, the projection of the RX on the plane containing the ITX, lies outside $\mathcal{S}$. Here, $\mathsf{O}$ represents the origin $(0,0,0)$ and $\mathbb{P}_{\mathrm{RX}}$ is $(-\ell \, \mathrm{sin} \gamma, 0, 0)$. The set $\mathcal{S}$ is a disc of radius $R$. In this figure, $p$ and $\Omega$ denote the distance and the angle in a polar coordinate system centered at $\mathbb{P}_{\mathrm{RX}}$. 
\normalsize}
\end{figure}

 We consider the near field system in Fig. \ref{fig:outside_sideview} where $\gamma>0$ and $\ell\, \mathrm{sin} \gamma >R$\footnote{The robust beamformer for $\gamma<0$ and $|\ell\, \mathrm{sin} \gamma| >R$ can be obtained by flipping the designed phase profile about the $y-$axis.}. To simplify $\tilde{g}(f)$ in \eqref{eq:gtild_def_1}, we use a polar coordinate system that is centered around the projection $\mathbb{P}_{\mathrm{RX}}$ instead of the origin $\mathsf{O}$. We define $p$ as the distance between $\mathbb{P}_{\mathrm{RX}}$ and a coordinate $(x,y)\in \mathcal{S}$, i.e., 
\begin{equation}
\label{eq:polar_p_defn}
p=\sqrt{(x+\ell\mathrm{sin}\gamma)^2+y^2}.
\end{equation} 
From the right-angled triangle formed by the RX, $\mathbb{P}_{\mathrm{RX}}$ and $(x,y)$, we observe that
\begin{equation}
\ell(x,y)=\sqrt{p^2+\ell^2 \mathrm{cos}^2 \gamma}.
\end{equation}
For the standard beamformer in \eqref{eq:phi_fc}, the phase applied at $(x,y)$ is proportional to the distance $\ell(x,y)$. Now, this distance is same for all the ITX coordinates that are equidistant from the projection $\mathbb{P}_{\mathrm{RX}}$, i.e., points with the same $p$. A set of such coordinates is marked by an arc within the ITX in Fig. \ref{fig:outside_sideview}. Similar to the standard phase profile $\phi_{\std}(x,y)$ which has the same phase at all these locations, the phase profile $\psi_{\mathrm{des}}(x,y)$ is assumed to be constant at all the coordinates that are equidistant from $\mathbb{P}_{\mathrm{RX}}$. Such an assumption simplifies the $2\mathrm{D}$-phase profile design problem to a $1\mathrm{D}$-function optimization problem. A more sophisticated approach could optimize the $2\mathrm{D}$-phase profile directly, but we defer this to future work.
\par  Now, we express $\tilde{g}(f)$ in \eqref{eq:gtild_def_1} using the polar coordinate system centered at $\mathbb{P}_{\mathrm{RX}}$. We define $\Omega$ as the angle made by the line joining $(x,y)$ and $\mathbb{P}_{\mathrm{RX}}$ with the $x$ axis. We note that $\Omega=\mathrm{tan}^{-1}(y/(x+\ell \mathrm{sin} \gamma))$. For the coordinates in $\mathcal{S}$, the minimum and the maximum values of $p$ are defined as $p_{1}=\ell \mathrm{sin} \gamma - R$ and $p_{2}=\ell \mathrm{sin} \gamma +R$. We use $\Omega_{\mathrm{min}}(p)$ and $\Omega_{\mathrm{max}}(p)$ to denote the smallest and the largest angles associated with points in $\mathcal{S}$ which are at a distance of $p$ from $\mathbb{P}_{\mathrm{RX}}$. These angles are measured in the anti-clockwise direction with the $x-$axis. To simplify $\tilde{g}(f)$, we first substitute $\ell(x,y)=\sqrt{p^2+\ell^2 \mathrm{cos}^2 \gamma}$ in \eqref{eq:gtild_def_1}. Then, we change $\dx \dy$ to $p \mathrm{d}p  \dOmg$ by using the Jacobian of the transformation relating the variables. The simplified integral is then 
\begin{align}
\label{eq:pout_gtilde_1}
\tilde{g}(f)&=\int_{p=p_1}^{p_2}\int_{\Omega=\Omega_{\mathrm{min}}(p)}^{\Omega_{\mathrm{max}}(p)}  \frac{1}{ 2 \pi \sqrt{p^2+\ell^2 \mathrm{cos}^2 \gamma}} e^{\imj \psi_{\mathrm{des}}(x,y)} e^{-\imj \omega \sqrt{p^2+\ell^2 \mathrm{cos}^2 \gamma}} p \mathrm{d}p  \dOmg \\
\label{eq:pout_gtilde_1b}
&=\int_{p=p_1}^{p_2}\frac{\Omega_{\mathrm{max}}(p)-\Omega_{\mathrm{min}}(p)}{ 2 \pi \sqrt{p^2+\ell^2 \mathrm{cos}^2 \gamma}} e^{\imj \psi_{\mathrm{des}}(x,y)} e^{-\imj \omega \sqrt{p^2+\ell^2 \mathrm{cos}^2 \gamma}} p \mathrm{d}p.
\end{align}
To further simplify the integral in \eqref{eq:pout_gtilde_1b}, we define a variable
\begin{equation}
u=\sqrt{p^2+\ell^2 \mathrm{cos}^2 \gamma}, 
\end{equation}
$u_1=\sqrt{p_1^2+\ell^2 \mathrm{cos}^2 \gamma}$ and $u_2=\sqrt{p_2^2+\ell^2 \mathrm{cos}^2 \gamma}$. Then, $\mathrm{d}u=p \mathrm{d}p /\sqrt{p^2+\ell^2 \mathrm{cos}^2 \gamma}$. The assumption that $\psi_{\mathrm{des}}(x,y)$ only varies with $p$, equivalently $u$, allows us to define a $1\mathrm{D}$ function $\psi(u)$ such that $\psi(u)=\psi_{\mathrm{des}}(x,y)$. The angle made by the arc in $\mathcal{S}$ at a distance of $p$ from $\mathbb{P}_{\mathrm{RX}}$ is $\Omega_{\mathrm{max}}(p)-\Omega_{\mathrm{min}}(p)$. We define a real positive function
\begin{equation}
\label{eq:a_u_defn}
a(u)=\frac{\Omega_{\mathrm{max}}(p)-\Omega_{\mathrm{min}}(p)}{2\pi}.
\end{equation}
The integral in \eqref{eq:pout_gtilde_1b} can now be expressed as
\begin{align}
\label{eq:pout_gtilde_2}
\tilde{g}(f)&=\int_{u=u_1}^{u_2} a(u) e^{\imj \psi(u)} e^{-\imj \omega u} \mathrm{d}u.
\end{align}
We observe that $a(u)$ induces an amplitude modulation effect over $e^{\imj \psi(u)}$. A closed form expression of $a(u)$ is given by \eqref{eq:explicit_amp} in Appendix-B. The objective of misfocus robust beamforming is to construct a $1\mathrm{D}$-phase profile $ \psi(u)$ that leads to an approximately flat $|\tilde{g}(f)|^2$ for $f\in [\fc-B/2, \fc+B/2]$.
\par We discuss why a linear FMCW chirp-based solution is not well suited for misfocus robust beamforming when the RX is not along the boresight direction. Similar to our derivation for the boresight setting, we define $\hat{g}(\omega)=\tilde{g}(\fc + c \omega/(2\pi))$, and use \eqref{eq:pout_gtilde_2} to write
\begin{align}
\label{eq:pout_ghat_1}
\hat{g}(\omega)&=\int_{u=-\infty}^{\infty} a(u) e^{\imj \psi(u)} \mathbb{I}_{u_1,u_2} e^{-\imj \omega u} \mathrm{d}u.
\end{align}
Now, $\psi(u)$ must be designed to achieve an approximately flat $|\hat{g}(\omega)|^2$ over $\omega\in [-\pi B/c],\pi B/c]$. Setting $e^{\imj \psi(u)} \mathbb{I}_{u_1,u_2}$ to a linear FMCW chirp, as in the boresight scenario, does not result in a flat $|\hat{g}(\omega)|^2$ due to the amplitude modulation effect induced by $a(u)$. The stationary phase method characterizes the impact of this amplitude modulation on $\hat{g}(\omega)$ under the assumption that $a(u)$ varies slowly when compared to $\psi(u)$. We use $\omega_u$ to denote the instantaneous frequency at $u$, i.e., $\omega_u=\psi'(u)$. The stationary phase method approximates $|\hat{g}(\omega_u)|^2$ as \cite{cook2012radar}
\begin{equation}
\label{eq:psi_2_eqn_1}
|\hat{g}(\omega_u)|^2 \approx \frac{2 \pi a^2(u)}{|\psi''(u)|}.
\end{equation} 
For a linear FMCW chirp, we observe that $\psi''(u)$ is constant. In such a case, the spectral magnitude $|\hat{g}(\omega_u)|$ is proportional to the amplitude modulation. Therefore, a linear FMCW-based construction for $\psi(u)$ does not result in the desired ``flat'' $|\hat{g}(\omega)|^2$ when $a(u)$ varies over $[u_1,u_2]$.
\par We derive the frequency profile of a non-linear FMCW chirp that achieves robustness to misfocus. Under the assumption that $a(u)$ varies slowly when compared to $\psi(u)$, the instantaneous frequency of $a(u) e^{\imj \psi(u)} \mathbb{I}_{u_1,u_2}$ is $\psi'(u)$. The goal of InFocus is to design a $\psi(u)$ such that the spectrum of $|\hat{g}(\omega)|^2$ is contained within $[-\pi B/c, \pi B/c]$ and is approximately uniform. To this end, we assume that $\omega_u$, the instantaneous frequency of $a(u) e^{\imj \psi(u)} \mathbb{I}_{u_1,u_2}$, is a continuous function that increases from $\psi'(u_1)=- \pi B/c$ to $\psi'(u_2)= \pi B/c$. The increase, however, can be non-linear and depends on $a(u)$. We assume that $\psi''(u)>0\, \forall u \in [u_1,u_2]$. From \eqref{eq:psi_2_eqn_1}, we observe that the stationary phase method relates $a(u)$ and $\psi(u)$ as 
\begin{equation}
\label{eq:psi_2_eqn_2}
\psi''(u)\approx \frac{2 \pi a^2(u)}{|\hat{g}(\omega_u)|^2 }.
\end{equation}
For $u\in [u_1, u_2]$, the instantaneous frequency $\omega_u \in [-\pi B/c, \pi B/c]$. It can be observed from \eqref{eq:psi_2_eqn_2} that a phase profile which achieves a flat $|\hat{g}(\omega)|^2$ for $\omega \in [-\pi B/c, \pi B/c]$ satisfies
\begin{equation}
\label{eq:psi_2_eqn_3}
\psi''(u)=\kappa a^2(u),
\end{equation}
for some positive constant $\kappa$. The instantaneous frequency in \eqref{eq:psi_2_eqn_3} can be determined using $\psi'(u)=\int_{u_1}^{u}\psi''(u) \mathrm{d}u$, $\psi'(u_1)=-\pi B/c$ and $\psi'(u_2)=\pi B/c$. The solution is given by 
\begin{equation}
\label{eq:psi_1_eqn_1}
\psi'(u)=\frac{2 \pi B \int_{u_1}^{u}a^2(u) \mathrm{d}u}{c \int_{u_1}^{u_2}a^2(u) \mathrm{d}u} - \frac{\pi B}{c}.
\end{equation}
In this paper, we compute the integral of $a^2(u)$ in \eqref{eq:psi_1_eqn_1} through numerical integration. The non-uniform nature of $a^2(u)$ over $[u_1, u_2]$ results in a non-linear frequency profile $\psi'(u)$. 
\par
\begin{figure}[h!]
\centering
\subfloat[Amplitude modulation function.]{\includegraphics[trim=1cm 6.25cm 2.5cm 7.5cm,clip=true,width=4.5cm, height=4.5cm]{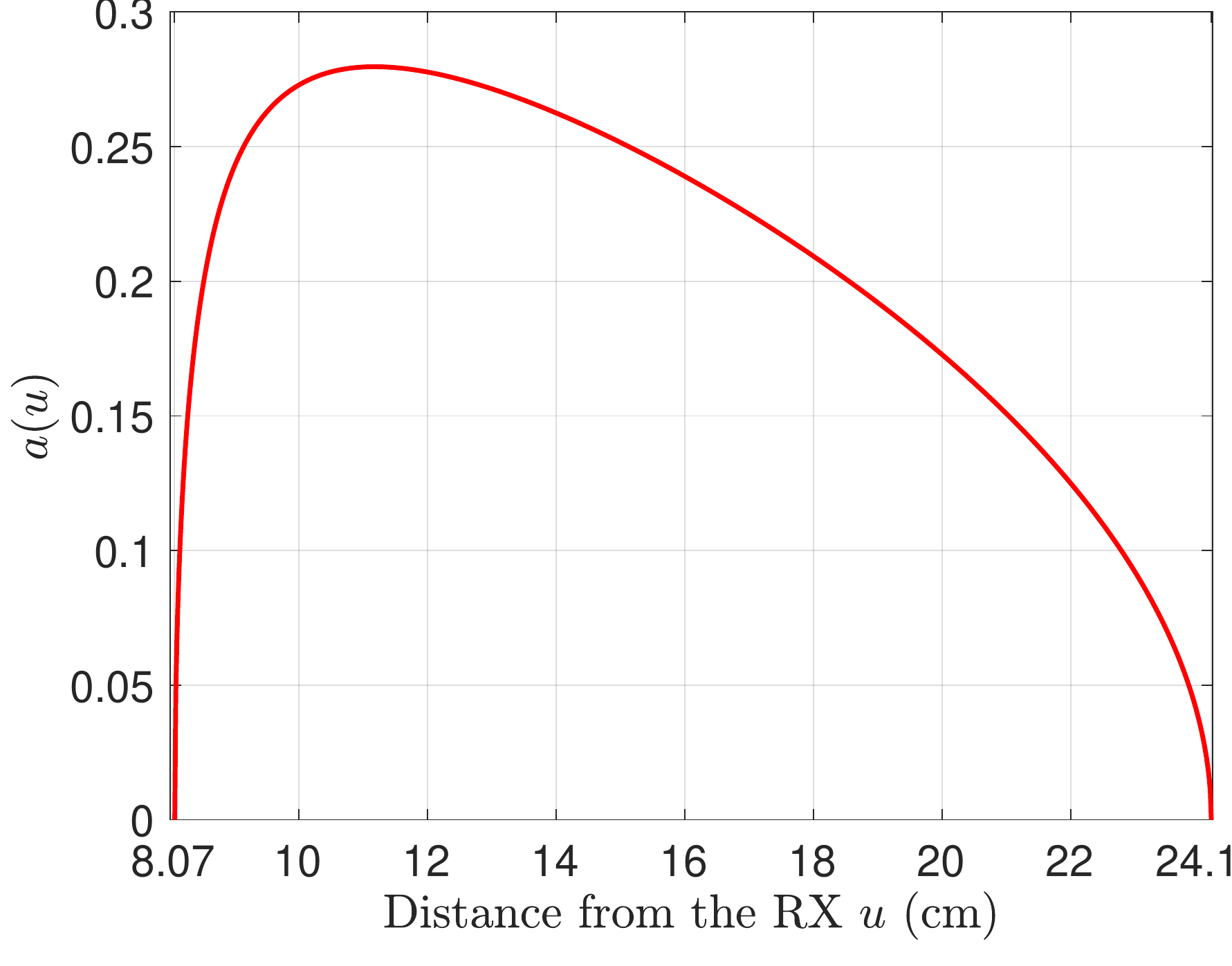}\label{fig:outside_amplitude}}
\:\:\:
\subfloat[Instantaneous frequency $\psi'(u)$.]{\includegraphics[trim=1cm 6.25cm 2.5cm 7.5cm,clip=true,width=4.5cm, height=4.5cm]{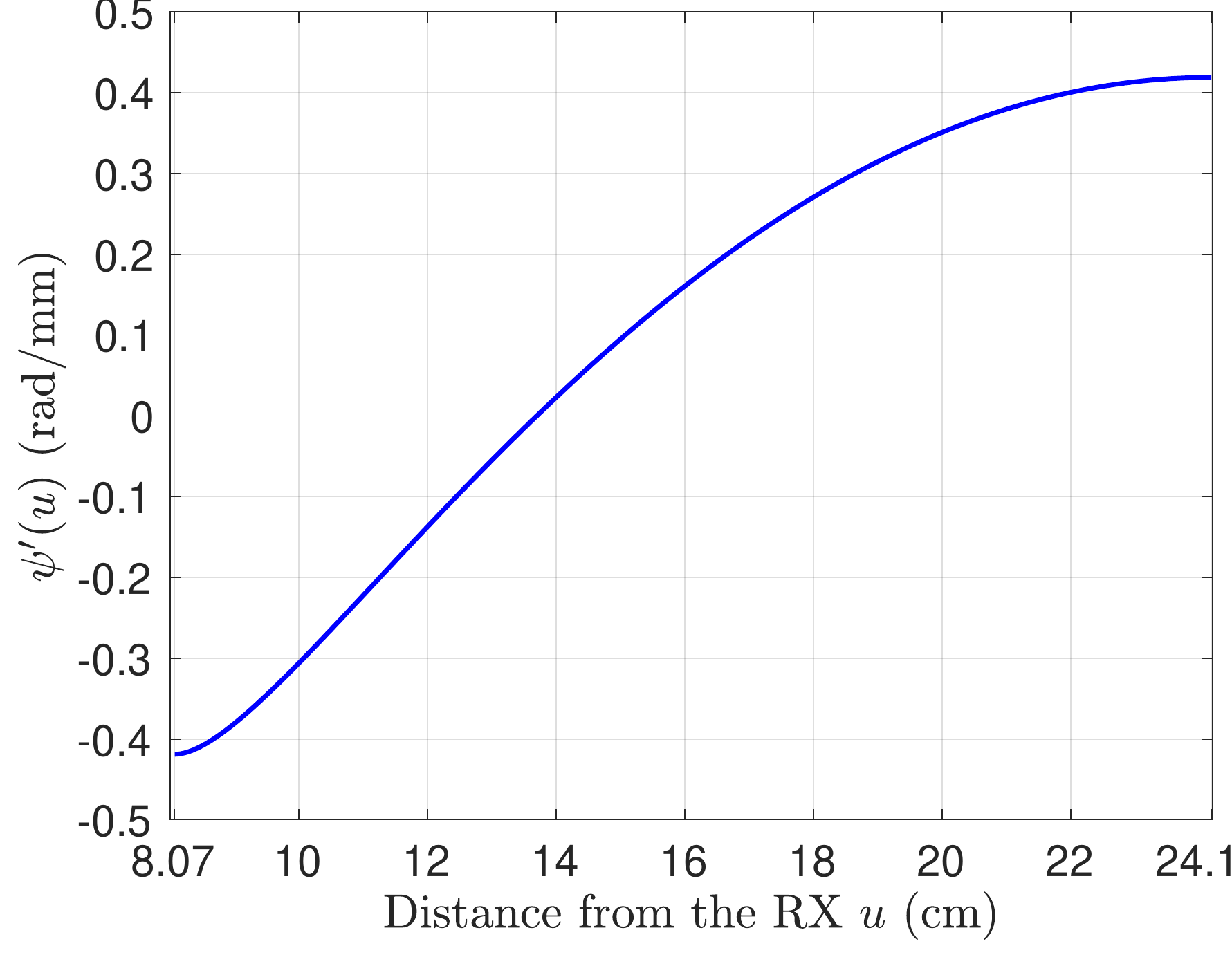}\label{fig:outside_freq_u}}
\:\:\:
\subfloat[Designed phase profile $\psi(u)$.]{\includegraphics[trim=1cm 6.25cm 2.5cm 7.5cm,clip=true,width=4.5cm, height=4.5cm]{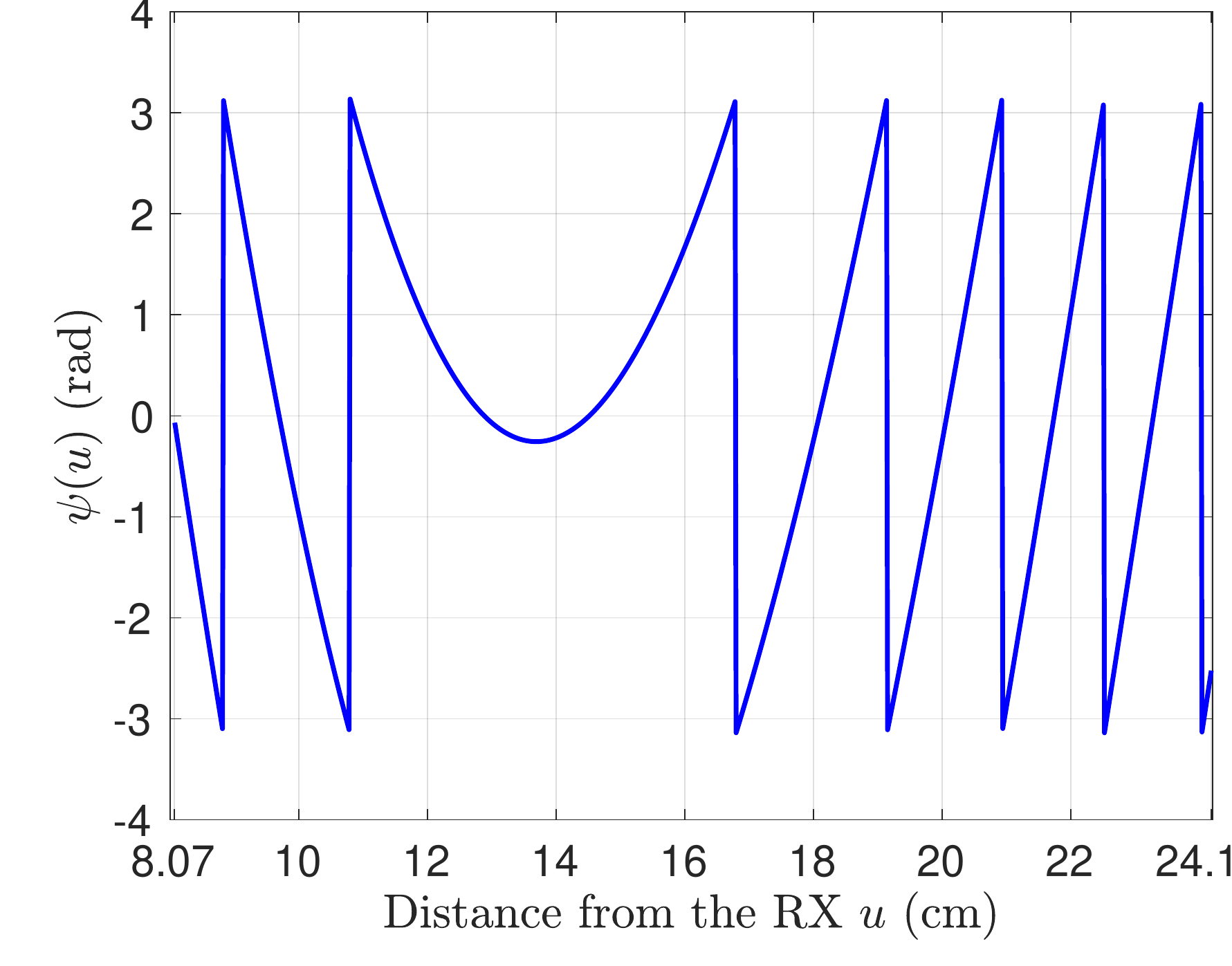}\label{fig:outside_phase_u}}
\caption{Here, we show the amplitude modulation function $a(u)$ for a near field scenario with $\ell=15\, \mathrm{cm}$, $R= 10 \, \mathrm{cm}$ and $\gamma= 60 \degree$. In this example, $B=40\, \mathrm{GHz}$ and $\fc= 300\, \mathrm{GHz}$. We observe that $\psi'(u)$,  i.e., the instantaneous frequency of the designed chirp, is a non-linear function of $u$. The phase profile $\psi(u)$ is the integral of $\psi'(u)$. 
\normalsize}
\end{figure}

 We derive the phase profile $\psi(s)$ from \eqref{eq:psi_1_eqn_1} for the near field system in Fig. \ref{fig:outside_sideview}. The solution to the differential equation in \eqref{eq:psi_1_eqn_1}, i.e., 
\begin{equation}
\psi(u)=\int_{u_1}^{u} \psi'(u) \mathrm{d}u,
\end{equation}
is computed using numerical integration. For a near field system with $\ell=15\, \mathrm{cm}$, $R=10\, \mathrm{cm}$ and $\gamma=60 \degree$, it can be noticed that $\ell\, \mathrm{sin} \gamma > R$ and $\mathbb{P}_{\mathrm{RX}} \notin \mathcal{S}$. Here, $u_1=8.07\, \mathrm{cm}$ and $u_2=24.18\, \mathrm{cm}$. The amplitude modulation function $a(u)$ for this example is shown in Fig. \ref{fig:outside_amplitude}. The second derivative of the designed phase profile, i.e., $\psi''(u)$, is proportional to $a^2(u)$ by the stationary phase equation in \eqref{eq:psi_2_eqn_3}. The instantaneous frequency of the chirp, i.e., $\psi'(u)$ and the phase profile  $\psi(u)$ are shown in Fig. \ref{fig:outside_freq_u} and Fig. \ref{fig:outside_phase_u}. It can be observed from Fig. \ref{fig:outside_freq_u} that the rate of change of the instantaneous frequency is small when the amplitude modulation function is low. Due to this slow increase, the dwell time of the chirp at this frequency is longer. The longer dwell time at such frequencies helps compensate for the low amplitude scaling and achieves a flat frequency spectrum in the desired range.
\par
\begin{figure}[h!]
\centering
\subfloat[Phase profile $\psi_{\mathrm{des}}(x,y)$.]{\includegraphics[trim=1cm 6.25cm 2.5cm 7.5cm,clip=true,width=4.5cm, height=4.5cm]{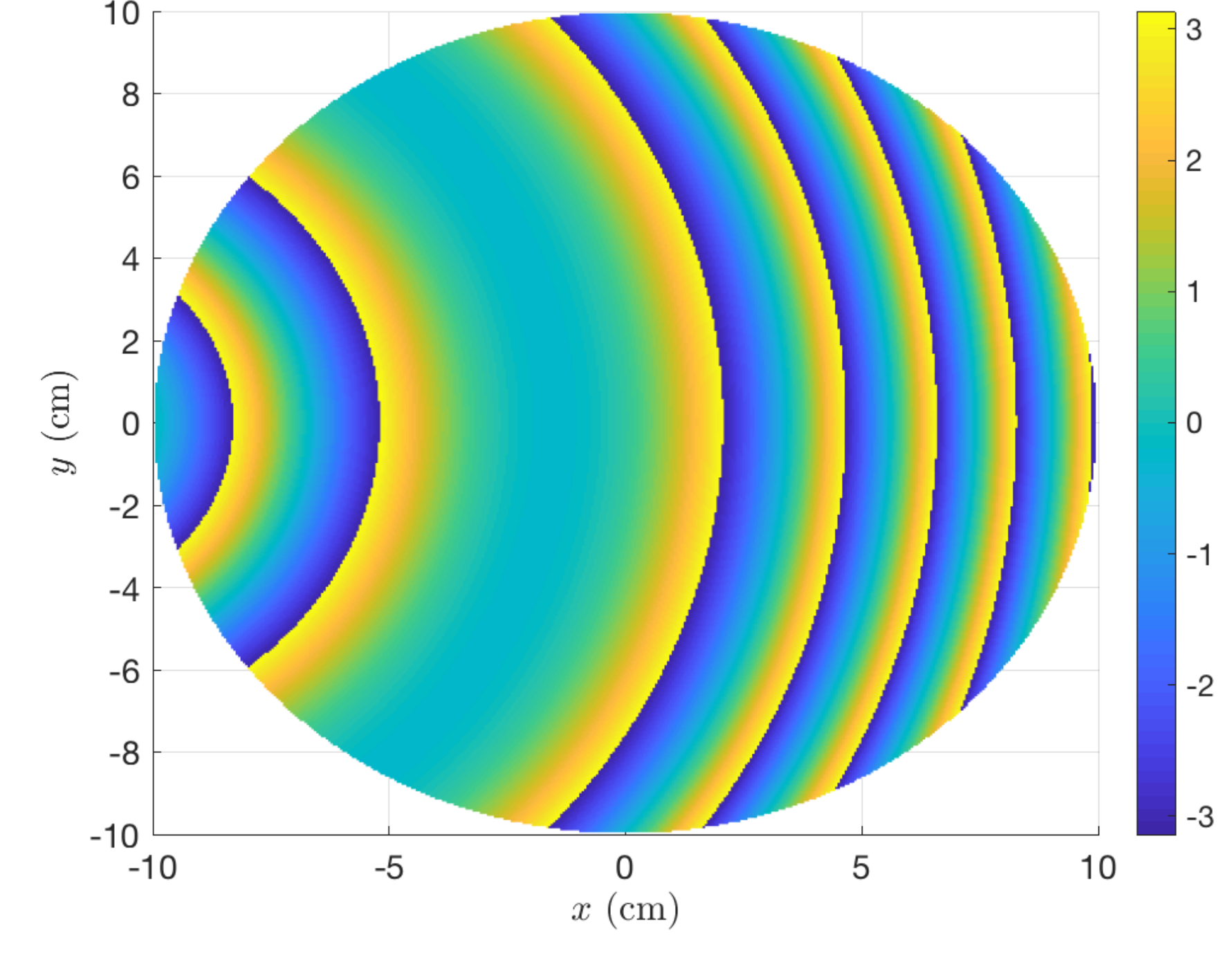}\label{fig:outside_psi_des}}
\:\:\:
\subfloat[$\phi_{\std}(x,y)+\psi_{\mathrm{des}}(x,y)$.]{\includegraphics[trim=1cm 6.25cm 2.5cm 7.5cm,clip=true,width=4.5cm, height=4.5cm]{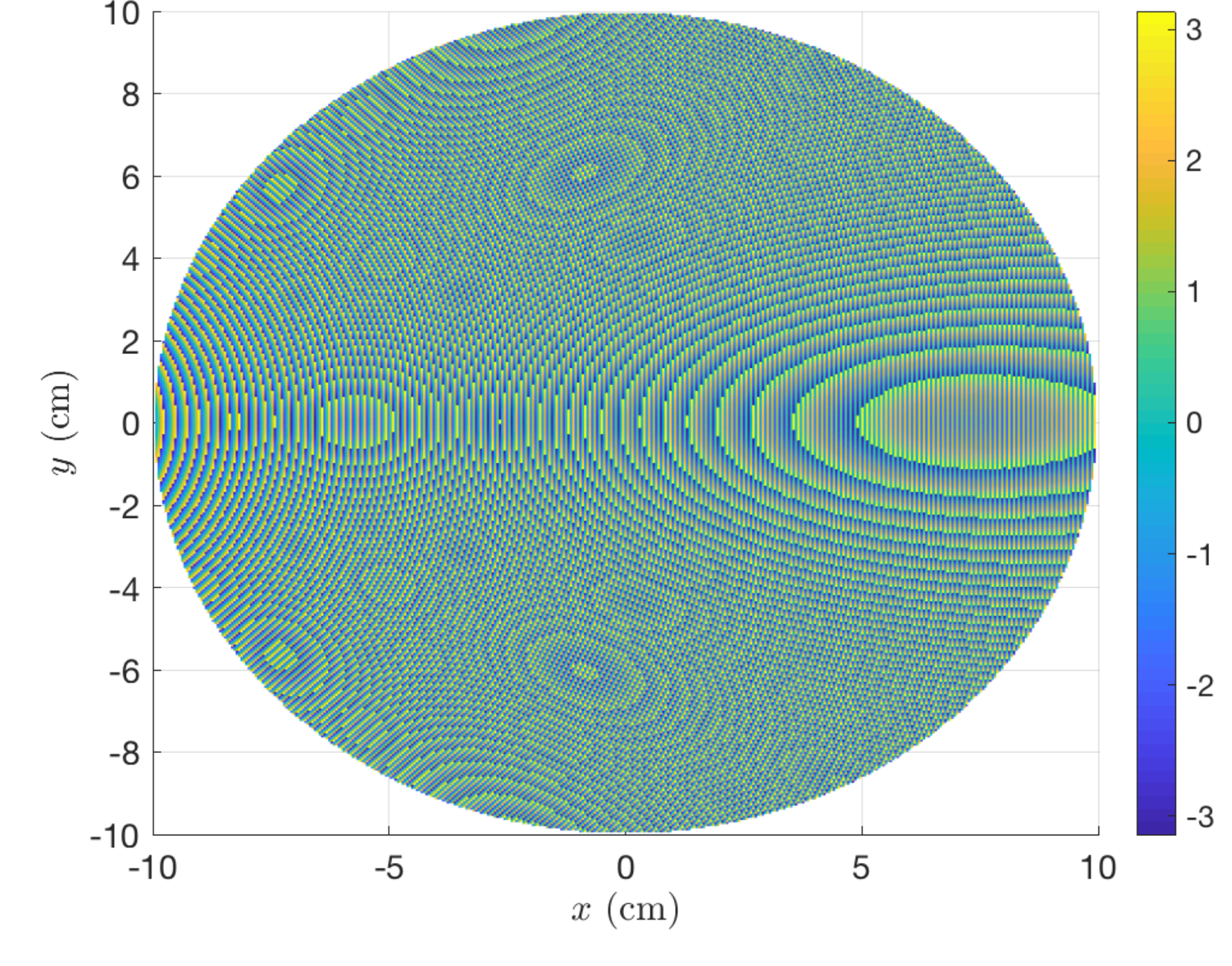}\label{fig:outside_phase_applied}}
\:\:\:
\subfloat[$20\,\mathrm{log}_{10}|g(f)|$ with frequency.]{\includegraphics[trim=1cm 6.25cm 2.5cm 7.5cm,clip=true,width=4.5cm, height=4.5cm]{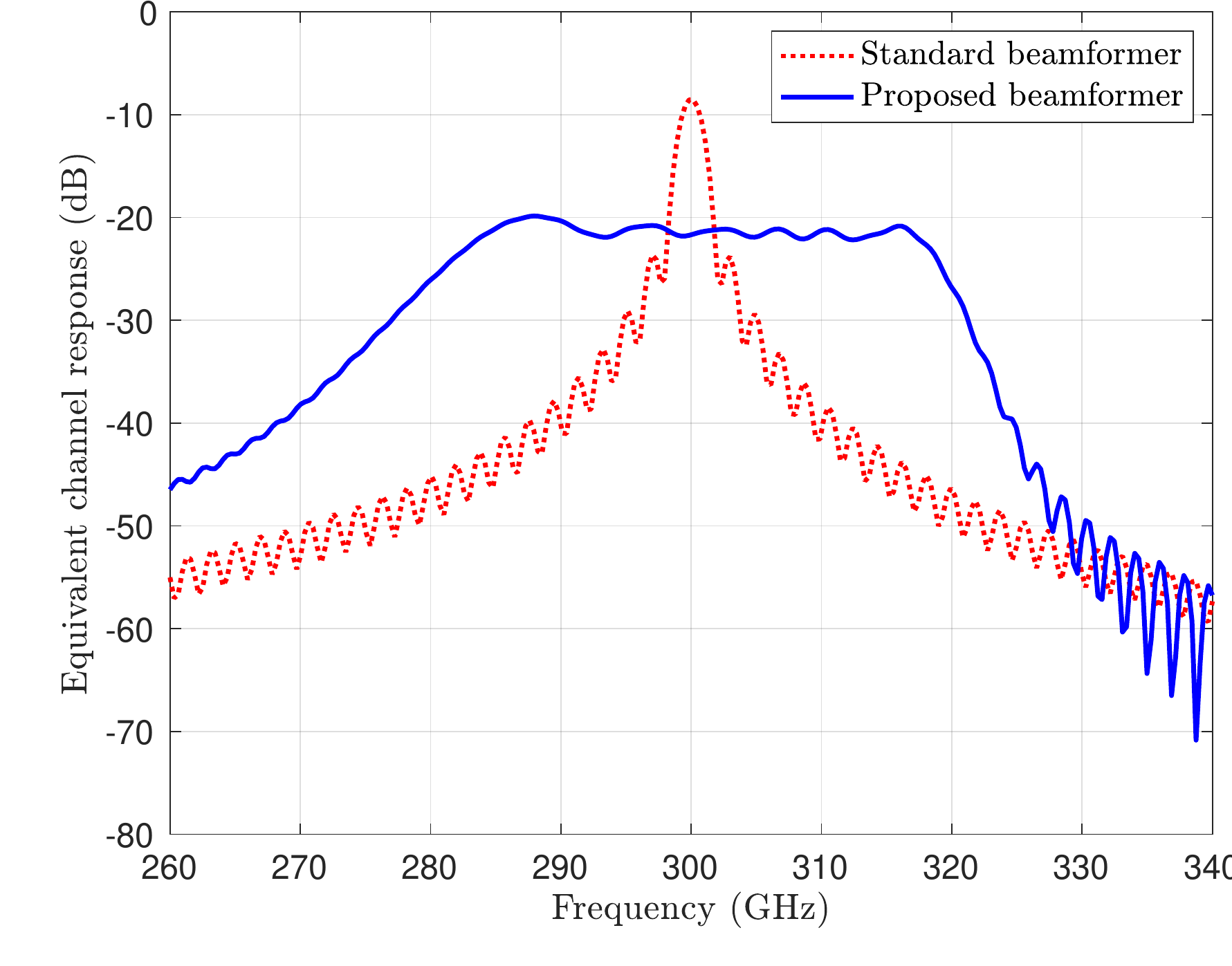}\label{fig:outside_freq_resp}}
\caption{ The $2\mathrm{D}$-phase profile $\psi_{\mathrm{des}}(x,y)$ designed with InFocus for a near field scenario where $\mathbb{P}_{\mathrm{RX}} \notin \mathcal{S}$ is shown in Fig. \ref{fig:outside_psi_des}. When the TX applies the phase profile in Fig. \ref{fig:outside_phase_applied}, the RX observes the frequency domain channel in Fig. \ref{fig:outside_freq_resp}. InFocus achieves an approximately flat channel response over the desired bandwidth of $40\, \mathrm{GHz}$.
\normalsize}
\end{figure}

 Now, we demonstrate the performance of the beamformer associated with the designed $1\mathrm{D}$ function $\psi(u)$. The $2\mathrm{D}$ phase profile $\psi_{\mathrm{des}}(x,y)$ is given by 
\begin{equation}
\label{eq:map_1d_2d}
\psi_{\mathrm{des}}(x,y)=\psi(\sqrt{(x+\ell \mathrm{sin}\gamma)^2+y^2+\ell^2 \mathrm{cos}^2\gamma}), 
\end{equation}
and the phase profile applied at the TX is $\phi(x,y)=\phi_{\std}(x,y)+\psi_{\mathrm{des}}(x,y)$. To illustrate our design, we consider a $40\, \mathrm{GHz}$ bandwidth system operating at $\fc= 300 \,\mathrm{GHz}$. We use $\ell=15\, \mathrm{cm}$, $R=10\, \mathrm{cm}$ and $\gamma=60 \degree$. In Fig. \ref{fig:outside_psi_des}, we show the designed phase profile $\psi_{\mathrm{des}}(x,y)$. The proposed phase profile $\phi_{\std}(x,y)+\psi_{\mathrm{des}}(x,y)$ is shown in Fig.  \ref{fig:outside_phase_applied} and the frequency response of the equivalent SISO channel is shown in Fig.  \ref{fig:outside_freq_resp}. We observe from Fig. \ref{fig:outside_freq_resp} that the proposed phase profile achieves an approximately flat beamforming gain over the desired bandwidth.
\subsubsection{Solution when $\mathbb{P}_{\mathrm{RX}} \in \mathcal{S}$}\label{sec:projection_inside}
\begin{figure}[h!]
\vspace{-3mm}
\centering
\subfloat[A near field scenario where $\mathbb{P}_{\mathrm{RX}} \in \mathcal{S}$.]{\includegraphics[trim=0cm 0cm 0cm 0cm,clip=true,width=7.5cm, height=3.5cm]{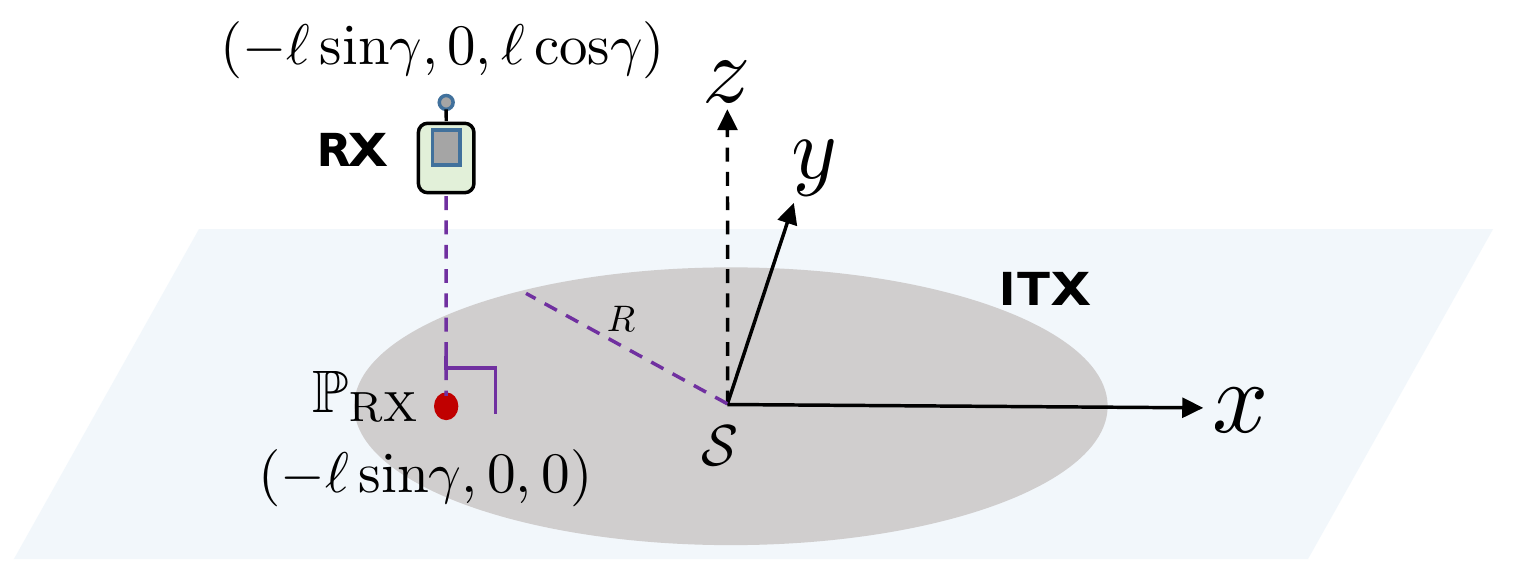}\label{fig:inside_sideview}}
\:\: \:\: \:\:
\subfloat[Top view of the system in Fig. \ref{fig:inside_sideview}.]{\includegraphics[trim=0cm 0cm 0cm 0cm,clip=true,width=5.5cm, height=3.5cm]{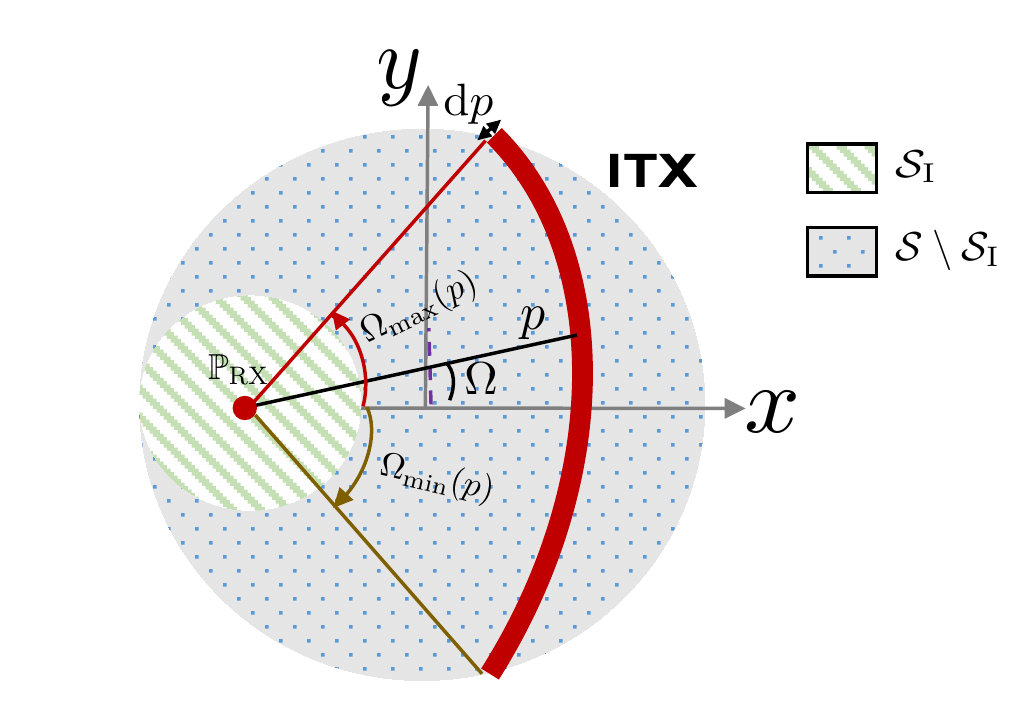}\label{fig:inside_topview}}
\caption{A scenario where $\mathbb{P}_{\mathrm{RX}}$, the projection of the RX on the plane containing the ITX, lies inside $\mathcal{S}$ which is a disc of radius $R$. The set $\mathcal{S}_{\mathrm{I}}$ contains the ITX coordinates which are within a distance of $R- \ell \, \mathrm{sin} \gamma $ from $\mathbb{P}_{\mathrm{RX}}$.
\normalsize}
\end{figure}

We now construct the misfocus robust phase profile for a near field scenario in Fig. \ref{fig:inside_sideview} where $\mathbb{P}_{\mathrm{RX}} \in \mathcal{S}$. In this scenario, $0<\ell\, \mathrm{sin} \gamma <R$. Similar to our assumption in Sec. \ref{sec:projection_outside}, we assume that the phase profile $\psi_{\mathrm{des}}(x,y)$ is constant for all the ITX coordinates that are equidistant from $\mathbb{P}_{\mathrm{RX}}$. To construct a robust $\psi_{\mathrm{des}}(x,y)$, we first design a $1\mathrm{D}$-phase function $\psi(u)$. Here, $u$ represents the distance between an ITX coordinate and the RX, and $u \in [\ell, \sqrt{(R+\ell\,\mathrm{sin}\gamma)^2+\ell^2}]$ when $\mathbb{P}_{\mathrm{RX}} \in \mathcal{S}$. We split the set of ITX coordinates, i.e., $\mathcal{S}$, into $\mathcal{S}_{\mathrm{I}}$ and its complement $\mathcal{S} \setminus \mathcal{S}_{\mathrm{I}}$ as shown in Fig. \ref{fig:inside_topview}. Here, $\mathcal{S}_{\mathrm{I}}$ represents the set of all ITX coordinates that are within a distance of $R- \ell \, \mathrm{sin} \gamma $ from $\mathbb{P}_{\mathrm{RX}}$. We observe that the RX is along the boresight of the array corresponding to $\mathcal{S}_{\mathrm{I}}$, and the projection of the RX lies outside $\mathcal{S} \setminus \mathcal{S}_{\mathrm{I}}$. In this section, we show how splitting $\mathcal{S}$ into $\mathcal{S}_{\mathrm{I}}$ and its complement allows us to reuse the results in Sec. \ref{sec:boresight_section} and Sec. \ref{sec:projection_outside}. 
\par We obtain a compact representation of $\tilde{g}(f)$, an approximation of the equivalent SISO channel. The integral in \eqref{eq:gtild_def_1} can be evaluated over the two regions $\mathcal{S}_{\mathrm{I}}$ and $\mathcal{S} \setminus \mathcal{S}_{\mathrm{I}}$ as
\begin{equation}
\label{eq:gtild_in_1}
 \tilde{g}(f)=\underbrace{\int_{\mathcal{S}_{\mathrm{I}}} \frac{1}{ 2 \pi \ell(x,y)} e^{\imj \psi_{\mathrm{des}}(x,y)} e^{-\imj \omega \ell(x,y)} \dx \dy}_{T_1} + \underbrace{\int_{\mathcal{S} \setminus \mathcal{S}_{\mathrm{I}}} \frac{1}{ 2 \pi \ell(x,y)} e^{\imj \psi_{\mathrm{des}}(x,y)} e^{-\imj \omega \ell(x,y)} \dx \dy}_{T_2}.
\end{equation}
To simplify \eqref{eq:gtild_in_1}, we use a circular coordinate system with $\mathbb{P}_{\mathrm{RX}}$ as the center. The radius and the angle in this system are denoted by $p$ and $\Omega$. We observe that $\ell(x,y)=\sqrt{p^2+\ell^2 \mathrm{cos}^2 \gamma}$ and set $u=\sqrt{p^2+\ell^2 \mathrm{cos}^2 \gamma}$. Now, the first term $T_1$ in \eqref{eq:gtild_in_1} involves an integral over $ \mathcal{S}_{\mathrm{I}}$, i.e., a disc of radius $R- \ell \, \mathrm{sin} \gamma$. This integral has the same structure as \eqref{eq:gtild_def_2} and can be simplified to 
\begin{equation}
T_1=\int_{u=\ell\, \mathrm{cos} \gamma}^{\sqrt{\ell^2\mathrm{cos}^2 \gamma+(R- \ell \, \mathrm{sin} \gamma)^2}}e^{\imj \psi(u)} e^{-\imj \omega u } \mathrm{d}u.
\end{equation}
For the integral over $\mathcal{S} \setminus \mathcal{S}_{\mathrm{I}}$, it can be shown that the second term $T_2$ in \eqref{eq:gtild_in_1} takes the same form as \eqref{eq:pout_gtilde_1}. The limits of integration, however, are different as $p\in [R-\ell\, \mathrm{sin} \gamma, R+\ell\, \mathrm{sin} \gamma]$ for the ITX coordinates in $ \mathcal{S} \setminus \mathcal{S}_{\mathrm{I}}$. Using the same arguments in Sec. \ref{sec:projection_outside}, we express $T_2$ as
\begin{equation}
T_2=\int_{u=\sqrt{\ell^2\mathrm{cos}^2 \gamma+(R- \ell \, \mathrm{sin} \gamma)^2}}^{\sqrt{\ell^2\mathrm{cos}^2 \gamma+(R+ \ell \, \mathrm{sin} \gamma)^2}} a(u) e^{\imj \psi(u)} e^{-\imj \omega u } \mathrm{d}u.
\end{equation}
To express $\tilde{g}(f)$ in compact form, we define a new amplitude modulation function 
\begin{equation}
b(u)=\left\{
                \begin{array}{ll}
                  1 \;\;\; \; \;\;\; \; \;\;\; \;\;\;  \;\;\; \;\;\; \ell\, \mathrm{cos} \gamma \leq u \leq \sqrt{\ell^2\mathrm{cos}^2 \gamma+(R- \ell \, \mathrm{sin} \gamma)^2}\\
                  a(u) \;\;\; \;\;\;  \;\;\; \;\;\; \sqrt{\ell^2\mathrm{cos}^2 \gamma+(R- \ell \, \mathrm{sin} \gamma)^2} < u \leq \sqrt{\ell^2\mathrm{cos}^2 \gamma+(R+ \ell \, \mathrm{sin} \gamma)^2}
                \end{array}
              \right. .
\end{equation}
Substituting $T_1$ and $T_2$ in \eqref{eq:gtild_in_1}, we can express $\tilde{g}(f)$ as 
\begin{equation}
\label{eq:gtild_in_2}
\tilde{g}(f)=\int_{u=\ell\, \mathrm{cos} \gamma}^{\sqrt{\ell^2\mathrm{cos}^2 \gamma+(R+ \ell \, \mathrm{sin} \gamma)^2}} b(u) e^{\imj \psi(u)} e^{-\imj \omega u} \mathrm{d}u.
\end{equation}
As \eqref{eq:gtild_in_2} has the same structure as \eqref{eq:pout_gtilde_2}, the stationary phase method can be used to design $\psi(u)$, i.e., the phase profile of the non-linear FMCW chirp, so that $|\tilde{g}(f)|^2$ is ``uniform'' over the desired bandwidth.
\begin{figure}[h!]
\centering
\subfloat[Amplitude modulation function.]{\includegraphics[trim=1cm 6.25cm 2.5cm 7.5cm,clip=true,width=4.5cm, height=4.5cm]{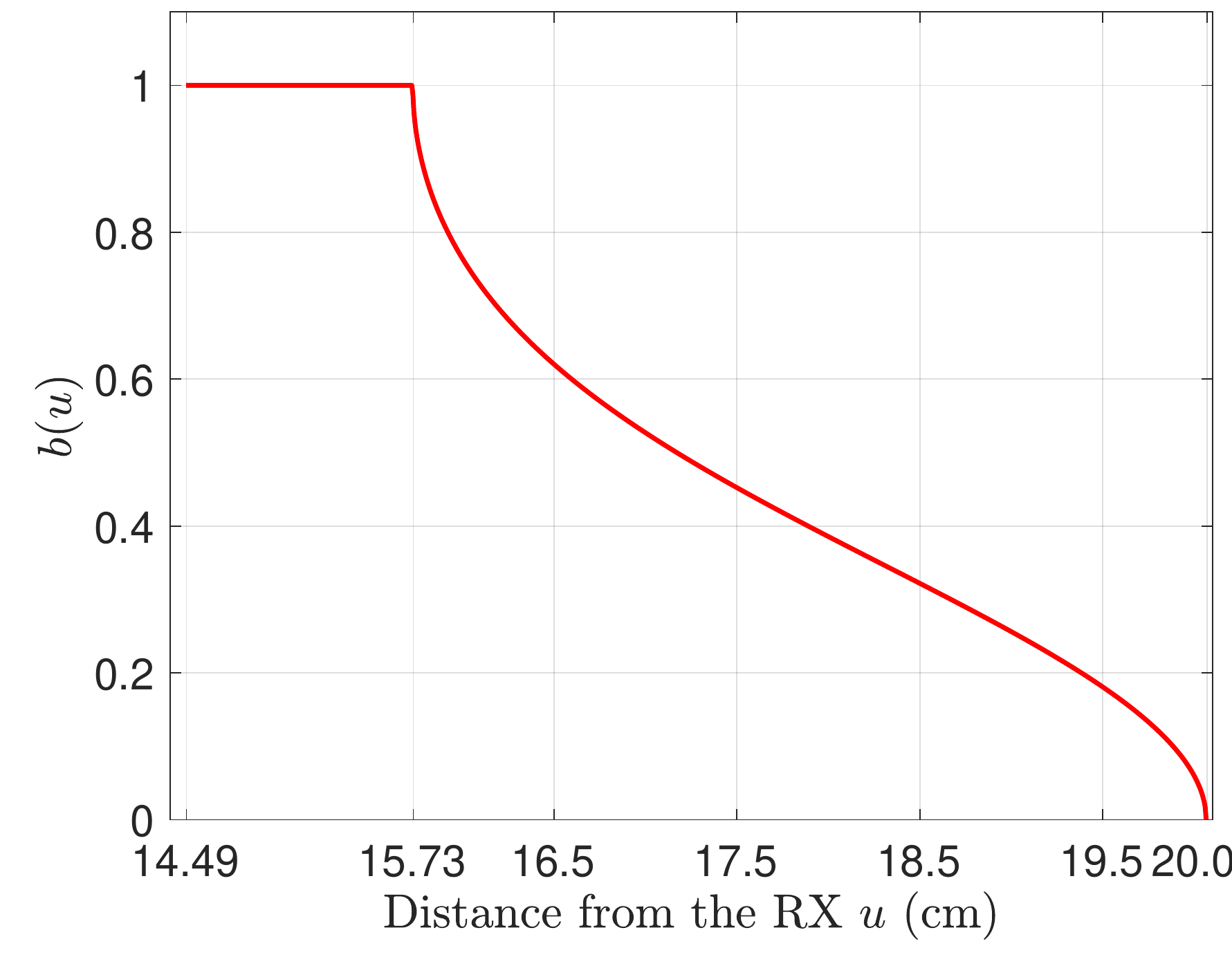}\label{fig:inside_amplitude}}
\:\:\:
\subfloat[Instantaneous frequency $\psi'(u)$.]{\includegraphics[trim=1cm 6.25cm 2.5cm 7.5cm,clip=true,width=4.5cm, height=4.5cm]{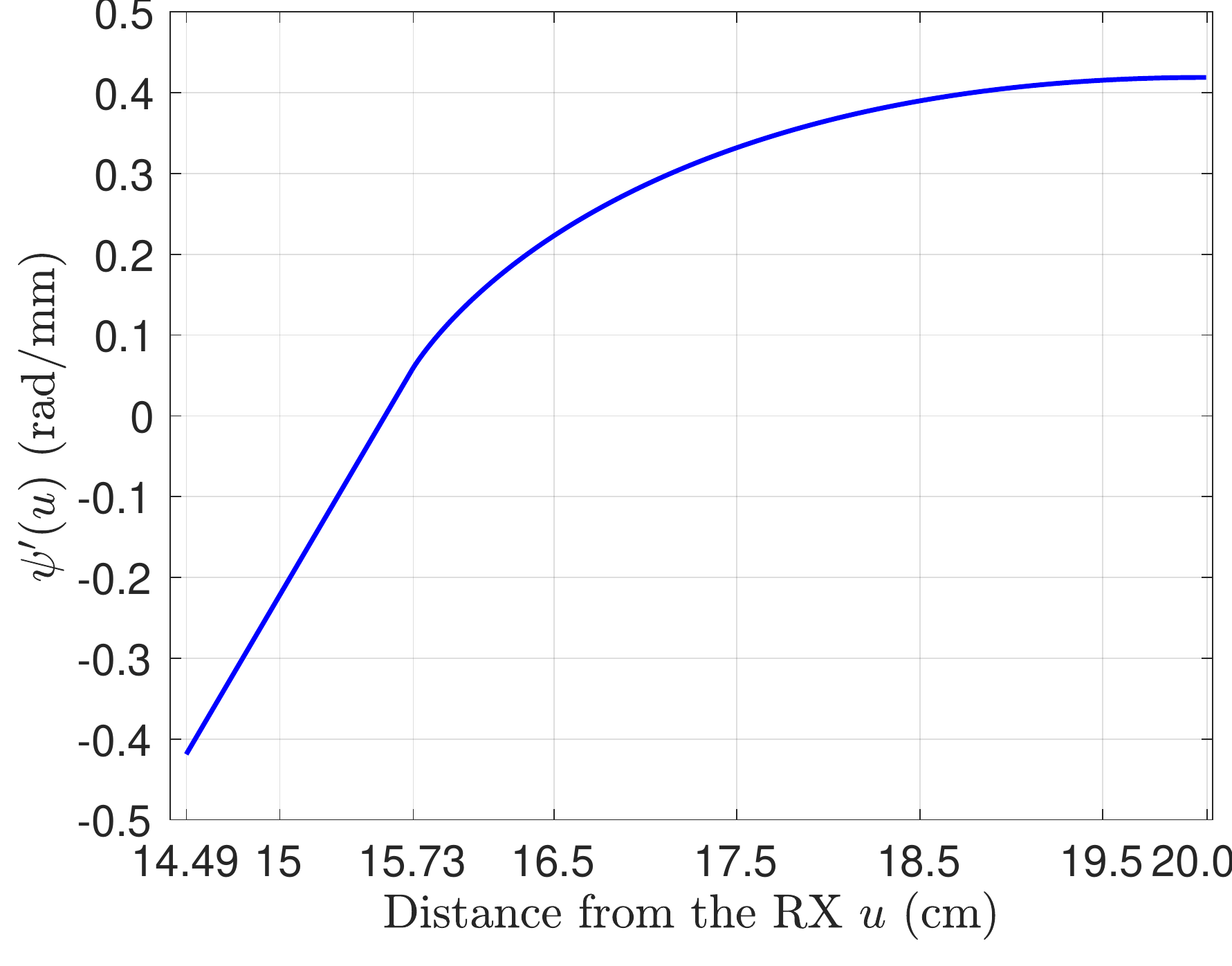}\label{fig:inside_freq_u}}
\:\:\:
\subfloat[Designed phase profile $\psi(u)$.]{\includegraphics[trim=1cm 6.25cm 2.5cm 7.5cm,clip=true,width=4.5cm, height=4.5cm]{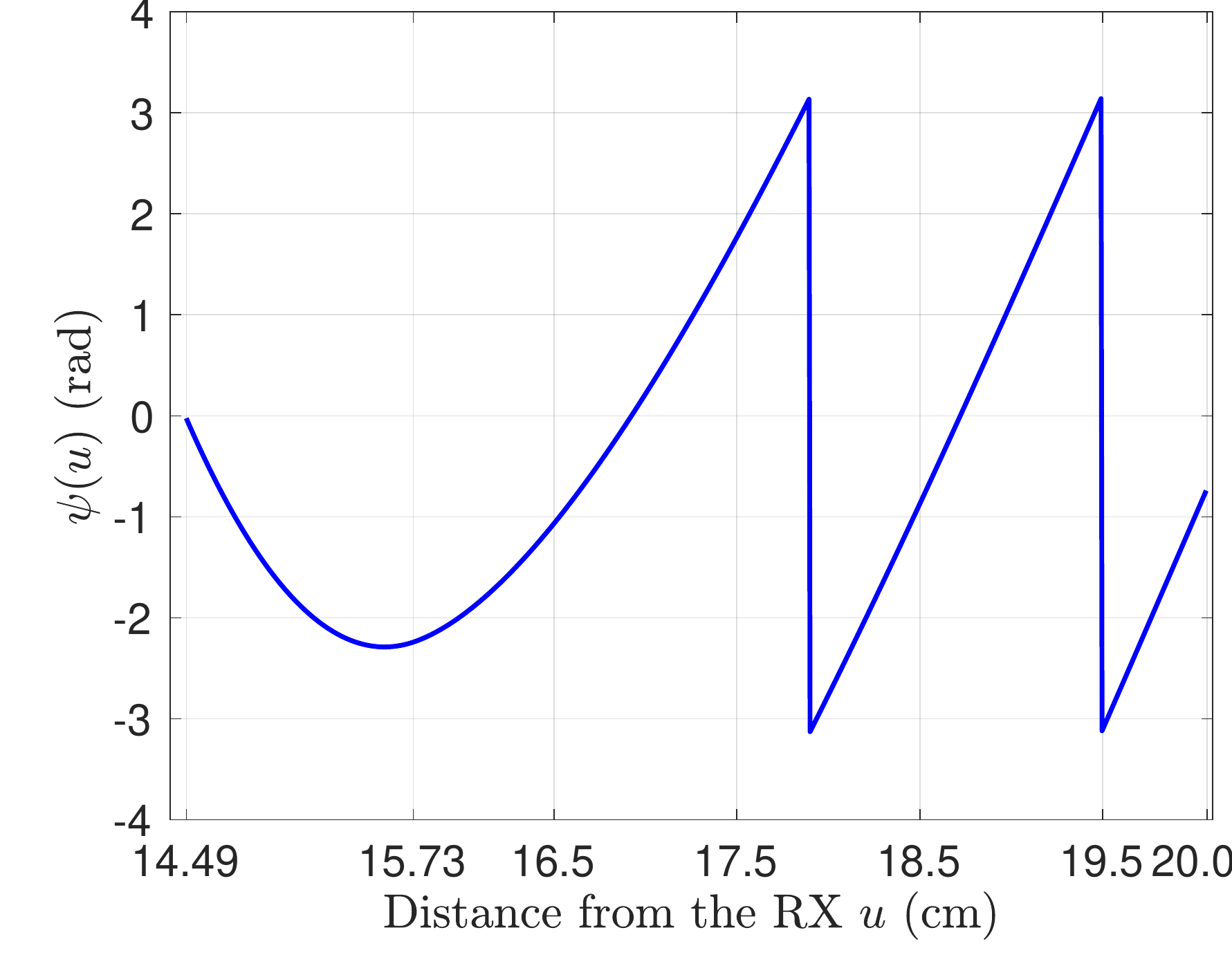}\label{fig:inside_phase_u}}
\caption{ The amplitude modulation function $b(u)$ is $1$ for the ITX coordinates that lie within $\mathcal{S}_{\mathrm{I}}$. In this example, $\ell=15\, \mathrm{cm}$, $\gamma= 15 \degree$ and $R=10\, \mathrm{cm}$. A bandwidth of $40\, \mathrm{GHz}$ is used at $\fc= 300\, \mathrm{GHz}$. The instantaneous frequency and the phase profile of the designed chirp are shown in Fig. \ref{fig:inside_freq_u} and Fig. \ref{fig:inside_phase_u}.
\normalsize}
\end{figure}
\par
\begin{figure}[h!]
\vspace{-2mm}
\centering
\subfloat[Phase profile $\psi_{\mathrm{des}}(x,y)$.]{\includegraphics[trim=1cm 6.25cm 2.5cm 7.5cm,clip=true,width=4.5cm, height=4.5cm]{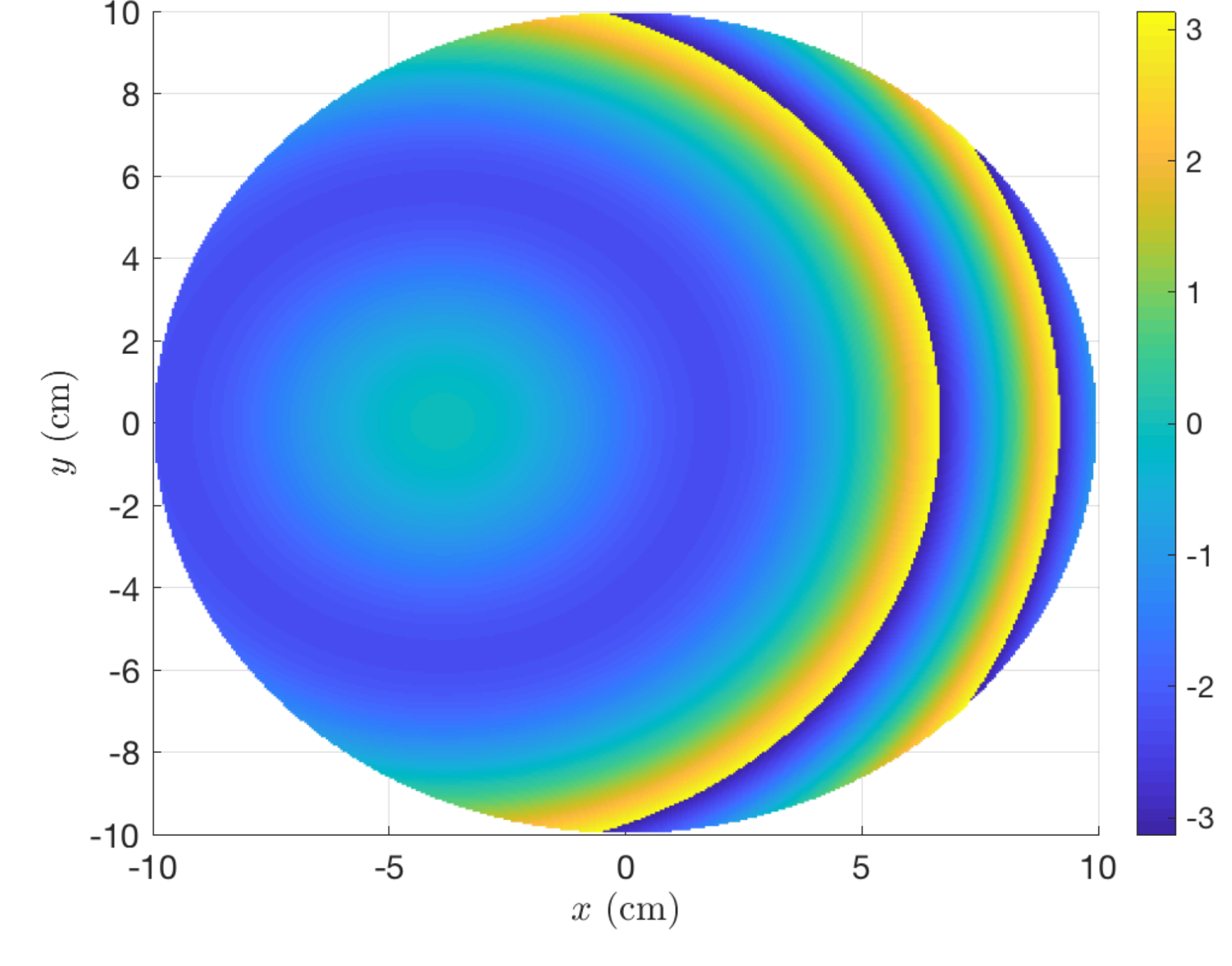}\label{fig:inside_psi_des}}
\:\:\:
\subfloat[$\psi_{\mathrm{des}}(x,y)+\phi_{\std}(x,y)$.]{\includegraphics[trim=1cm 6.25cm 2.5cm 7.5cm,clip=true,width=4.5cm, height=4.5cm]{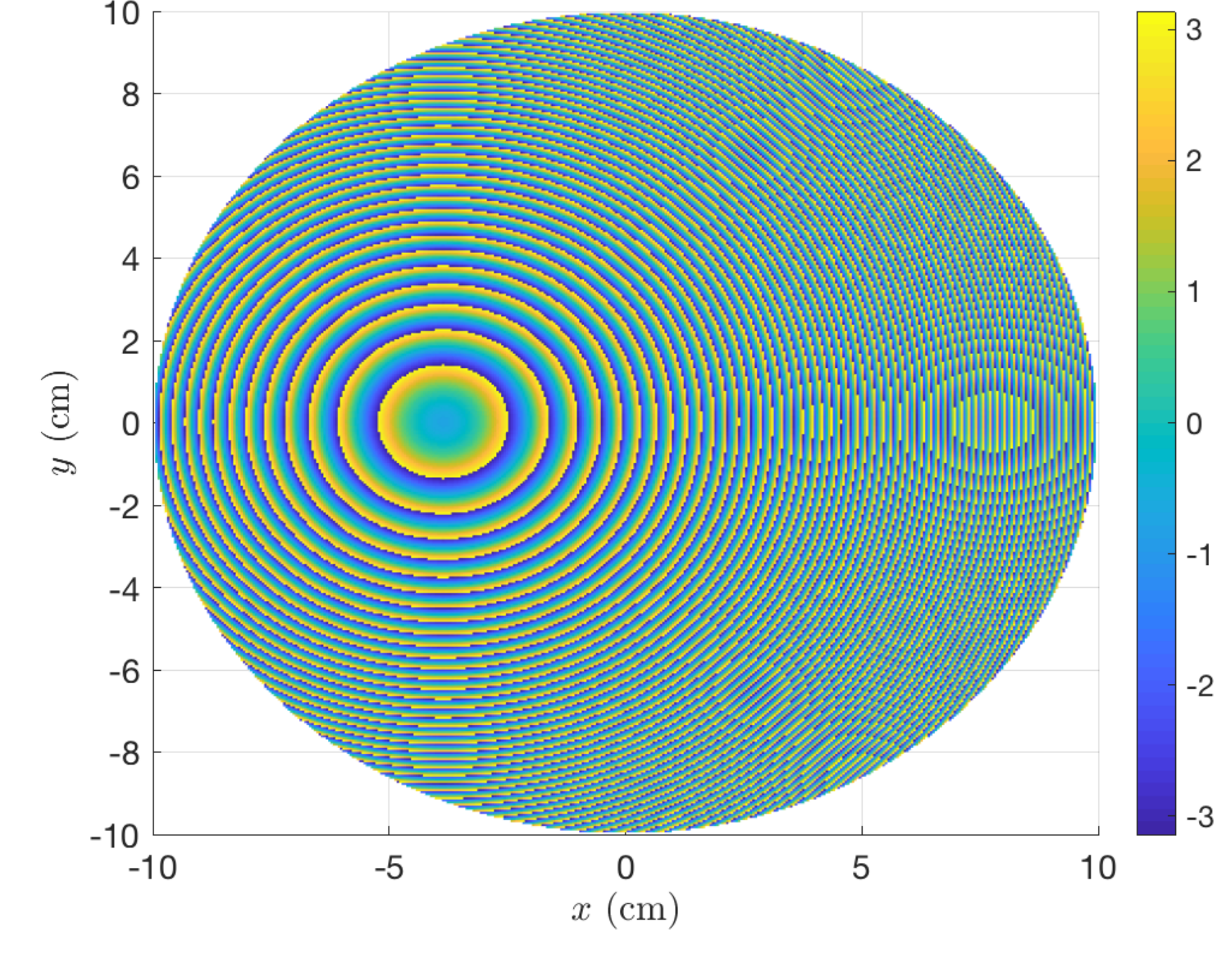}\label{fig:inside_phase_applied}}
\:\:\:
\subfloat[$20\,\mathrm{log}_{10}|g(f)|$ with frequency.]{\includegraphics[trim=1cm 6.25cm 2.5cm 7.5cm,clip=true,width=4.5cm, height=4.5cm]{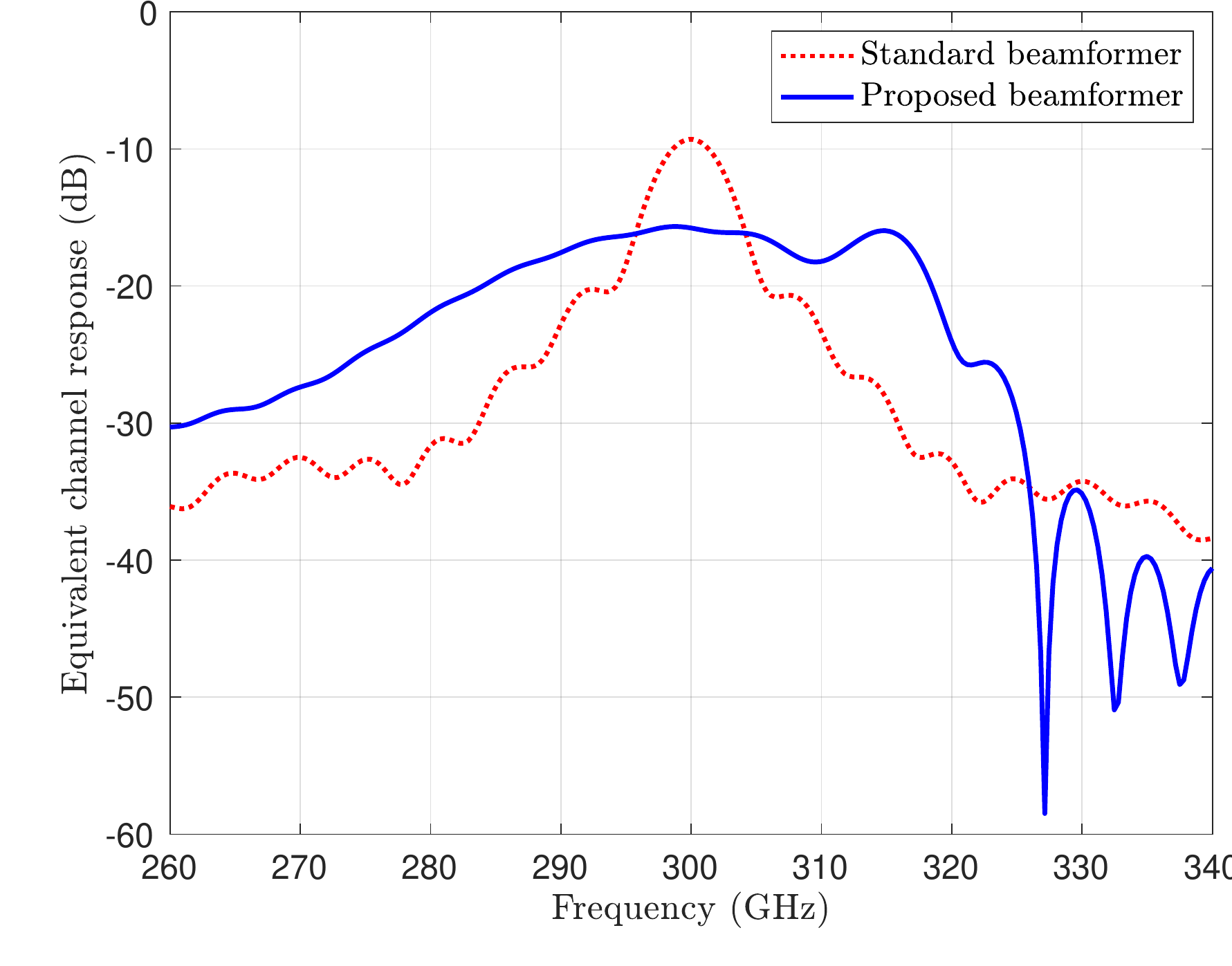}\label{fig:inside_freq_resp}}
\caption{An example of the discrete chirp-based phase profile designed with InFocus is shown in Fig. \ref{fig:inside_psi_des} for $\Delta=0.5\, \mathrm{mm}$. Here, $B=40\, \mathrm{GHz}$, $\fc=300\, \mathrm{GHz}$, $\ell=15\, \mathrm{cm}$, $\gamma= 15 \degree$ and $R=10\, \mathrm{cm}$. For such parameters, $\mathbb{P}_{\mathrm{RX}} \in \mathcal{S}$. The equivalent channel gains achieved with InFocus and the standard beamformer are shown in Fig. \ref{fig:inside_freq_resp}.}
\end{figure}

 We now describe the chirp signal designed with the stationary phase method. In this method, the second derivative of $\psi(u)$ is proportional to $b^2(u)$. From \eqref{eq:gtild_in_2}, we observe that the chirp signal with a phase profile of $\psi(u)$ starts at $u=\ell \, \mathrm{cos} \gamma$ and ends at $u=\sqrt{\ell^2\mathrm{cos}^2 \gamma+(R+ \ell \, \mathrm{sin} \gamma)^2}$. We consider a near field scenario with $\ell=15\, \mathrm{cm}$, $\gamma= 15^{\degree}$, and $R=10\, \mathrm{cm}$. In this scenario, $\mathbb{P}_{\mathrm{RX}} \in \mathcal{S}$ and the corresponding amplitude modulation function is shown in Fig. \ref{fig:inside_amplitude}. We plot the instantaneous frequency of the chirp in Fig. \ref{fig:inside_freq_u} and the phase function $\psi(u)$ in Fig. \ref{fig:inside_phase_u}. We use a bandwidth of $40\, \mathrm{GHz}$ around $\fc= 300\, \mathrm{GHz}$ to derive the phase function in Fig. \ref{fig:inside_phase_u}. The 2D-phase profile $\psi_{\mathrm{des}}(x,y)$ associated with the designed chirp is computed using \eqref{eq:map_1d_2d} and is shown in Fig. \ref{fig:inside_psi_des}. When the phase profile in Fig. \ref{fig:inside_phase_applied} is applied at the TX, it can be observed from  Fig. \ref{fig:inside_freq_resp} that the equivalent SISO channel $g(f)$ is approximately constant over the desired frequency band.
\section{Achievable rate with InFocus} \label{sec:simulations}
\par In this section, we describe the simulation setup and explain how to compute the achievable rate with the equivalent SISO channel obtained after beamforming. Then, we study the rate achieved with InFocus and standard beamforming as a function of the RX location, the operating bandwidth and the resolution of phase shifters.
\par We consider a near field system in Fig. \ref{fig:LOS} with a circular planar array of radius $R=10\, \mathrm{cm}$ at the TX and a single antenna RX. We assume that the location of the RX, equivalently the channel $h(x,y,f)$, is known to the TX. We use a carrier frequency of $\fc=300\, \mathrm{GHz}$. The spacing between the antenna elements at the TX is $\Delta=\lambda_{\mathrm{c}}/2$, which is $0.5\, \mathrm{mm}$.  For the half-wavelength spaced circular planar array at the TX, the number of antennas is $\Ntx=124,980$. The total power transmitted by the TX array is set to $1\, \mathrm{mW}$. We use $q$ to denote the resolution of the RF phase shifters at the TX. The phase shift alphabet has $2^q$ uniformly spaced angles in $[0, 2\pi)$ defined by the set $\mathbb{Q}_q=\{0, 2\pi /2^q, 4\pi /2^q, \cdots, 2 \pi (2^q-1)/ 2^q\}$. The phase profiles derived with standard beamforming and InFocus take continuous values in $[0, 2\pi)$. The entries of these phase profiles are quantized to the nearest element in the set $\mathbb{Q}_q$ and the quantized phase shifts are applied to the TX array for beamforming. The equivalent SISO channels with InFocus and standard beamforming are calculated using \eqref{eq:eqsiso}.
\par The achievable rate corresponding to an equivalent SISIO channel is computed using the procedure in \cite{noise_psd}. In this procedure, the wideband channel over $f\in [\fc-B/2, \fc+B/2]$ is first split into $\Nsub$ sub-bands. We define $\{f_k\}^{\Nsub}_{k=1}$ as $\Nsub$ equally spaced frequencies in $[\fc-B/2, \fc+B/2]$. The power allocated over the $k^{\mathrm{th}}$ sub-band is defined as $\eta_k$ and the corresponding power density is $\eta_k \Nsub/B$. The total transmit power is defined as $\eta=\sum^{\Nsub}_{k=1} \eta_k$. The equivalent channel gain for a sub-band centered at $f$ is $|g(f)|^2$ where $g(f)$ is defined in \eqref{eq:eqsiso}. We use the frequency selective thermal noise model discussed in \cite{noise_psd}. We define $n(f)$ as the noise power spectral density (PSD), $\hslash$ as the Planck's constant, $k_{\mathrm{btz}}$ as the Boltzmann's constant, and $T$ as the system temperature. The noise PSD is then \cite{noise_psd}
\begin{equation}
n(f)=\frac{ \hslash f}{\mathrm{exp}\left( \frac{\hslash f}{k_{\mathrm{btz}} T}\right)-1}.
\end{equation}
The achievable rate corresponding to the equivalent SISO system is expressed as \cite{noise_psd}
\begin{equation}
R=\frac{B}{\Nsub} \sum_{k=1}^{\Nsub} \mathrm{log}_2\left(1+ \frac{\eta_k \Nsub |g(f_k)|^2}{n(f_k)B}\right).
\end{equation}
In our simulations, we use $\Nsub=512$, $T=290\, \mathrm{Kelvin}$, $\hslash=6.625 \times 10^{-34}\, \mathrm{Joule}.\mathrm{sec}$ and $k_{\mathrm{btz}}=1.3806\times 10^{-23}\, \mathrm{Joule}/\mathrm{Kelvin}$. A transmit power of $\eta=1\, \mathrm{mW}$ is distributed across different sub-bands using water filling-based power allocation to maximize the rate. As the equivalent SISO channel with standard beamforming has a large gain at frequencies close to $\fc$, the water filling method allocates higher power around $\fc$ when compared to other frequencies. The equivalent SISO channel with InFocus, however, has a ``constant'' gain over the desired frequency band. In this case, the water filling technique achieves ``uniform'' power allocation over the desired bandwidth. 
\par
\begin{figure}[h!]
\vspace{-3mm}
\centering
\subfloat[Achievable rate with the transceiver distance $\ell$.]{\includegraphics[trim=1.5cm 6.5cm 2cm 7.5cm,clip=true,width=0.45 \textwidth]{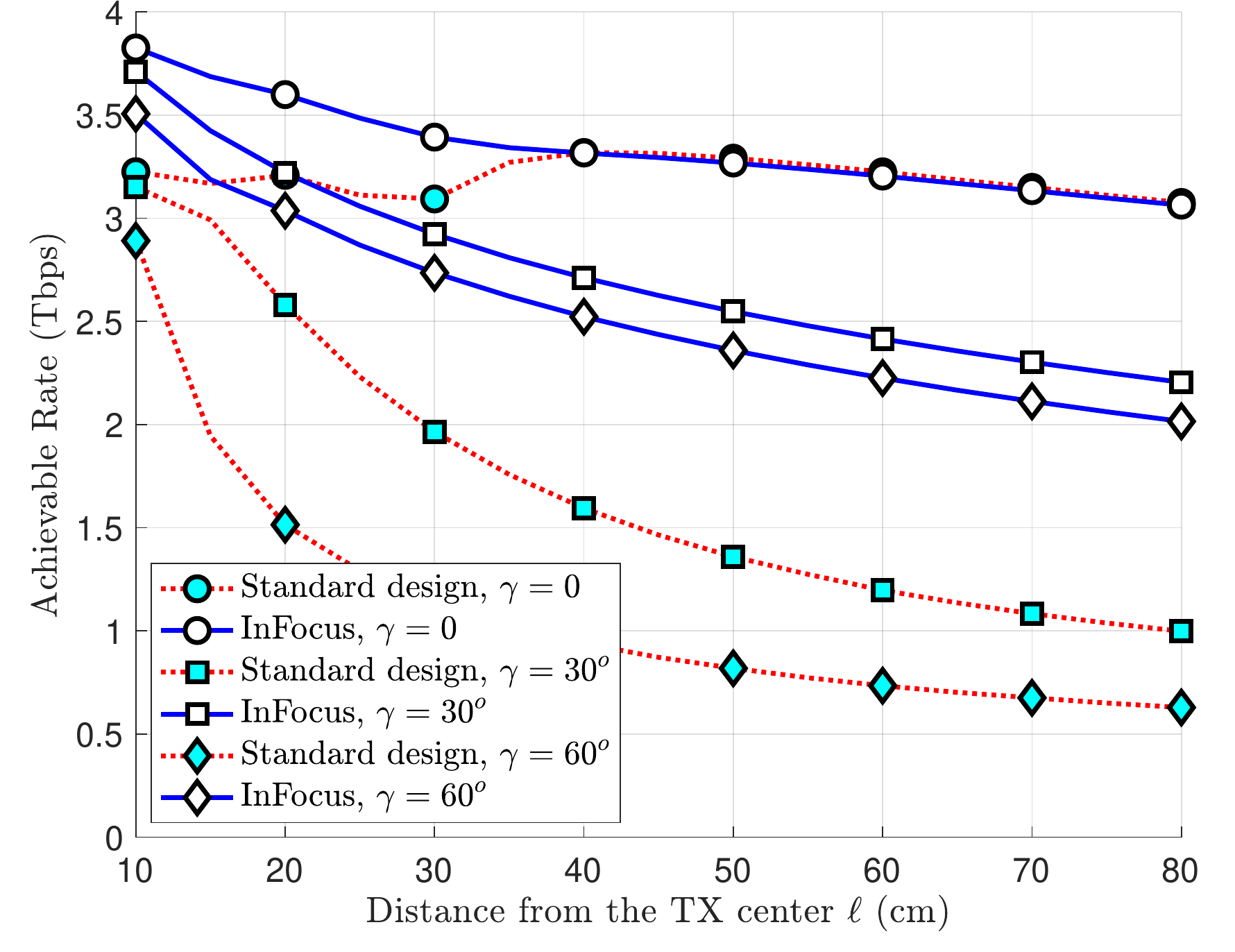}\label{fig:rate_dist}}
\subfloat[Achievable rate with angle $\gamma$.]{\includegraphics[trim=1.5cm 6.5cm 2cm 7.5cm,clip=true,width=0.45 \textwidth]{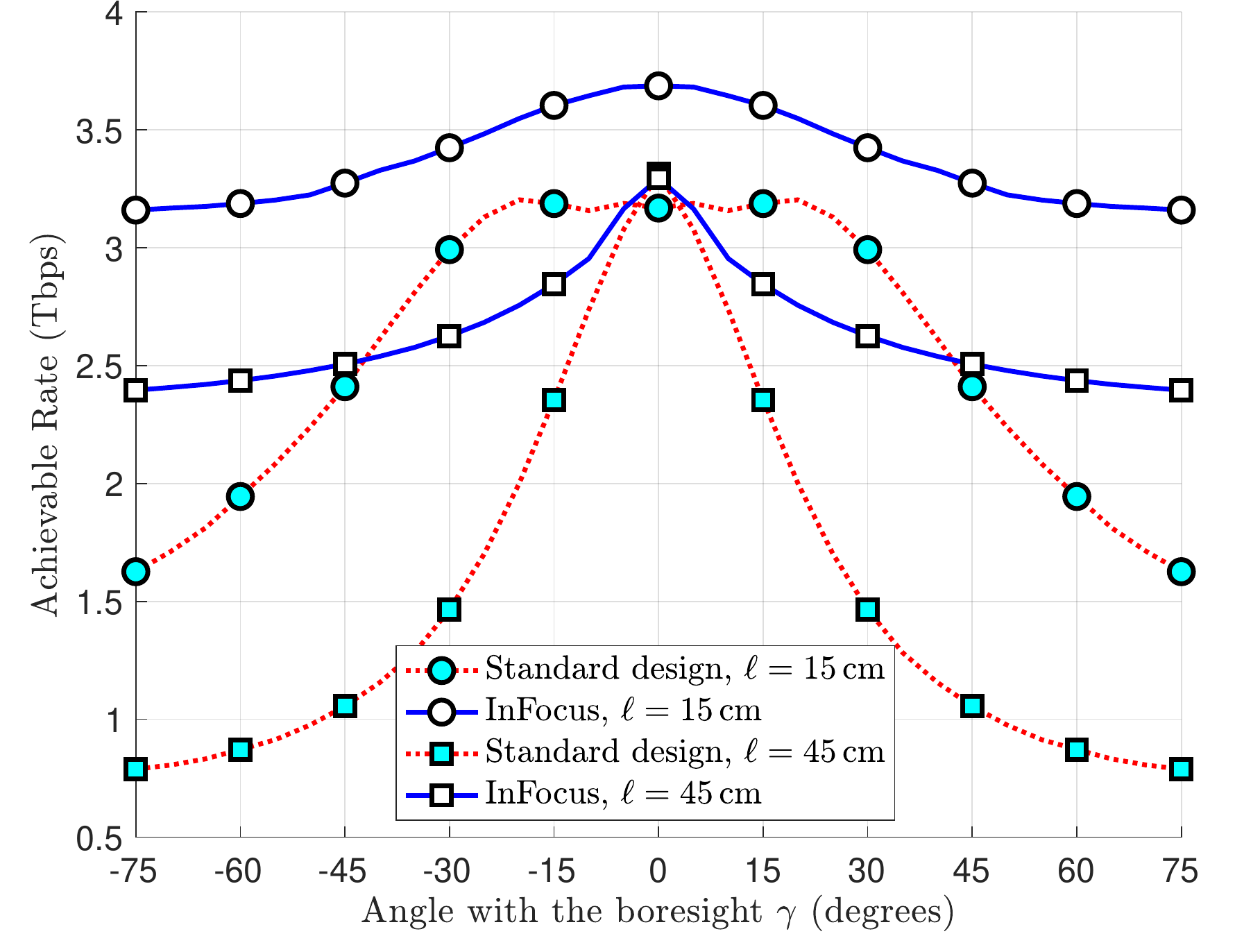}\label{fig:rate_angle}}
\caption{We observe from Fig. \ref{fig:rate_dist} that InFocus achieves a higher rate than standard beamforming for $\gamma=30  \degree$ and $60  \degree$. In a boresight setting, standard beamforming achieves the same rate as InFocus for $\ell\geq 40\, \mathrm{cm}$. This is because beam squint, the analogue of misfocus in the far field, does not occur along the boresight direction. Fig. \ref{fig:rate_angle} shows that InFocus performs better than standard beamforming for all angles in $[-75 \degree, 75 \degree]$.}
\vspace{-2mm}
\end{figure}

 We now investigate the rate achieved with InFocus as a function of the RX location. We use $B= 40 \, \mathrm{GHz}$ and set the resolution of the phase shifters to $q=2$ bits. In a boresight scenario where $\gamma=0 \degree$, we observe from Fig. \ref{fig:rate_dist} that standard beamforming results in a lower rate than InFocus for $\ell\leq 35\, \mathrm{cm}$ due to the misfocus effect in the near field regime. The rate with both the techniques, however, is the same for $\ell \geq 40\, \mathrm{cm}$ due to the reduced misfocus effect at larger distances. For large distances, the misfocus effect is same as the beam squint effect which does not occur in the boresight direction \cite{mingming_gc}. For angles $\gamma=30 \degree$ and $60 \degree$, the rate with InFocus is higher than the one achieved by the standard design at all distances. The superior performance achieved with InFocus at large distances makes it promising for misfocus and beam squint robust transmission in wideband systems. From Fig. \ref{fig:rate_angle}, we note that standard design performs poor for a large $\gamma$. The phase profile constructed with InFocus allows an efficient use of the operating bandwidth than the standard design for all $\gamma \in [-75 \degree, 75 \degree]$.
\par
\begin{figure}[h!]
\vspace{-3mm}
\centering
\subfloat[$20\, \mathrm{log}_{10}|g(f)|$ with frequency.]{\includegraphics[trim=1.5cm 6.5cm 2cm 7.5cm,clip=true,width=0.45 \textwidth]{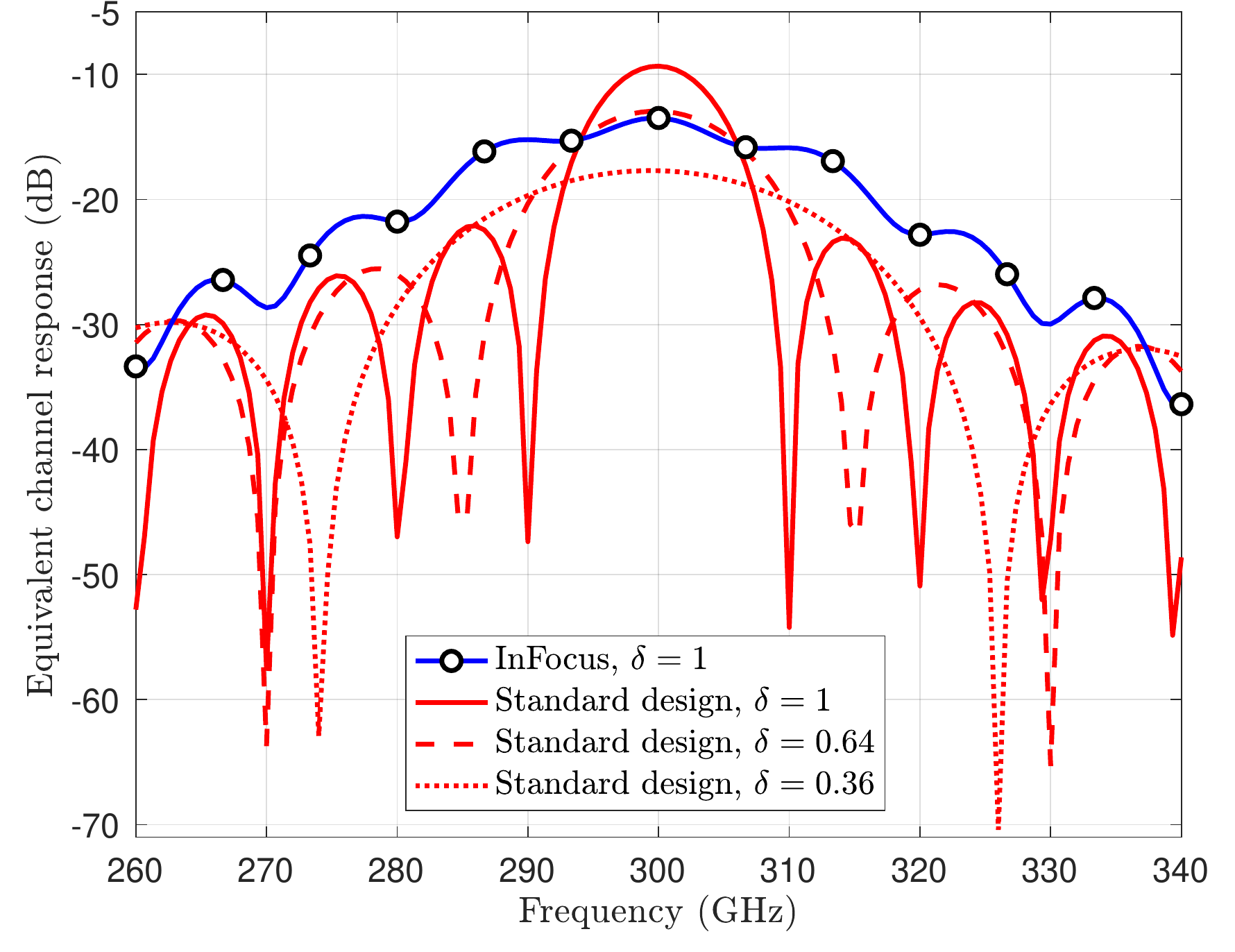}\label{fig:equiv_chan_thin}}
\subfloat[Rate with the fraction of active antennas $\delta$.]{\includegraphics[trim=1.5cm 6.5cm 2cm 7.5cm,clip=true,width=0.45 \textwidth]{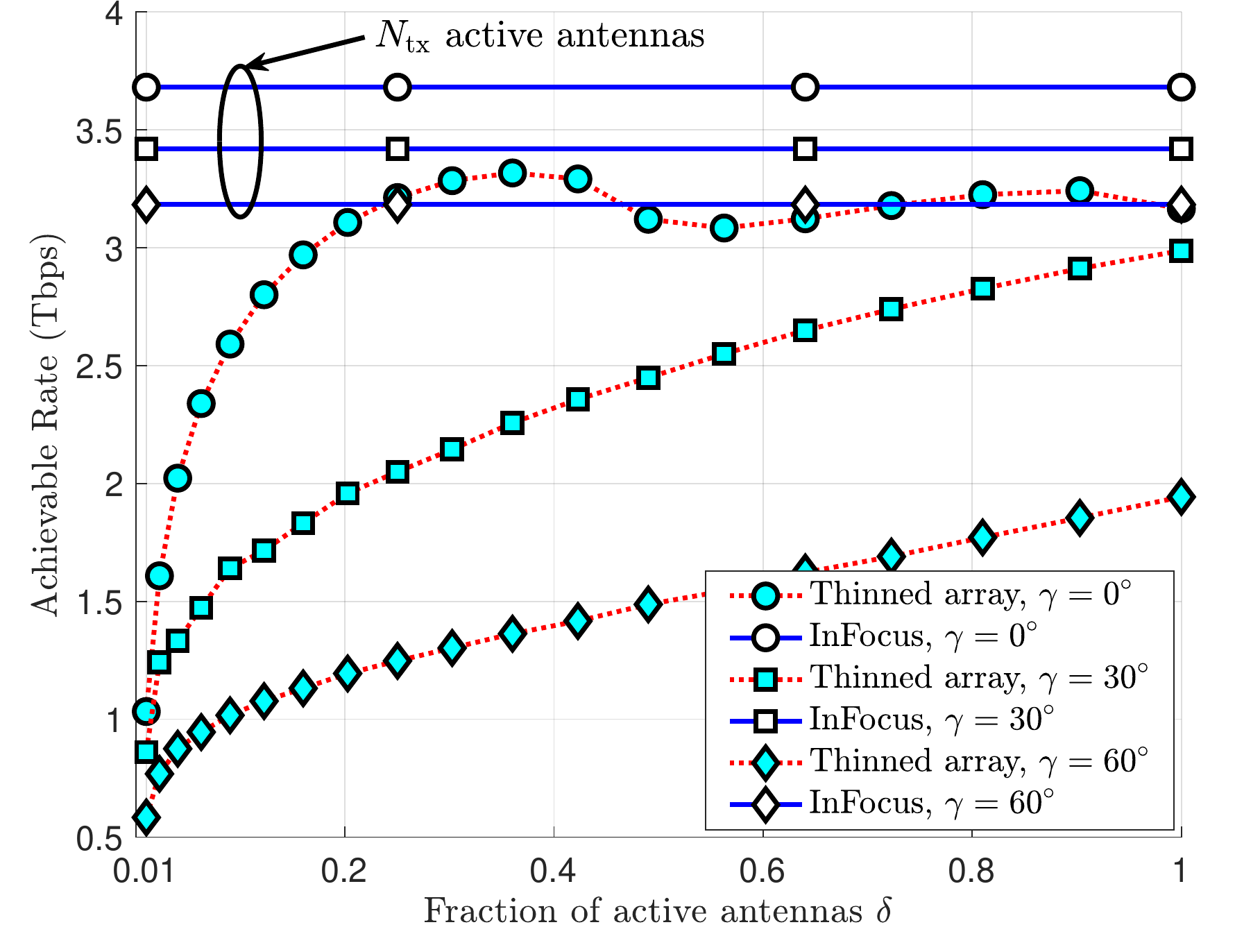}\label{fig:rate_thin}}
\caption{In Fig. \ref{fig:equiv_chan_thin}, we consider a boresight scenario with $\ell=15\, \mathrm{cm}$, $B=40\, \mathrm{GHz}$ and $q=2$ bits. Standard beamformer achieves a reasonable gain over the desired bandwidth when the fraction of active antennas is $\delta=0.36$. The gain, however, is less than that achieved with InFocus under the per-antenna power constraint. Fig. \ref{fig:rate_thin} shows that InFocus achieves a higher rate than standard beamforming using a thinned array for any $\delta$.}
\vspace{-2mm}
\end{figure}

 Now, we discuss a performance benchmark based on standard beamforming with a thinned array. Thinning is a technique where a set of antennas in an array are turned off to reduce the effective aperture. We study standard beamforming with a radially thinned array. In this configuration, the antennas outside a disc of radius $r$ are switched off and the standard phase profile $\phi_{\std}(x,y)$ is applied for the active antennas. We define $\delta$ as the fraction of antennas that are active in the thinned array. Here, $\delta \approx r^2/R^2$. Under the per-antenna power constraint, the magnitude of the beamforming weights at the active antennas is $1/\sqrt{\Ntx}$, and the norm of the beamformer is $\delta$. We observe that a smaller $\delta$ corresponds to a smaller aperture. Although reducing the aperture mitigates misfocus, it results in a lower beamforming gain at $\fc$ as shown in Fig. \ref{fig:equiv_chan_thin}. The poor gain when compared to the full aperture scenario is due to a lower total transmit power under the per-antenna power constraint. We observe from Fig. \ref{fig:rate_thin} that the thinned array-based approach results in a lower rate than InFocus for any $\delta$. InFocus performs better as it activates all the antennas while achieving robustness to misfocus.
\par
\begin{figure}[h!]
\vspace{-3mm}
\centering
\subfloat[Rate with the operating bandwidth $B$.]{\includegraphics[trim=1.5cm 6.5cm 2cm 7.5cm,clip=true,width=0.45 \textwidth]{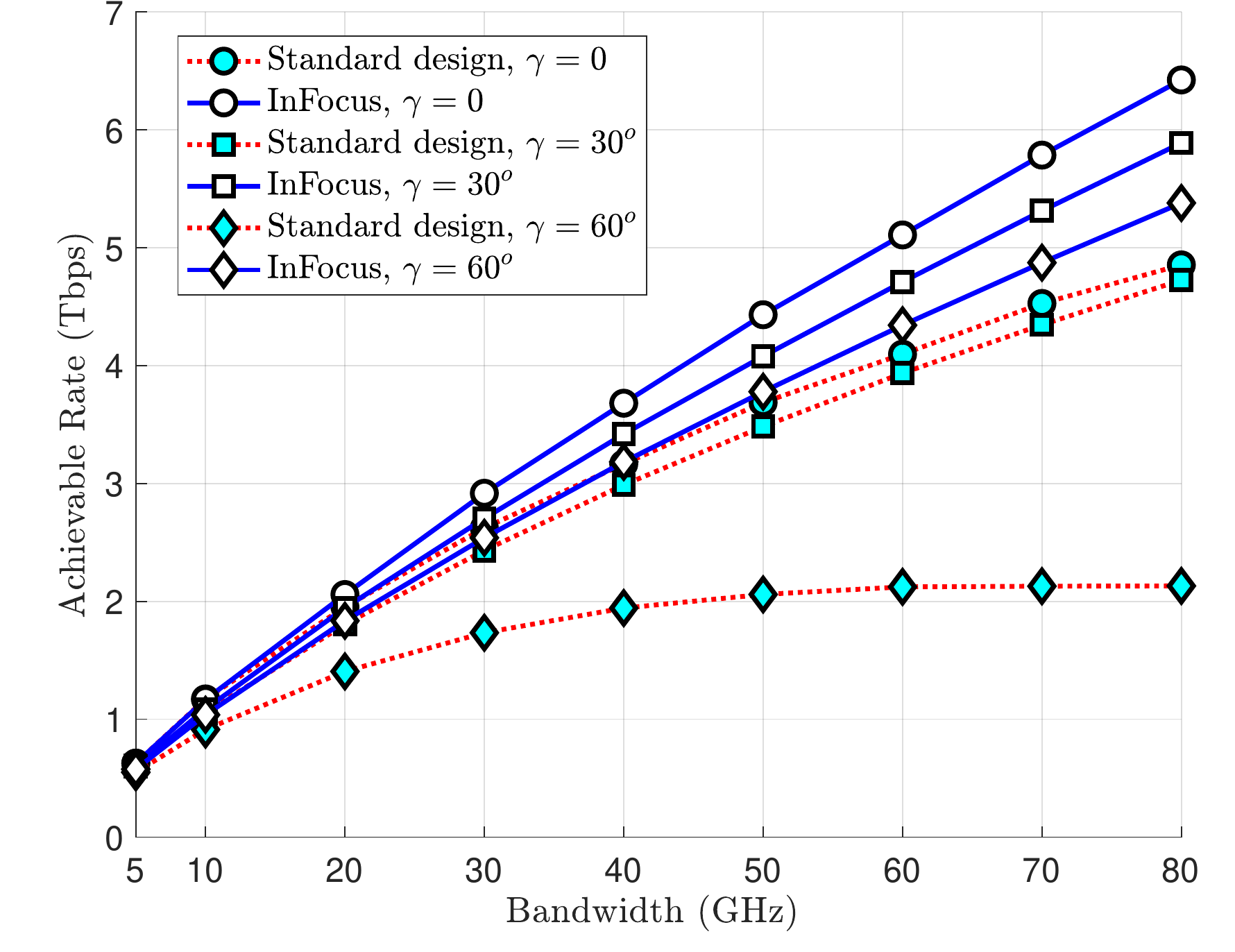}\label{fig:rate_bandwidth}}
\subfloat[Rate with the resolution of phase shifters.]{\includegraphics[trim=1.5cm 6.5cm 2cm 7.5cm,clip=true,width=0.45 \textwidth]{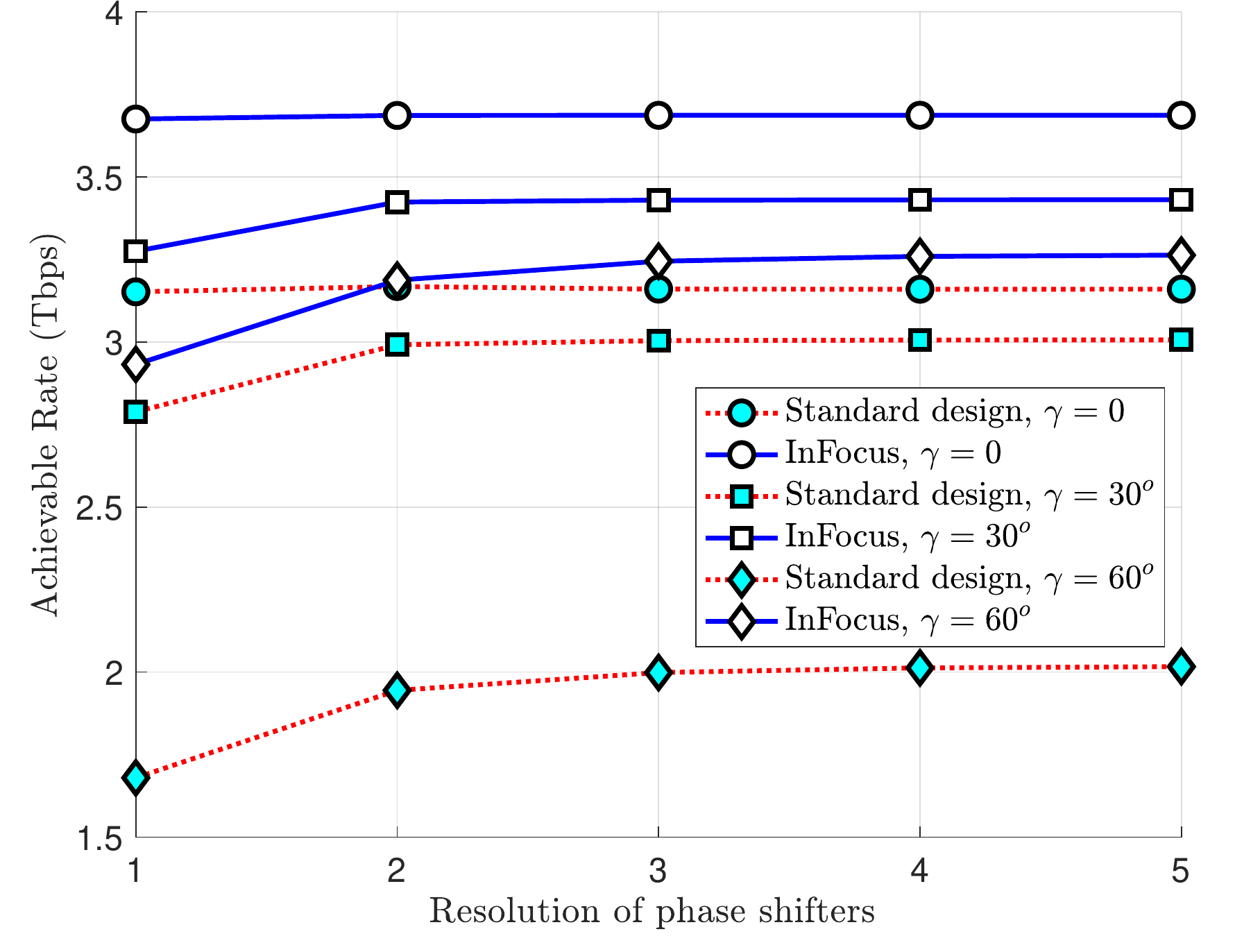}\label{fig:rate_vs_psresln}}
\caption{In this example, we use $\ell=15\, \mathrm{cm}$ and $\gamma\in \{0 \degree, 30 \degree, 60 \degree\}$. InFocus adapts the phase profile, equivalently the beam, according to the operating bandwidth and performs better than the standard design as seen in Fig. \ref{fig:rate_bandwidth}. From Fig. \ref{fig:rate_vs_psresln}, we observe that InFocus achieves a higher rate even under coarse phase quantization.}
\end{figure}

 We would like to highlight that InFocus adapts its beam according to the operating bandwidth. For example, the phase profile corresponding to \eqref{eq:psi_opt1D} linearly increases with the bandwidth $B$. Standard beamforming, however, designs a beam that is agnostic to the bandwidth and suffers from the misfocus effect. In Fig. \ref{fig:rate_bandwidth}, we plot the achievable rate as a function of the operating bandwidth for a near field system with $q=2$ bit phase shifters at the TX  and $\ell=15\, \mathrm{cm}$. As the misfocus effect is prominent in systems operating over wide bandwidths, the standard beamforming method performs poor at such bandwidths. We observe from Fig. \ref{fig:rate_vs_psresln} that two-bit phase shifters are sufficient to achieve a reasonable rate. Furthermore, InFocus-based beams performs better than the standard design even with one- or two-bit phased arrays. We believe that InFocus marks an important step towards achieving high speed data transmission using phase shifter-based arrays.
\vspace{-3mm}
\section{Conclusions and future work}\label{sec:concl_future}
\par Near field beams focus the RF signals in a spatial region instead of a direction. The use of massive phased arrays to realize near field beams, however, results in a misfocus effect in wideband systems with standard center frequency-based beamforming. Such an issue arises because the region of focus changes with the frequency of the RF signal. In this paper, we studied the misfocus effect for receivers in the boresight direction of the transmit array. Furthermore, we proposed a spatial FMCW chirp-based beam that achieves robustness to misfocus in a boresight setting. We also extended our design to scenarios where the receiver does not lie along the boresight direction. Our extension used the stationary phase method which resulted in a non-linear spatial FMCW chirp for robustness to misfocus. Beamforming with the proposed design achieves a uniform gain over a wide bandwidth and a higher rate than the standard design. 
\par InFocus solves an important problem in near field LoS systems under certain assumptions. These assumptions include perfect polarization alignment between the TX and the RX, the use of a single antenna RX, perfect channel state information, and the absence of reflectors in the propagation environment. In future, we will relax these assumptions to develop new techniques for misfocus compensation in richer propagation scenarios.
\section*{Appendix}
\subsection{Equivalent channel with standard beamforming}
We simplify $g_{\mathrm{a}, \std} (f)$, an approximation of the equivalent SISO channel with standard beamforming. As the integrand in \eqref{eq:ga_f_std1} is independent of the angle $\theta$, we can integrate over this angle to write 
\begin{equation}
\label{eq:app_gaf_1}
g_{\mathrm{a}, \std} (f)=\frac{c}{\sqrt{\pi} R f \Delta} \int_{r=0}^R\frac{1}{\sqrt{r^2+\ell^2}}e^{-\imj \frac{2 \pi (f-\fc)  \sqrt{r^2+\ell^2}}{c} } r \dr. 
\end{equation}
We now use the definition of $\omega$ in \eqref{eq:defn_omega} and $s=\sqrt{r^2+\ell^2}$. The integral in \eqref{eq:app_gaf_1} is then
\begin{equation}
\label{eq:app_gaf_2}
g_{\mathrm{a}, \std} (f)=\frac{c}{\sqrt{\pi} R f \Delta}\int_{s=\ell}^{\sqrt{R^2+\ell^2}}e^{-\imj \omega s} \ds.
\end{equation}
The integral in \eqref{eq:app_gaf_2} is the Fourier transform of a rectangular function which is $1$ for $s \in [\ell, \sqrt{R^2+\ell^2}]$. The Fourier transform of this function can be expressed in terms of the $\mathrm{sinc}$ function defined as $\mathrm{sinc}(x)= \mathrm{sin}(x)/ x$. We note that 
\begin{equation}
\label{eq:rect_sinc}
\int_{s=a}^{b}e^{-\imj \omega s} \ds=(b-a)e^{-\imj \frac{\omega (a+b)}{2}}\mathrm{sinc}\left(\frac{\omega(b-a)}{2}\right). 
\end{equation}
We put these observations in \eqref{eq:app_gaf_2} to obtain the result in \eqref{eq:ga_f_std3}.
\subsection{Chirp amplitude modulation function $a(u)$}
A closed form expression for $a(u)=(\Omega_{\mathrm{max}}(p)-\Omega_{\mathrm{min}}(p))/(2\pi)$ can be computed using geometry. We observe from Fig. \ref{fig:outside_topview} that a circle of radius $p$ around $\mathbb{P}_{\mathrm{RX}}$ intersects $\mathcal{S}$ at an arc. The angle made by this arc at $\mathbb{P}_{\mathrm{RX}}$ is $\Omega_{\mathrm{max}}(p)-\Omega_{\mathrm{min}}(p)$. In this section, we first compute this angle in terms of $p$ and then write $a(u)$ as a function of $u=\sqrt{p^2+ \ell^2 \mathrm{cos}^2 \gamma}$.
\par The angle $\Omega_{\mathrm{max}}(p)$ can be found from the triangle $\mathsf{O} \mathbb{P}_{\mathrm{RX}} \mathsf{Q}$ shown in Fig. \ref{fig:outside_topview}. The lengths of the sides of this triangle are $\overline{\mathsf{O} \mathbb{P}_{\mathrm{RX}}}= \ell \, \mathrm{sin} \gamma$, $\overline{\mathbb{P}_{\mathrm{RX}} \mathsf{Q}}=p$ and $\overline{\mathsf{OQ}}=R$. The cosine of the angle at the vertex $\mathbb{P}_{\mathrm{RX}}$ is then \cite{cosinerule}
\begin{equation}
\label{eq:cosine_rule}
\mathrm{cos} \, \Omega_{\mathrm{max}}(p)= \frac{p^2+ \ell^2 \, \mathrm{sin}^2 \gamma -R^2}{2p \ell\, \mathrm{sin} \gamma}.
\end{equation}
We notice from Fig. \ref{fig:outside_topview} that $\Omega_{\mathrm{min}}(p)=-\Omega_{\mathrm{max}}(p)$ by symmetry. The amplitude modulation function is then $a(u)=\Omega_{\mathrm{max}}(p)/ \pi$. Putting this observation together with $p=\sqrt{u^2- \ell^2 \mathrm{cos}^2 \gamma}$ and the result in \eqref{eq:cosine_rule}, we can express $a(u)$ as
\begin{equation}
\label{eq:explicit_amp}
a(u)=\frac{1}{\pi}\mathrm{cos}^{-1}\left( \frac{u^2- \ell^2 \mathrm{cos}^2 \gamma+ \ell^2 \, \mathrm{sin}^2 \gamma -R^2}{2 \ell\, \mathrm{sin} \gamma \sqrt{u^2- \ell^2 \mathrm{cos}^2 \gamma}}\right).
\end{equation}
An example of the amplitude modulation function in \eqref{eq:explicit_amp} is shown in Fig. \ref{fig:outside_amplitude}.
\bibliographystyle{IEEEtran}
\bibliography{refs}
\end{document}